\documentclass[11pt, letterpaper]{article}
\usepackage{amsfonts,amsmath,amsopn,amssymb,amsthm,bbm,latexsym,mathrsfs,pstricks,verbatim}

\let\n\noindent

\font\small=cmr8

\font\tenmsy=msbm10
\font\sevenmsy=msbm10 at 7pt
\font\fivemsy=msbm10 at 5pt
\newfam\msyfam 
\textfont\msyfam=\tenmsy
\scriptfont\msyfam=\sevenmsy
\scriptscriptfont\msyfam=\fivemsy
\def\blackB{\fam\msyfam\tenmsy}
\def\Z{{\blackB Z}}

\def\NN{{\blackB N}}

\let\e\epsilon
\let\s\sigma

\let\R\rangle
\let\l\left
\let\r\right

\def\z{{\cal Z}}
\let\lf\lfloor
\let\rf\rfloor
\let\lc\lceil
\let\rc\rceil

\def\mb{{\bar m}}
\let\La\Lambda

\let\la\lambda
\let\La\Lambda

\def\y{{\infty}}

\let\Rw\Rightarrow
\def\l{{\left}}
\def\r{{\right}}

\def\rw{\rightarrow}
\def\lrw{\leftrightarrow}

\def\R{\rangle}
\def\frac#1#2{{#1 \over #2}}

\def\w{{w^*}}
\def\wb{{\bar w}}
\let\ka\kappa

\font\small=cmr8

\begin{document}

\vskip18pt

\title{\vskip60pt {\bf New path description for the  ${\cal M} (k+1,2k+3)$ \\ models and the dual $\z_k$ graded parafermions}}

\vskip18pt


\smallskip
\author{ \bf{P. Jacob and P.
Mathieu}\thanks{patrick.jacob@durham.ac.uk,
pmathieu@phy.ulaval.ca.  This work is supported by EPSRC (PJ) NSERC (PM).} \\ 
\\
Department of Mathematical Sciences, \\University of Durham, Durham, DH1 3LE, UK\\
and\\
D\'epartement de physique, de g\'enie physique et d'optique,\\
Universit\'e Laval,
Qu\'ebec, Canada, G1K 7P4.
}

\vskip .2in
\bigskip
\date{August 2007}

\maketitle

\let\Rw\Rightarrow
\let\rw\rightarrow
\let\l\left
\let\r\right
\let\s\sigma
\let\ka\kappa
\let\de\delta

\def\M{{\cal M}}
\def\SM{{\cal SM}}

 \let\g\gamma

\def\lp{{\bar {{\rm P}}^{[k-\frac12]}}}

\def\LP{{ {\rm P}^*}^{[k]}}


\vskip0.3cm
\centerline{{\bf ABSTRACT}}
\vskip18pt
We present a new path description for the states of the  non-unitary ${\cal M} (k+1,2k+3)$ models. This description differs from the one induced by the Forrester-Baxter solution, in terms of configuration sums, of their restricted-solid-on-solid model. The proposed path representation is actually very similar to the one underlying  the unitary minimal models $\M(k+1,k+2)$, with an analogous 
 Fermi-gas  interpretation. This interpretation leads to fermionic expressions for the finitized ${\cal M} (k+1,2k+3)$ characters, whose infinite-length limit represent  new  fermionic characters for the
   irreducible modules. The ${\cal M} (k+1,2k+3)$ models are also shown to be related to the $\z_k$ graded parafermions via a ($q\lrw q^{-1}$) duality transformation.



\section{Introduction}

A fermionic character formula in conformal field theory reflects a description of the space of states in terms of quasi-particles subject to restriction rules, a description which directly  accounts
for their manifest positivity \cite{KKMMa,KKMMb}. Originally, an important source of inspiration for many conjectured fermionic characters comes from the representation of the conformal model under consideration as the scaling limit of a spin chain or a statistical model.  In the goal of providing intrinsic conformal-field-theoretical derivation of these fermionic formulae, such representations 
do provide important hints. 

A particular example we have in mind is the representation of the unitary minimal models $\M(k+1,k+2)$ in terms of  the Andrews-Baxter-Forrester restricted-solid-on-solid (RSOS) model \cite{ABF}.\footnote{The identification of these RSOS models with the unitary minimal models has first been pointed out in \cite{Huse}. The parameter $r$ defining the different models in \cite{ABF} is $k+2$.} The solution   
via the corner-transfer-matrix method leads to the representation of every state, in a finitized version of an appropriate  irreducible Virasoro module, in terms of a configuration. Instead of the configuration itself, we can consider its contour, which defines an integer-lattice path.\footnote{
Let us recall that the configurations are those of  height variables $\s_i\in \{0,1,\cdots ,k\}$, where $i$ ranges from 0 to $L$, the finitization parameter. Adjacent heights  are subject to the restriction $|\s_i-\s_{i+1}|=1$ and the two boundary values $\s_0$ and $ \s_L$ are fixed. The configuration-sum then takes the form
\begin{equation}
X_{\s_0,\s_L}(q)= \sum_{\substack{ \s_1,\cdots , \s_{L-1}=0\\|\s_i-\s_{i+1}|=1}}^k  q^{\sum_{i=1}^{L-1} w(i)}\;,  
\end{equation}
where the weight function $w(i)$ is 0 if $i$ is a local extremum and $i/2$ otherwise (cf. below). If we plot all the vertices $(i,\s_i)$ of a given configuration and link adjacent vertices by an edge, we obtain a path  we call  a RSOS path \cite{OleJS,FLPW,FW}. In the following, we trade the pair $(i,\s_i)$ for $(x,y)$.
}
 This leads then to  a RSOS path representation of every basis state in the conformal theory.
As demonstrated in \cite{OleJS}, this path description embodies a natural one-dimensional Fermi-gas description that leads  to a fermionic expression of the characters \cite{OleJS}.  The quasi-particle description underlying this  path description of the unitary models should  capture essential aspects of the yet to be framed conformal-field-theoretical quasi-particle formalism. 

Along that line, we can ask to which extend such path descriptions are known for the other minimal models.
There is actually  a lattice path description of the space of states for all minimal models $\M(p',p)$ $(p>p')$ via the generalization of the Andrews-Baxter-Forrester  RSOS models due to  Forrester and Baxter \cite{FB}.\footnote{References \cite{Rig,Nak} are early studies of the relation between these generalized RSOS models and the  non-unitary minimal models.}
Again, these paths are  simply the contour description of the Forrester-Baxter RSOS configurations. More explicitly, a RSOS path is a sequence of North-East (NE) and South-East (SE) edges lying within the strip $x\geq 0$ and $0\leq y\leq p-2$ of the integer $(x,y)$ lattice. The weight of a path, related to the conformal dimension of the corresponding state,  is the sum of the weight of all its vertices.  The weighted sum of all paths of length $L$ (with specified boundary conditions) provides thus a finitized expression of the character of the irreducible Virasoro modules (specified by the path  boundary conditions). The full characters are recovered in the limit $L\rw \y$.


For the Forrester-Baxter models, the weight function turns out to be a rather complicated expression. Indeed,
 each vertex -- of coordinates $(x,y)$  -- contributes $x/2$ to the weight except at local extrema, where the weight is rather $\pm x\lfloor ( y_\pm)(p-p')/p\rfloor$ (with the $+$ for a minimum and the $-$ for a maximum), with $y_-=y$ and $y_+=y+2$ \cite{FB}.
The relative complexity of the weight function for the generic case makes this path representation  hard to manage combinatorially \cite{FLPW,FW,Wel}.  

When considered from the point of view of the general case,  the  $\M(k+1,k+2)$ models present a radical simplification:
the local extrema have zero weight so that only the straight-up and the  straight-down segments  contribute to the weight of the path. And every weight-contributing vertex contributes simply to $x/2$ (independently of their height).\footnote{There is also a simple  path description for the  $\M(2,2k+1)$ models but it is not formulated directly in terms of RSOS paths of \cite{FB}; it is expressed  in terms of a different type of paths, the so-called Bressoud paths \cite{BreL}, to be  introduced below.}

Let us stick to  the unitary case for a moment. 
Note first  that if all vertices had contributed to the same value $x/2$, the path generating function would have been trivial. Now, that some vertices do not contribute to the weight suggests a very simple and natural way of defining dual paths in terms of a dual definition of the weight. This dual weight function is defined by setting the weight of the vertices at  local extrema equal to $x/2$  and   that  of all the others equal to 0. In the path generating function, where the weight is coded in the exponent of a formal variable $q$, the duality transformation amounts to interchanging $q$ by $1/q$ \cite{ABF}. Somewhat remarkably, in this very sense, the finitized $\M(k+1,k+2)$ models are dual to a finitized version of the $\z_k$ parafermionic theory \cite{ZF} -- see e.g., \cite{BMlmp,OleJS, Kyoto,FWa}.
In other words, the dual to the $\M(k+1,k+2)$ quasi-particles are of the  parafermionic type (following the parafermionic path interpretation of \cite{Path}).


Quite surprisingly, we have found  that there exists a simple deformation of the   $\M(k+1,k+2)$ path  description that describes  the $\M(k+1,2k+3)$ models.
%
These new paths share the key simplifying  property of the  $\M(k+1,k+2)$ ones which is that only the straight-up and  straight-down segments need to be considered for the evaluation of the weight.
However, they differ from their unitary relatives in that they are defined on a lattice with {\it half-integer spacing}. A constraint on the peaks, which  are forced to have integer coordinates, make these new paths different from the  rescaled version of the ones pertaining to the unitary case.

These paths are described more precisely in the next section. However, we  stress at once  that  they differ radically  from the  standard RSOS paths appropriate to this class of non-unitary models \cite{FB,FW}. In addition, they are presented in a somewhat ad hoc fashion, without any underlying statistical model representative.


As pointed out previously,  the  $\M(k+1,k+2)$ characters  are related to the $\z_k$ parafermionic models by a  duality transformation.  The instrumental properties of the $\M(k+1,k+2)$  path weight function underlying this result have been identified, which are that not all vertices contribute and the  contributing ones do it  uniformly as $x/2$.   But these features are also present for the new $\M(k+1,2k+3)$ paths. Their dual version are then expected to be well-defined conformal theories of the  parafermionic type. And this is indeed so: the  characters of the $\M(k+1,2k+3)$ models are shown to be dual to those of the $\z_k$ graded parafermionic models (introduced in  \cite{CRS} and whose characters have been obtained in \cite{JM, BFJM, JMO}). This is an  unexpected duality.

The structural similarity between the path description of the $\M(k+1,2k+3)$ and the $\M(k+1,k+2)$ models allows us to generalize rather immediately the combinatorial  analysis of the latter models by Warnaar \cite{OleJS,OleJSb}  and derive the generating function for the former paths in the form of a positive multiple sum. This happens to generate novel fermionic expressions.



The article is organized as follows. The new paths are introduced in Sect. 2. Their generating function is calculated in Sect. 3 for the special class of paths that pertains to the vacuum module.  The resulting generating function is the candidate finitized character of the vacuum module for the $\M(k+1,2k+3)$ in fermionic form. Given that this differs from other previously obtained fermionic forms for these characters, this identification needs to be substantiated. This has been done first by $q$-expanding  the infinite length limit and comparing, to high order,  the result with the usual  bosonic characters implementing the subtraction of the singular vectors (see e.g., \cite{CFT}). A more direct test amounts to evaluating the central charge by the asymptotic form of the resulting femionic expression \cite{Nahm,KKMMa,KKMMb}. This computation is presented in Sect. 3.7. These results are then generalized to the other irreducible modules in Sect. 4. The different modules are singled out by suitable boundary conditions on the paths. 
Section 5 is devoted to the analysis of the dual models. By implementing the transformation $q\rw q^{-1}$ directly at the level of the (vacuum) character, we readily recover the (vacuum) character of the $\z_k$ graded parafermionic model.  This remarkable duality transformation between the $\M(k+1,2k+3)$ and graded parafermionic characters is then generalized (in the rest of this section) to a precise one-to-one correspondence between the dual paths and the parafermionic states. Complementary combinatorial results are reported in Appendix A. Finally, in Appendix B, we identify still another class of conformal models whose states are naturally represented by paths on a half-integer lattice: these are the superconformal minimal models ${\cal SM}(2,4\ka)$.

\section{Paths for the finitized $\M(k+1,2k+3)$ models}

\subsection{Description of the finitized $\M(k+1,2k+3)$  paths}

The paths representing the $\M(k+1,2k+3)$ models, called $\M^{[k]}$ paths,  are defined as follows:

\n {\bf  $\M^{[k]}$ paths} {\it  are defined on  a lattice with {\it  half-integer spacing}, in the first quadrant  of the $(x,y)$ plane and within the rectangle $0\leq y \leq k$, (where $k$ is a positive integer and $0\leq x \leq L$). 
%
An edge from $x$ to $x+\tfrac12$ is either up (NE or +) or down (SE or $-$). An essential constraint is that the $x$ and $y$ coordinates of any peak must  both  be {\it  integral}. In addition, the initial height $y_0$ is forced to be integer (where $y_x$ stands for the height at $x$). Every path of fixed length $L$ is supposed to end at the height $y_L\in\{\tfrac12,\tfrac32,\cdots,k-\tfrac12\} $, with a SE edge. 
In the following, we set $y_0=a$ and $y_L=b$ and denote paths with boundary conditions $(a,b)$ as $\M^{[k]}_{(a,b)}$. The {\it weight} $w$  of a $\M^{[k]}$ path is}
\begin{equation}\label{weig}
w= \sum_{x=\frac12}^{L-\frac12} w(x)\qquad \text{where} \qquad w(x)= \frac{x}2\left|\, y_{x+\frac12}-y_{x-\frac12}\, \right|.
\end{equation} 

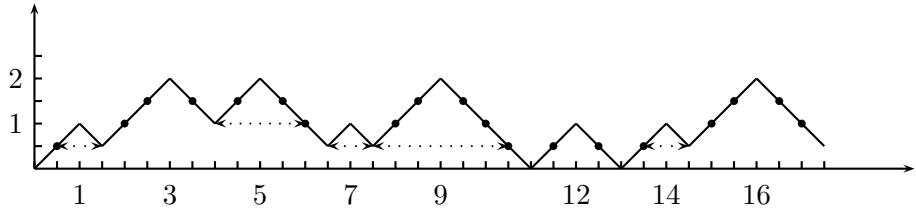
\begin{figure}[ht]
\caption{{\footnotesize An example of a $\M^{[2]}$ path starting at $a=y_0=0$ and ending at $b=y_L=\frac12$ with $L=\frac{35}2$. Recall that the upper index 2 in $\M^{[2]}$ refers to the maximal height (i.e., maximal value of $y$). The position (both the $x$ and $y$ coordinates) of the peaks are integers but there are no constraint on the position of the local minima. The weight of this path is half the sum of the $x$-coordinate of the  indicated dots ($w=\frac{345}4)$. The charge of the  peaks from left to right is $\frac12, 2,1,\frac 12, \frac32, 1,\frac12,2.$ A dotted line indicates the line from which the height must be measured to give the charge. The total charge is thus  $m= 9$. There are three charge complexes in the augmented path (that terminates on the $x$-axis): these are delimitated by the $x$ positions: 0, 11, 13 and 18.}}\label{fig1}
\begin{center}
\begin{pspicture}(0,0)(12.5,3)
\psline{->}(0.3,0.3)(0.3,2.5) \psline{->}(0.3,0.3)(12.0,0.3)
\psset{linestyle=dotted} \psline{<->}(0.6,0.6)(1.2,0.6)
\psline{<->}(4.2,0.6)(4.8,0.6) \psline{<->}(4.8,0.6)(6.6,0.6)
\psline{<->}(2.7,0.9)(3.9,0.9) \psline{<->}(8.4,0.6)(9.0,0.6)
\psset{linestyle=solid}
\psline{-}(0.3,0.3)(0.3,0.4) \psline{-}(0.6,0.3)(0.6,0.4)
\psline{-}(0.9,0.3)(0.9,0.4) \psline{-}(1.2,0.3)(1.2,0.4)
\psline{-}(1.5,0.3)(1.5,0.4) \psline{-}(1.8,0.3)(1.8,0.4)
\psline{-}(2.1,0.3)(2.1,0.4) \psline{-}(2.4,0.3)(2.4,0.4)
\psline{-}(2.7,0.3)(2.7,0.4) \psline{-}(3.0,0.3)(3.0,0.4)
\psline{-}(3.3,0.3)(3.3,0.4) \psline{-}(3.6,0.3)(3.6,0.4)
\psline{-}(3.9,0.3)(3.9,0.4) \psline{-}(4.2,0.3)(4.2,0.4)
\psline{-}(4.5,0.3)(4.5,0.4) \psline{-}(4.8,0.3)(4.8,0.4)
\psline{-}(5.1,0.3)(5.1,0.4) \psline{-}(5.4,0.3)(5.4,0.4)
\psline{-}(5.7,0.3)(5.7,0.4) \psline{-}(6.0,0.3)(6.0,0.4)
\psline{-}(6.3,0.3)(6.3,0.4) \psline{-}(6.6,0.3)(6.6,0.4)
\psline{-}(6.9,0.3)(6.9,0.4) \psline{-}(7.2,0.3)(7.2,0.4)
\psline{-}(7.5,0.3)(7.5,0.4) \psline{-}(7.8,0.3)(7.8,0.4)
\psline{-}(8.1,0.3)(8.1,0.4) \psline{-}(8.4,0.3)(8.4,0.4)
\psline{-}(8.7,0.3)(8.7,0.4) \psline{-}(9.0,0.3)(9.0,0.4)
\psline{-}(9.3,0.3)(9.3,0.4) \psline{-}(9.6,0.3)(9.6,0.4)
\psline{-}(9.9,0.3)(9.9,0.4) \psline{-}(10.2,0.3)(10.2,0.4)
\psline{-}(10.5,0.3)(10.5,0.4) \psline{-}(10.8,0.3)(10.8,0.4)

\rput(0.9,-0.05){{\small $1$}} \rput(2.1,-0.05){{\small $3$}}
\rput(3.3,-0.05){{\small $5$}} \rput(4.5,-0.05){{\small
$7$}}\rput(5.7,-0.05){{\small $9$}} \rput(7.5,-0.05){{\small $12$}}
\rput(8.7,-0.05){{\small $14$}}\rput(9.9,-0.05){{\small $16$}}
 \psline{-}(0.3,0.6)(0.4,0.6)
\psline{-}(0.3,0.9)(0.4,0.9) \psline{-}(0.3,1.2)(0.4,1.2)
\psline{-}(0.3,1.5)(0.4,1.5) \psline{-}(0.3,1.8)(0.4,1.8)

\rput(0.05,0.9){{\small $1$}} \rput(0.05,1.5){{\small $2$}}
\psline{-}(0.3,0.3)(0.6,0.6) \psline{-}(0.6,0.6)(0.9,0.9)

\psline{-}(0.9,0.9)(1.2,0.6)

\psline{-}(1.2,0.6)(1.5,0.9) \psline{-}(1.5,0.9)(1.8,1.2)
\psline{-}(1.8,1.2)(2.1,1.5)

\psline{-}(2.1,1.5)(2.4,1.2) \psline{-}(2.4,1.2)(2.7,0.9)

\psline{-}(2.7,0.9)(3.0,1.2) \psline{-}(3.0,1.2)(3.3,1.5)

\psline{-}(3.3,1.5)(3.6,1.2) \psline{-}(3.6,1.2)(3.9,0.9)
\psline{-}(3.9,0.9)(4.2,0.6)

\psline{-}(4.2,0.6)(4.5,0.9)

\psline{-}(4.5,0.9)(4.8,0.6)

\psline{-}(4.8,0.6)(5.1,0.9) \psline{-}(5.1,0.9)(5.4,1.2)
\psline{-}(5.4,1.2)(5.7,1.5)

\psline{-}(5.7,1.5)(6.0,1.2)\psline{-}(6.0,1.2)(6.3,0.9)
\psline{-}(6.3,0.9)(6.6,0.6) \psline{-}(6.6,0.6)(6.9,0.3)

\psline{-}(6.9,0.3)(7.2,0.6) \psline{-}(7.2,0.6)(7.5,0.9)

\psline{-}(7.5,0.9)(7.8,0.6) \psline{-}(7.8,0.6)(8.1,0.3)

\psline{-}(8.1,0.3)(8.4,0.6) \psline{-}(8.4,0.6)(8.7,0.9)

\psline{-}(8.7,0.9)(9.0,0.6)

\psline{-}(9.0,0.6)(9.3,0.9) \psline{-}(9.3,0.9)(9.6,1.2)
\psline{-}(9.6,1.2)(9.9,1.5)

\psline{-}(9.9,1.5)(10.2,1.2) \psline{-}(10.2,1.2)(10.5,0.9)
\psline{-}(10.5,0.9)(10.8,0.6)
\psset{dotsize=3pt}
\psdots(0.6,0.6)(1.5,0.9)(1.8,1.2)(2.4,1.2)(3.0,1.2)(3.6,1.2)
(3.9,0.9)(5.1,0.9)(5.4,1.2)(6.0,1.2)(6.3,0.9)(6.6,0.6)(7.2,0.6)
(7.8,0.6)(8.4,0.6)(9.3,0.9)(9.6,1.2)(10.2,1.2)(10.5,0.9)

\end{pspicture}
\end{center}
\end{figure}

The crucial simplifying aspect of these paths, compared with a typical RSOS path for a non-unitary minimal model, is that the cusps (local extrema) in the path do not contribute to its weight. An example is displayed in Fig. \ref{fig1}.


The objective is to determine the generating functions of the  $\M^{[k]}_{(a,b)}$ paths graded by their weight, which is encoded in the exponent of a formal variable $q$.
These will be the candidate of  the finitized characters  for the $\M(k+1,2k+3)$ irreducible modules, the different modules  being  specified by the path boundary conditions.

\subsection{Paths of the vacuum module; charge of a path}

An important characteristic of a path is its {\it charge content}, that is, the charge assignment of each of its peaks. 
To  introduce this notion, it suffices to consider paths starting from the origin. Let us first extend the path with a sequence of $2y_L$ SE edges to make the path reach the horizontal axis and refer to this extension as the augmented path. 

 The charge of a peak with  coordinates $(x,y)$ is the largest number $c$ such that we can find two vertices $(x',y-c)$ and $(x'',y-c)$ on the augmented path  with $x'<x<x''$ and such that between these two vertices there are no peak of height larger than $y$ and every peak of height equal to $y$ is located at its right \cite{BP}. Figure \ref{fig1} illustrates this definition.

Denote by $n_j$ the number of peaks of charge $j$.
The charge $m$ of a path is the sum of the charge of all its peaks
\begin{equation} 
m=\sum_{j=\frac12,1, \frac32, \ldots, k} j n_j\ .\end{equation} 
In the summation, it has been made explicit that the sum runs over all integer and half-integer entries between $\frac12$ and $k$. {\it An increment of $\tfrac12$ in summations will always be assumed from now on unless specifically indicated. }
The total charge is related to the length $L$ of the path  (ending at height $b$) as 
\begin{equation} 
L= 2m-b,\end{equation} (so that $2m$ is the length of the augmented path).

 \subsection{Paths and  gas of charged particles}
 
 In the terminology of \cite{OleJS}, a peak of charge $j$ is interpreted as a {\it particle of charge} $j$.
 Such a particle is a localized object of finite size, its diameter being  twice its charge. This size-charge relationship is clear when the particle is isolated, that is, when it is delimitated by two vertices on the horizontal axis. In that case, a particle of charge $j$ is represented by a triangle of height $j$ so that the distance between the two vertices on the horizontal axis is $2j$ --  which distance  is referred to as the particle diameter. But this charge-diameter relation is also preserved (in a `composite' way) within a  charge complex. A charge complex is  a sequence of peaks that is delimitated by two vertices on the horizontal axis of the augmented path \cite{OleJS}. For instance, for the path in Fig. \ref{fig1}, there are three charge complexes.

\n We have already indicated how we can read off the charges of the different particles of a complex. If the total charge of the complex  is $n$, its diameter is $2n$. 
 
 The notion of particle is  convenient because it embodies the idea of size finiteness  (while the peak refers either to the whole triangle or its center position). It also allows us to reinterpret the generating function for all paths with given boundaries as the grand-canonical partition function for a gas of charged particles. The energy $E$  is then related to the weight $w$  through the variable $q$ as follows \cite{OleJS}:
\begin{equation} -\beta E = w \ln q,
\end{equation}
with $\beta $ is the Boltzmann constant  times the temperature.

\section{Generating function for paths in the  vacuum module}
\subsection{The ground-state configuration}


To simplify the presentation, we first consider 
  those paths that contribute to the finitized vacuum module. Such paths start at vertical position $a= 0 $ and end at height $b=\tfrac12$.

The objective of this section is to evaluate the generating  function of such  paths with fixed charge content, that is, fixed values of the individual $n_j$, and graded by their weight. Our analysis follows closely that of \cite{OleJS}. The first step amounts to determine the {\it minimal-weight configuration}  and calculate its weight. This is the subject of the next  subsection. This minimal-weight configuration  is defined to be  the configuration of lowest weight with all $n_j$ fixed. The second step (considered in Section 3.3) is to determine the number of configurations that  do contribute for a given charge content, and calculate their weight. Ultimately, we will sum over all possible values of $n_j$  for a given fixed total charge $\sum_j j n_j$. This yields the generating function for finite paths.  The generating function for  infinite paths is obtained by taking  the limit where the  total charge becomes infinite.

\begin{figure}[ht]
\caption{{\footnotesize The ground-state path in the vacuum module of all $\M(k+1,2k+3)$ models (here for $L=\frac{17}2$). The single vertex contributing to the weight is indicated by a dot. The charge content is $n_1= 1$ and $n_{\frac12}=7$.}} \label{fig2}
\begin{center}
\begin{pspicture}(0,0)(11.5,3)
\psline{->}(0.5,0.5)(0.5,2.5) \psline{->}(0.5,0.5)(10.0,0.5)
\psset{linestyle=solid}
\psline{-}(0.5,0.5)(0.5,0.6) \psline{-}(1.0,0.5)(1.0,0.6)
\psline{-}(1.5,0.5)(1.5,0.6) \psline{-}(2.0,0.5)(2.0,0.6)
\psline{-}(2.5,0.5)(2.5,0.6) \psline{-}(3.0,0.5)(3.0,0.6)
\psline{-}(3.5,0.5)(3.5,0.6) \psline{-}(4.0,0.5)(4.0,0.6)
\psline{-}(4.5,0.5)(4.5,0.6) \psline{-}(5.0,0.5)(5.0,0.6)
\psline{-}(5.5,0.5)(5.5,0.6) \psline{-}(6.0,0.5)(6.0,0.6)
\psline{-}(6.5,0.5)(6.5,0.6) \psline{-}(7.0,0.5)(7.0,0.6)
\psline{-}(7.5,0.5)(7.5,0.6) \psline{-}(8.0,0.5)(8.0,0.6)
\psline{-}(8.5,0.5)(8.5,0.6) \psline{-}(9.0,0.5)(9.0,0.6)
\rput(1.5,0.25){{\small $1$}}\rput(2.5,0.25){{\small
$2$}}\rput(3.5,0.25){{\small $3$}}\rput(4.5,0.25){{\small
$4$}}\rput(5.5,0.25){{\small $5$}}\rput(6.5,0.25){{\small
$6$}}\rput(7.5,0.25){{\small $7$}}\rput(8.5,0.25){{\small $8$}}
\psline{-}(0.5,1.0)(0.6,1.0) \psline{-}(0.5,1.5)(0.6,1.5)
\psline{-}(0.5,2.0)(0.6,2.0) \rput(0.25,1.5){{\small $1$}}
\psline{-}(0.5,0.5)(1.0,1.0) \psline{-}(1.0,1.0)(1.5,1.5)
\psline{-}(1.5,1.5)(2.0,1.0) \psline{-}(2.0,1.0)(2.5,1.5)
\psline{-}(2.5,1.5)(3.0,1.0) \psline{-}(3.0,1.0)(3.5,1.5)
\psline{-}(3.5,1.5)(4.0,1.0) \psline{-}(4.0,1.0)(4.5,1.5)
\psline{-}(4.5,1.5)(5.0,1.0) \psline{-}(5.0,1.0)(5.5,1.5)
\psline{-}(5.5,1.5)(6.0,1.0) \psline{-}(6.0,1.0)(6.5,1.5)
\psline{-}(6.5,1.5)(7.0,1.0) \psline{-}(7.0,1.0)(7.5,1.5)
\psline{-}(7.5,1.5)(8.0,1.0) \psline{-}(8.0,1.0)(8.5,1.5)
\psline{-}(8.5,1.5)(9.0,1.0)
\psset{dotsize=3pt} \psdots(1.0,1.0)

\end{pspicture}
\end{center}
\end{figure}
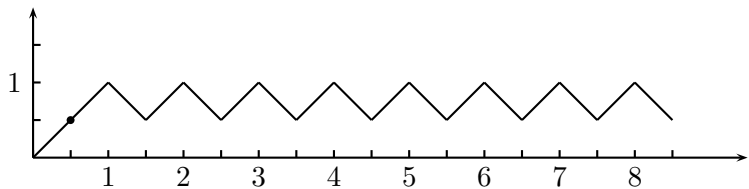

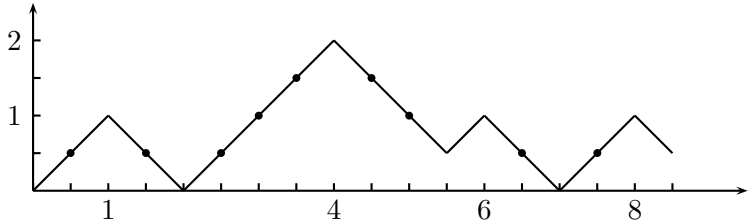
\begin{figure}[ht]
\caption{{\footnotesize A $\M^{[2]}$ path with charges (from left to right) $1,2,\frac12,\frac12$. The vertices contributing to the weight are indicated by dots. The relative weight of this path (relative to the ground state of Fig. \ref{fig2})  is $17$. }} \label{fig3}
\begin{center}
\begin{pspicture}(0,0)(11.5,3.5)
\psline{->}(0.5,0.5)(0.5,3.0) \psline{->}(0.5,0.5)(10.0,0.5)
\psset{linestyle=solid}
\psline{-}(0.5,0.5)(0.5,0.6) \psline{-}(1.0,0.5)(1.0,0.6)
\psline{-}(1.5,0.5)(1.5,0.6) \psline{-}(2.0,0.5)(2.0,0.6)
\psline{-}(2.5,0.5)(2.5,0.6) \psline{-}(3.0,0.5)(3.0,0.6)
\psline{-}(3.5,0.5)(3.5,0.6) \psline{-}(4.0,0.5)(4.0,0.6)
\psline{-}(4.5,0.5)(4.5,0.6) \psline{-}(5.0,0.5)(5.0,0.6)
\psline{-}(5.5,0.5)(5.5,0.6) \psline{-}(6.0,0.5)(6.0,0.6)
\psline{-}(6.5,0.5)(6.5,0.6) \psline{-}(7.0,0.5)(7.0,0.6)
\psline{-}(7.5,0.5)(7.5,0.6) \psline{-}(8.0,0.5)(8.0,0.6)
\psline{-}(8.5,0.5)(8.5,0.6) \psline{-}(9.0,0.5)(9.0,0.6)
\rput(1.5,0.25){{\small $1$}}\rput(4.5,0.25){{\small
$4$}}\rput(6.5,0.25){{\small $6$}}\rput(8.5,0.25){{\small $8$}}

\psline{-}(0.5,1.0)(0.6,1.0) \psline{-}(0.5,1.5)(0.6,1.5)
\psline{-}(0.5,2.0)(0.6,2.0) \psline{-}(0.5,2.5)(0.6,2.5)
\rput(0.25,1.5){{\small $1$}}\rput(0.25,2.5){{\small $2$}}
\psline{-}(0.5,0.5)(1.0,1.0) \psline{-}(1.0,1.0)(1.5,1.5)
\psline{-}(1.5,1.5)(2.0,1.0) \psline{-}(2.0,1.0)(2.5,0.5)
\psline{-}(2.5,0.5)(3.0,1.0) \psline{-}(3.0,1.0)(3.5,1.5)
\psline{-}(3.5,1.5)(4.0,2.0) \psline{-}(4.0,2.0)(4.5,2.5)
\psline{-}(4.5,2.5)(5.0,2.0) \psline{-}(5.0,2.0)(5.5,1.5)
\psline{-}(5.5,1.5)(6.0,1.0) \psline{-}(6.0,1.0)(6.5,1.5)
\psline{-}(6.5,1.5)(7.0,1.0) \psline{-}(7.0,1.0)(7.5,0.5)
\psline{-}(7.5,0.5)(8.0,1.0) \psline{-}(8.0,1.0)(8.5,1.5)
\psline{-}(8.5,1.5)(9.0,1.0)

\psset{dotsize=3pt} \psdots(1.0,1.0)(2.0,1.0)(3.0,1.0)(3.5,1.5)
(4.0,2.0)(5.0,2.0)(5.5,1.5)(7.0,1.0)(8.0,1.0)

\end{pspicture}
\end{center}
\end{figure}

Before plunging into the details of the analysis just outlined, let us point out that the weight of the different configurations appropriate to a specific module (here, the vacuum module) must always be evaluated with respect to the weight of the module's {\it ground-state}, which we now define.
Among all possible configurations of a fixed length, that is, irrespective of the individual values of the $n_j$ but for a fixed value of $\sum_j j n_j$, there is one configuration with lowest weight. It is called the ground state. The ground state corresponding to the vacuum module for all $\M(k+1,2k+3)$ models is presented in Fig. \ref{fig2}. 
The actual number of  peaks of charge $\frac12$  is determined by the length of the path, or equivalently, its total charge. With $L= \frac{17}2$, the charge content of the ground state is $n_1= 1$ and $n_{\frac12}=7$. The ground-state path has only one  vertex contributing to its weight, which is the one at $x=\frac12$    and it  contributes for $\frac14$. When comparing the weight of a path to that of the ground state, we need to subtract the weight of the latter. 
An example is given in Fig. \ref{fig3}.


Let us point out at once that  for these boundary conditions, $(a,b)=(0,\frac12)$,  the requirements of an integral $x$-coordinate for a peak position  implies that 
\begin{equation} \label{petcon}
 n_{j-\frac12}\not=0 \quad (j\in\NN)\qquad\Rw\qquad  {\rm min}\, (n_j,n_{j+1},\cdots, n_k)\geq 1. \end{equation} 
 In other words, if there is at least one peak of given half-integer charge $j-\tfrac12$, there must be at least one peak of integer charge $\geq j$. In particular, it is impossible to define a path (with $a=0$) only with peaks of   half-integer charge.
For the ground state, it  means that there must be one peak of charge 1 before the sequence of peaks of charge $\frac12$ as illustrated in Fig. \ref{fig2}.

\subsection{The minimal-weight configuration}


We now characterize the minimal-weight configuration.
 The configuration of lowest possible weight, with  all $n_j$ fixed,  is the one whose longest straight-up and straight-down segments are as close as  possible to the origin.  This immediately implies  that this configuration is the one with peaks as far apart as possible, that is, forming charge complexes by themselves whenever this is allowed by the integrality of the peak positions, and ordered, from left to right, by decreasing values of the charge.

\begin{figure}[ht]
\caption{{\footnotesize The minimal-weight configuration for $k=2$ with $(n_{\frac12},n_1,n_{\frac32}, n_2)= (2,2,1,2)$. Observe that the path does not reach the $x$ axis in-between particles of charge 2 and $\frac32$ (at $x=\frac{15}2$) and in-between particles of charge 1 and $\frac12$ (at $x=\frac{29}2$). The relative weight of this path is 62, in agreement with the result of eq. (\ref{defW}).}} \label{fig4}
\begin{center}
\begin{pspicture}(0,0)(14.4,2.8)

\psline{->}(0.4,0.4)(0.4,2.4) \psline{->}(0.4,0.4)(14.0,0.4)

\psline{-}(0.4,0.4)(0.4,0.5) \psline{-}(0.8,0.4)(0.8,0.5)
\psline{-}(1.2,0.4)(1.2,0.5) \psline{-}(1.6,0.4)(1.6,0.5)
\psline{-}(2.0,0.4)(2.0,0.5) \psline{-}(2.4,0.4)(2.4,0.5)
\psline{-}(2.8,0.4)(2.8,0.5) \psline{-}(3.2,0.4)(3.2,0.5)
\psline{-}(3.6,0.4)(3.6,0.5) \psline{-}(4.0,0.4)(4.0,0.5)
\psline{-}(4.4,0.4)(4.4,0.5) \psline{-}(4.8,0.4)(4.8,0.5)
\psline{-}(5.2,0.4)(5.2,0.5) \psline{-}(5.6,0.4)(5.6,0.5)
\psline{-}(6.0,0.4)(6.0,0.5) \psline{-}(6.4,0.4)(6.4,0.5)
\psline{-}(6.8,0.4)(6.8,0.5) \psline{-}(7.2,0.4)(7.2,0.5)
\psline{-}(7.6,0.4)(7.6,0.5) \psline{-}(8.0,0.4)(8.0,0.5)
\psline{-}(8.4,0.4)(8.4,0.5) \psline{-}(8.8,0.4)(8.8,0.5)
\psline{-}(9.2,0.4)(9.2,0.5) \psline{-}(9.6,0.4)(9.6,0.5)
\psline{-}(10.0,0.4)(10.0,0.5) \psline{-}(10.4,0.4)(10.4,0.5)
\psline{-}(10.8,0.4)(10.8,0.5) \psline{-}(11.2,0.4)(11.2,0.5)
\psline{-}(11.6,0.4)(11.6,0.5) \psline{-}(12.0,0.4)(12.0,0.5)
\psline{-}(12.4,0.4)(12.4,0.5) \psline{-}(12.8,0.4)(12.8,0.5)
\psline{-}(13.2,0.4)(13.2,0.5) \psline{-}(13.6,0.4)(13.6,0.5)
\rput(2.0,0.2){{\small $2$}} \rput(5.2,0.2){{\small$6$}}
\rput(7.6,0.2){{\small$9$}} \rput(10.0,0.2){{\small$12$}}
\rput(11.6,0.2){{\small$14$}} \rput(12.4,0.2){{\small$15$}}
\rput(13.2,0.2){{\small$16$}}
\psline{-}(0.4,0.8)(0.5,0.8) \psline{-}(0.4,1.2)(0.5,1.2)
\psline{-}(0.4,1.6)(0.5,1.6) \psline{-}(0.4,2.0)(0.5,2.0)
\rput(0.2,1.2){{\small $1$}}\rput(0.2,2.0){{\small $2$}}

\psline{-}(0.4,0.4)(0.8,0.8) \psline{-}(0.8,0.8)(1.2,1.2)
\psline{-}(1.2,1.2)(1.6,1.6) \psline{-}(1.6,1.6)(2.0,2.0)
\psline{-}(2.0,2.0)(2.4,1.6) \psline{-}(2.4,1.6)(2.8,1.2)
\psline{-}(2.8,1.2)(3.2,0.8) \psline{-}(3.2,0.8)(3.6,0.4)
\psline{-}(3.6,0.4)(4.0,0.8) \psline{-}(4.0,0.8)(4.4,1.2)
\psline{-}(4.4,1.2)(4.8,1.6) \psline{-}(4.8,1.6)(5.2,2.0)
\psline{-}(5.2,2.0)(5.6,1.6) \psline{-}(5.6,1.6)(6.0,1.2)
\psline{-}(6.0,1.2)(6.4,0.8) \psline{-}(6.4,0.8)(6.8,1.2)
\psline{-}(6.8,1.2)(7.2,1.6) \psline{-}(7.2,1.6)(7.6,2.0)
\psline{-}(7.6,2.0)(8.0,1.6) \psline{-}(8.0,1.6)(8.4,1.2)
\psline{-}(8.4,1.2)(8.8,0.8) \psline{-}(8.8,0.8)(9.2,0.4)
\psline{-}(9.2,0.4)(9.6,0.8) \psline{-}(9.6,0.8)(10.0,1.2)
\psline{-}(10.0,1.2)(10.4,0.8) \psline{-}(10.4,0.8)(10.8,0.4)
\psline{-}(10.8,0.4)(11.2,0.8) \psline{-}(11.2,0.8)(11.6,1.2)
\psline{-}(11.6,1.2)(12.0,0.8) \psline{-}(12.0,0.8)(12.4,1.2)
\psline{-}(12.4,1.2)(12.8,0.8) \psline{-}(12.8,0.8)(13.2,1.2)
\psline{-}(13.2,1.2)(13.6,0.8)

\psset{dotsize=3pt}
\psdots(0.8,0.8)(1.2,1.2)(1.6,1.6)(2.4,1.6)(2.8,1.2)
(3.2,0.8)(4.0,0.8)(4.4,1.2)(4.8,1.6)(5.6,1.6)(6.0,1.2)(6.8,1.2)
(7.2,1.6)(8.0,1.6)(8.4,1.2)(8.8,0.8)(9.6,0.8)(10.4,0.8)(11.2,0.8)

\end{pspicture}
\end{center}
\end{figure}
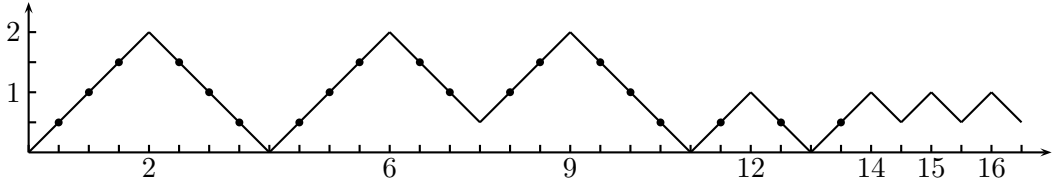

More explicitly, the minimal-weight configuration is the one with
 all the $n_k$ particles of charge $k$  at the left followed by the sequence of $n_k-\frac12$ particles of charge $k-\frac12$ and so on, ending with the sequence of $n_\frac12$ charge-$\tfrac12$ particles. Note that the initial vertex of the leftmost particle of  charge $k-\tfrac12$ is not on the $x$-axis but at the vertex with coordinates $(2kn_k-\tfrac12,\, \tfrac12)$.  Similarly, the initial and final vertices of particles of charges $k-\tfrac12$ have height $\tfrac12$.
 After the sequence of peaks of charge $k-\frac12$, the path reaches the $x$-axis.  It then describes the set of  particles with  charge $k-1$ and so on.
 When a sequence of peaks of integer charge is followed by a sequence of peaks of half-integer charge, the latter sequence together with the last peak of the integer charge sequence form a charge complex. In other words, the path reaches the $x$-axis only when the charge of the peaks change from integer to integer or half-integer to integer. This is neatly illustrated in Fig. \ref{fig4}. 




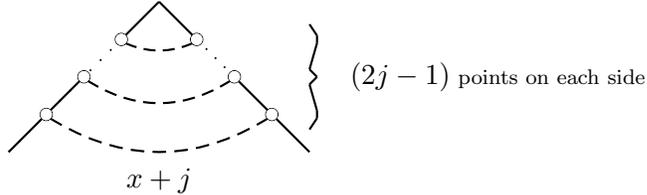
\begin{figure}[ht]
\caption{{\footnotesize An isolated particle of charge $j$ with peak at horizontal position at $x_0+j$.  Pairs of opposite points with respect to the center contribute to $x_0+j$ to the weight. }} \label{fig5}
\begin{center}
\begin{pspicture}(0,0)(5.0,3.0)

\psline{-}(0.5,0.5)(1.0,1.0) \psline{-}(1.0,1.0)(1.5,1.5)
\psline{-}(2.0,2.0)(2.5,2.5) \psline{-}(2.5,2.5)(3.0,2.0)
\psline{-}(3.5,1.5)(4.0,1.0) \psline{-}(4.0,1.0)(4.5,0.5)
\psset{linestyle=dotted}\psset{dotsep=4.5pt}
\psline{-}(0.5,0.5)(2.0,2.0) \psline{-}(3.0,2.0)(3.5,1.5)
\psset{linestyle=dashed}
\parabola(1.05,0.95)(2.5,0.5)
\parabola(1.55,1.45)(2.5,1.15)
\parabola(2.05,1.95)(2.5,1.85)

\psset{dotsize=5pt}\psset{dotstyle=o}
\psdots(1.0,1.0)(1.5,1.5)(2.0,2.0)(3.0,2.0)(3.5,1.5)(4.0,1.0)

\rput(2.5,0.125){{\small $x+j$}} \rput(7,1.5){{\small $(2j-1)$
points on each side}}

\psset{linestyle=solid} \psline{-}(4.5,0.8)(4.6,0.95)
\psline{-}(4.6,0.95)(4.6,1.05) \psline{-}(4.6,1.05)(4.5,1.4)
\psline{-}(4.5,1.4)(4.6,1.5) \psline{-}(4.6,1.5)(4.5,1.6)
\psline{-}(4.5,1.6)(4.6,1.95) \psline{-}(4.6,1.95)(4.6,2.05)
\psline{-}(4.6,2.05)(4.5,2.2)

\end{pspicture}
\end{center}
\end{figure}

Let us determine the weight of this minimal-weight  configuration. An isolated particle of charge $j$ that starts at $x_0$ (with the understanding  that $j+x_0$ is integer) has weight 
\begin{equation} 
 w(j,x_0)= (2j-1)(j+x_0) .\end{equation} 
 This result is easily verified: the contribution of any pair of vertices in the straight-up and straight-down segments that are symmetric with respect to the center of the triangle  contributes to $j+x_0$ to the weight and there are $2j-1$ such pairs (see Fig. \ref{fig5}). For a sequence of $n_j$ particles of charge $j$ (each particle having diameter $2j$), again starting at $x_0$, this is 
\begin{equation} 
\sum_{i=0}^{n_j-1} (2j-1) ( j+ 2 j i + x_0) = j(2j-1) n_j^2+(2j-1)n_j \,x_0 .\end{equation}
The value of $x_0$ is fixed by the number of particles of charge higher that $j$, namely
\begin{equation} 
x_0 =  - \e_j  + \sum _{l= j+\tfrac12}^k 2l n_l, \qquad {\rm with} \qquad \e_j= j-\lf j \rf, \end{equation}
where $\lf a \rf$ is the largest integer smaller than $a$ (while $\lc a\rc$, to be used later, is the smallest integer larger than $a$).
In other words,  $\e_j = \frac12 $ if $j$ is half integer and 0 otherwise.


Let us denote the  weight of the minimal-weight  configuration of the finitized $\M(k+1,2k+3)$ model as compared to the ground state by ${\cal W}^{(k)}$. If all particles were isolated (so that every particle would form a charge complex by itself), this weight would  be   the sum of the weight of all the particles, that is,
\begin{equation} 
\sum_{j=1} ^k[j(2j-1) n_j^2+(2j-1)n_j x_0 ]\;.
\end{equation}
Note that $n_\frac12$ does not contribute to this expression. But we need to introduce a correction factor  $\Delta$ that  takes care of the fact that   particles with half-integer charge have their initial and final vertices at height $\frac12$ and  also subtract  the ground-state contribution. We thus have:
\begin{equation} 
 {\cal W}^{(k)}=\sum_{j=1} ^k[j(2j-1) n_j^2+(2j-1)n_j x_0 ] +\Delta. \end{equation}

We now derive the expression of  $\Delta$.   Subtracting   the ground-state contribution is simply taken into account by removing the contribution of the first straight-up segment (that is, the vertex at $x=\frac12$). Next, since the particles of half-integer charge do not start on the horizontal axis, one must  delete the contribution of the last  vertex in the straight-down segment of the rightmost adjacent particle with  integer charge.  Moreover,  the last (right-most) particle of half-integer charge has one extra contributing vertex due to the extra edge that is needed because the path must  reach the horizontal axis before the sequence of  isolated particles of lower integer charge. 

 
 Taking care of all these factors, the explicit expression of $\Delta$ is seen to be
\begin{align} 
\Delta=& \frac12\left\{ -\frac12 - \left(2kn_k-\frac12\right)  +\left(2kn_k+2(k-\frac12)n_{k-\frac12}-\frac12\right)  -\cdots -\left(2kn_k+\cdots +2n_1-\frac12\right)  \right\}\nonumber \\ =& \frac12 \left\{ \sum_{i=1}^k(-1)^{2i+1} \left [\sum_{j=i}^k2jn_j-\frac12\right]\right\} - \frac14.\end{align}
%
Because the factors $n_j$ with $j$ half-integer occur an even number of times and with alternating signs, their contribution add up to zero. All  $n_j$ factors with $j$ integer occur an odd number of times and the non-canceled factor is negative. We have thus
\begin{equation} 
\Delta= - \sum _{\substack{j=1\\j\in \NN}}^k j n_j .\end{equation}
(That only the number of particles of  integer-charge do contribute to $\Delta$ is indicated by the restriction $j\in \NN$ in the summation.)

Let us return to the evaluation of $ {\cal W}^{(k)}$. Using the above  expressions for $x_0$ and $\Delta$, one has
\begin{align}\label{defW}
 {\cal W}^{(k)}&= \sum_{j=1}^k \big\{ j(2j-1)n_j^2+(2j-1)n_j \big[-\e_j+\sum _{l= j+\frac12}^k 2l n_l \big]\ \big\}- \sum _{\substack{j=1\\j\in \NN}}^k j n_j \nonumber \\ 
&= \sum_{j=1}^k \big\{ j(2j-1)n_j^2+(2j-1)n_j \sum _{l= j+\frac12}^k 2l n_l\ \big\} -  \sum _{\substack{j=3/2\\ j\in \NN+\frac12}}^k (j-\tfrac12) n_j 
 - \sum _{\substack{j=1\\j\in \NN}}^k j n_j \nonumber \\ 
&= \sum_{i,j=1}^k n_iB_{ij}n_j - \sum_{j=1}^k \, \lf j\rf \, n_j.
\end{align}
In the last line, we have introduced the   symmetric matrix $B$ with entries
\begin{equation} \label{defB}
B_{ij}= (2i-1)j \quad {\rm for}~  i\leq j . \end{equation}
As already indicated, the weight  of this configuration is independent upon the number $n_\frac12$ of particles with charge $\frac12$.

\subsection{Counting the different  configurations for a fixed charge content}

For a fixed charge content, all possible contributions can be obtained from the minimal-weight  configuration by displacing the particles subject to the following two constraints \cite{OleJS}:

\begin{enumerate}

\item Configurations producing equivalent paths are identified.

\item  Particles of different charges can interpenetrate. 
\end{enumerate}

The first condition implies that  the ordering  of particles with the same charge remains unchanged. This can be interpreted as a hard-core repulsion between particles of the same type \cite{OleJS}. 

In relation with  the second condition, consider two particles of charge $l$ and $j$, with $l>j$. If the  particle with the  largest charge is at the left, their minimal distance (the separation between their peak position) is $2j$. However, if the largest charge  is at the right, this minimal distance is $2j+1$. Indeed, in this case, a distance of $2j$ would correspond to a configuration where the two peaks are at the same height, a configuration that has already been considered (and associated to the inverted ordering). Equivalently, the prescription previously given for determining the particle content of a path implies that whenever two peaks have the same height, it is the leftmost one which is attributed the largest charge. However, as far as the next-to-be-considered left-displacements of the smallest particle will be concerned, once the distance $2j$ is reached, before the subsequent move, it will be supposed that the two particles get interchanged so that the next move is counted from  the position of the leftmost peak {\cite{BreL,OleJS}.

The displacement of the particles are always by {\it two} half-integer steps and toward the left. These are subject to the following rules \cite{OleJS}:

\begin{enumerate}

\item Particles of charge $k$ remain fixed. 

\item Particles of charge $k-\tfrac12$ are moved one by one toward the left, starting from the leftmost  one up to the rightmost one. 

\item  The process is repeated for particles of  charge $k-1,\ldots, \frac12$ treated  successively.

\end{enumerate}

 If we label the particles of charge $j$ by $1,2,\cdots, n_j$, from left to right and denote the displacement of the $i$-th one by $\mu_i$, we have:
\begin{equation}\label{cond} \mu_{i}\geq \mu_{i+1} \qquad {\rm and  } \qquad \mu_1\leq p_j .\end{equation}
The first condition reflects the hard-core repulsion between identical particles.
The upper bound $p_j$ on the first displacement of the particle of charge $j$ is determined below.

Any displacement of 1 toward the left increases the weight of the path by 1. Indeed, the weight of the path before the move is 
\begin{equation}
w= \frac12\sum_{{\rm all ~ vertices}}x - \frac12\sum _{{\rm cusps}} x .
\end{equation}
Displacing a particle by 1 toward the left implies that the peak-position of that particle, as well as the minimum just before it, are both displaced toward the left by 1. Therefore, the effect of a particle  move by 1 toward the left  is thus:
\begin{equation}  \sum _{{\rm cusps}} x \rw \sum _{{\rm cusps}} x- 2\qquad \Rw \qquad w\rw w+1\;.
\end{equation} 
Different cases are illustrated in Figs \ref{fig6} and \ref{fig7}.


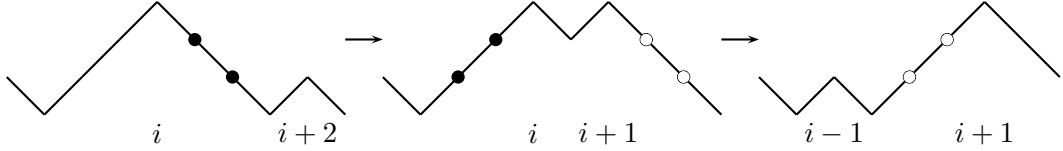
\begin{figure}[ht]
\caption{{\footnotesize Two displacements of distance 1 toward the left of a particle of charge $\frac12$ through a larger particle of half-integer charge, here $\frac32$. In the intermediate configuration, the left-most peak is the one of charge $\frac32$. However, before the next move, the particle identities are interchanged: the subsequent move is performed by considering the left-most peak to be the particle of charge $\frac12$. The vertices that are affected in evaluating the weight difference between each successive configurations are indicated as follows: black dots  are sent to open circles in the resulting configuration. This makes clear that the weight changes by 1 in each step. }} \label{fig6}
\begin{center}
\begin{pspicture}(0,0)(15.5,2.5)
\psset{linestyle=solid}
\psline{-}(0.5,1.0)(1.0,0.5) \psline{-}(1.0,0.5)(1.5,1.0)
\psline{-}(1.5,1.0)(2.0,1.5) \psline{-}(2.0,1.5)(2.5,2.0)
\psline{-}(2.5,2.0)(3.0,1.5) \psline{-}(3.0,1.5)(3.5,1.0)
\psline{-}(3.5,1.0)(4.0,0.5) \psline{-}(4.0,0.5)(4.5,1.0)
\psline{-}(4.5,1.0)(5.0,0.5)

\psline{-}(5.5,1.0)(6.0,0.5) \psline{-}(6.0,0.5)(6.5,1.0)
\psline{-}(6.5,1.0)(7.0,1.5) \psline{-}(7.0,1.5)(7.5,2.0)
\psline{-}(7.5,2.0)(8.0,1.5) \psline{-}(8.0,1.5)(8.5,2.0)
\psline{-}(8.5,2.0)(9.0,1.5) \psline{-}(9.0,1.5)(9.5,1.0)
\psline{-}(9.5,1.0)(10.0,0.5)

\psline{-}(10.5,1.0)(11.0,0.5) \psline{-}(11.0,0.5)(11.5,1.0)
\psline{-}(11.5,1.0)(12.0,0.5) \psline{-}(12.0,0.5)(12.5,1.0)
\psline{-}(12.5,1.0)(13.0,1.5) \psline{-}(13.0,1.5)(13.5,2.0)
\psline{-}(13.5,2.0)(14.0,1.5) \psline{-}(14.0,1.5)(14.5,1.0)

\psline{->}(5.0,1.5)(5.5,1.5) \psline{->}(10.0,1.5)(10.5,1.5)


\psset{dotsize=5pt}\psset{dotstyle=*}
\psdots(3.0,1.5)(3.5,1.0)(6.5,1.0)(7.0,1.5)

\psset{dotsize=5pt}\psset{dotstyle=o}
\psdots(9.0,1.5)(9.5,1.0)(12.5,1.0)(13.0,1.5)

\rput(2.5,0.25){{\small $i$}}\rput(4.5,0.25){{\small $i+2$}}
\rput(7.5,0.25){{\small $i$}}\rput(8.5,0.25){{\small $i+1$}}
\rput(11.5,0.25){{\small $i-1$}}\rput(13.5,0.25){{\small $i+1$}}
\end{pspicture}
\end{center}
\end{figure}


\begin{figure}[ht]
\caption{{\footnotesize Similar displacements (as described in Fig. (\ref{fig6})) of distance 1 toward the left of a particle of charge $\frac12$ through a larger particle of integer charge, here 2.  }} \label{fig7}
\begin{center}
\begin{pspicture}(0,0)(16.0,3.0)

\psline{-}(0.5,0.5)(1.0,1.0) \psline{-}(1.0,1.0)(1.5,1.5)
\psline{-}(1.5,1.5)(2.0,2.0) \psline{-}(2.0,2.0)(2.5,2.5)
\psline{-}(2.5,2.5)(3.0,2.0) \psline{-}(3.0,2.0)(3.5,1.5)
\psline{-}(3.5,1.5)(4.0,1.0) \psline{-}(4.0,1.0)(4.5,1.5)
\psline{-}(4.5,1.5)(5.0,1.0) \psline{-}(5.0,1.0)(5.5,0.5)
\psline{-}(6.0,0.5)(6.5,1.0) \psline{-}(6.5,1.0)(7.0,1.5)
\psline{-}(7.0,1.5)(7.5,2.0) \psline{-}(7.5,2.0)(8.0,2.5)
\psline{-}(8.0,2.5)(8.5,2.0) \psline{-}(8.5,2.0)(9.0,2.5)
\psline{-}(9.0,2.5)(9.5,2.0) \psline{-}(9.5,2.0)(10.0,1.5)
\psline{-}(10.0,1.5)(10.5,1.0) \psline{-}(10.5,1.0)(11.0,0.5)
\psline{-}(11.5,0.5)(12.0,1.0) \psline{-}(12.0,1.0)(12.5,1.5)
\psline{-}(12.5,1.5)(13.0,1.0) \psline{-}(13.0,1.0)(13.5,1.5)
\psline{-}(13.5,1.5)(14.0,2.0) \psline{-}(14.0,2.0)(14.5,2.5)
\psline{-}(14.5,2.5)(15.0,2.0) \psline{-}(15.0,2.0)(15.5,1.5)
\psline{-}(15.5,1.5)(16.0,1.0) \psline{-}(16.0,1.0)(16.5,0.5)

\psline{->}(5.5,1.5)(6.0,1.5) \psline{->}(11.0,1.5)(11.5,1.5)

\psset{dotsize=5pt}\psset{dotstyle=*}
\psdots(3.0,2.0)(3.5,1.5)(7.0,1.5)(7.5,2.0)

\psset{dotsize=5pt}\psset{dotstyle=o}
\psdots(9.5,2.0)(10.0,1.5)(13.5,1.5)(14.0,2.0)

\rput(2.5,0.25){{\small $i$}} \rput(4.5,0.25){{\small $i+2$}}
\rput(8.0,0.25){{\small $i$}} \rput(9.0,0.25){{\small $i+1$}}
\rput(12.5,0.25){{\small $i-1$}} \rput(14.5,0.25){{\small $i+1$}}

\end{pspicture}
\end{center}
\end{figure}
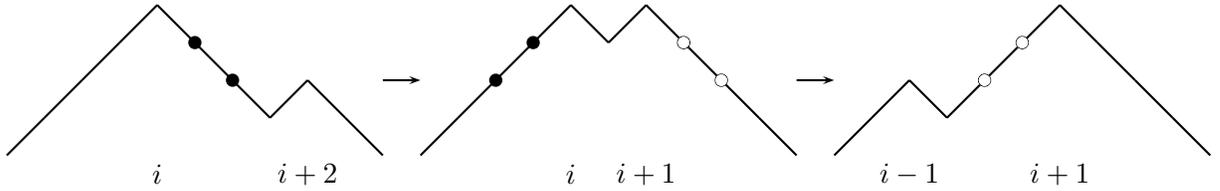

The combinatorial problem of counting the possible configurations 
while keeping track of their weight 
amounts to evaluating the following multiple summation \cite{OleJS}:
$$C(p_j,n_j)\equiv\sum_{\mu_1=0}^{p_j}\sum_{\mu_2=0}^{\mu_1}\cdots \sum_{\mu_{n_j=0}}^{\mu_{n_j-1} }q^{\mu_1+\cdots \mu_{n_j}}$$
This is indeed equivalent to $q$-enumerate the  partitions $(\mu_1,\cdots,\mu_{n_j})$ of $n= \mu_1+\cdots + \mu_{n_j}$  into at most $n_j$ (nonzero) parts, each part being at most equal to $p_j$ (i.e., $\mu_1\leq p_j$) and with each  partition being weighted by $q^n$. Denote  this number as ${\cal P}(p_j,n_j,n)$. Its generating function is well-known to be (cf.  \cite{Andr} Theorem 3.1):
\begin{equation}C(p_j,n_j)= \sum_{n\geq 0} {\cal P}(p_j,n_j,n)\, q^n= 
\begin{bmatrix}
p_j+n_j\\ n_j\end{bmatrix}
\end{equation} 
where
\begin{equation}
\begin{bmatrix}
a\\ b\end{bmatrix}=\frac{(q)_a}{(q)_{a-b}(q)_b} , \qquad {\rm with}\qquad (q)_a= (1-q)\cdots (1-q^a).
\end{equation} 

It remains to determine $p_j$, the maximal displacement of the leftmost particle $j$ through all particles of higher charges. Consider first the problem of displacing a particle of charge $j$ within  a particle of larger charge $l$. We start with the minimal configuration appropriate to this sequence of two particles and determine the number of possible moves of the charge-$j$ particle. For this we need to identify (i): all vertices within the straight-down part  of the triangle representing the  charge-$l$ particle at which  the starting vertex of the particle of charge $j$ could be located. Similarly, we must determine (ii): the number of vertices on the straight-up part of the charge-$l$ particle  where the charge-$j$ particle could terminate. 

\n (i):  In all cases (meaning all parities of $2l$ and $2j$) but one (treated in the following paragraph), there are $l-j$ available vertices on the down part of the triangle of height $l$ which are allowed candidates  for the starting vertex of the triangle of height $j$. Recall that two adjacent accessible vertices must differ by an integer. The difference $l-j$ takes into account the minimal distance between the two peak positions, which is $2j$ in this case.

\n  This counting is modified when $l$ is integer and $j$ half-integer. Then, in the minimal-weight configuration, the first  vertex of the charge-$j$ particle is already within the larger triangle. In that case, there are thus one point less available for a move. All cases are thereby accounted for if we replace $l-j$ by  $l-j-(2\e_l-1)2\e_j$. However, in view of generalization, is important to notice that this additional subtraction occurs only for the interpenetration of the rightmost integer-charge particle by the leftmost particle of  lower half-integer charge  and not for the remaining ones.

\n (ii): In all cases, there are $l-j$ available vertices on the straight-up part of the triangle of height $l$ which can coincide with the ending  vertex of particle of type $j$. Again, this takes into account the minimal distance between the particles when the lowest-charged one is at the left of the largest one; this distance is $2j+1$. In  this counting, we have also included the very first vertex of the larger triangle as a possibility (when it is allowed by the integral peak-position constraint). This takes into account the case where the particles are isolated (that is, they do not intepenetrate) but with their ordering interchanged.

The different situations are exemplified in Fig. \ref{fig8}.


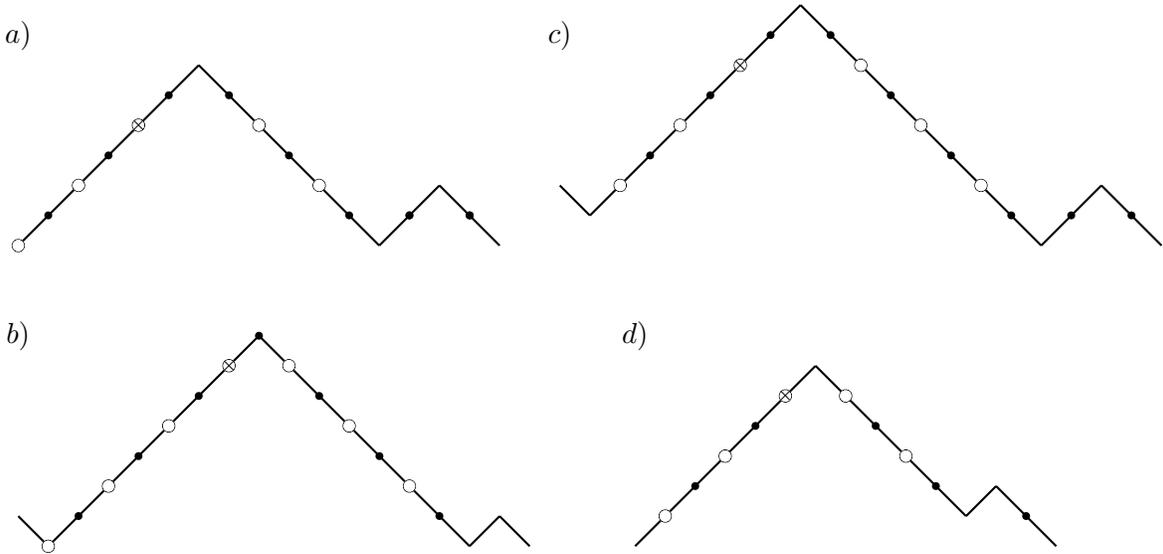
\begin{figure}[ht]
\caption{{\footnotesize The different vertices, within a larger particle of charge $l$, where  a particle of lower charge $j$ can start or finish. The starting (finishing) positions are on the straight-down (-up) part of the larger charge particle. An open circle indicates a point within the particle of charge $l$ where the starting or the final point of the particle of charge $j$ can be located. The crossed circle indicates a point that is excluded because  the resulting configuration has already been considered. The total number of possible configurations -- excluding the one that is drawn --  is the number of open circles and it is given by $2(l-j)-(2\e_l-1)2\e_j$. 
The different cases illustrate all the possible parities of $2l$ and $2j$. In case (a), $l$ and $j$ are both integers, $l=3$ and $j=1$, and the number of configuration is 4. In case  (b), the two particles have half-integer charge: $l=\frac72$ and $j=\frac12$. The number of configurations is $2(\frac72-\frac12)=6$. The case where $l$ is half-integer ($=\frac72$) and $j$ integer ($=1$) is illustrated  in (c), with $2(l-j)=5$. Finally, an example where $l$ is integer ($=3$) and $j$ half-integer ($=\frac12$) is illustrated in (d): the number of configurations is now modified to  $2(l-j)-1=4$. }} \label{fig8}
\begin{center}
\begin{pspicture}(0,0)(16.0,7.6)

\psline{-}(0.4,4.4)(0.8,4.8) \psline{-}(0.8,4.8)(1.2,5.2)
\psline{-}(1.2,5.2)(1.6,5.6) \psline{-}(1.6,5.6)(2.0,6.0)
\psline{-}(2.0,6.0)(2.4,6.4) \psline{-}(2.4,6.4)(2.8,6.8)
\psline{-}(2.8,6.8)(3.2,6.4) \psline{-}(3.2,6.4)(3.6,6.0)
\psline{-}(3.6,6.0)(4.0,5.6) \psline{-}(4.0,5.6)(4.4,5.2)
\psline{-}(4.4,5.2)(4.8,4.8) \psline{-}(4.8,4.8)(5.2,4.4)
\psline{-}(5.2,4.4)(5.6,4.8) \psline{-}(5.6,4.8)(6.0,5.2)
\psline{-}(6.0,5.2)(6.4,4.8) \psline{-}(6.4,4.8)(6.8,4.4)

\psset{dotsize=5pt}\psset{dotstyle=o}
\psdots(0.4,4.4)(1.2,5.2)(3.6,6.0)(4.4,5.2)(2.0,6.0)

\psset{dotsize=4pt}\psset{dotstyle=+}
\psset{dotangle=45}\psdots(2.0,6.0)

\psset{dotsize=3pt}\psset{dotstyle=*}
\psdots(0.8,4.8)(1.6,5.6)(2.4,6.4)(3.2,6.4)(4.0,5.6)(4.8,4.8)
(5.6,4.8)(6.4,4.8)

\rput(0.4,7.2){{\small $a)$}}

\psline{-}(0.4,0.8)(0.8,0.4) \psline{-}(0.8,0.4)(1.2,0.8)
\psline{-}(1.2,0.8)(1.6,1.2) \psline{-}(1.6,1.2)(2.0,1.6)
\psline{-}(2.0,1.6)(2.4,2.0) \psline{-}(2.4,2.0)(2.8,2.4)
\psline{-}(2.8,2.4)(3.2,2.8) \psline{-}(3.2,2.8)(3.6,3.2)
\psline{-}(3.6,3.2)(4.0,2.8) \psline{-}(4.0,2.8)(4.4,2.4)
\psline{-}(4.4,2.4)(4.8,2.0) \psline{-}(4.8,2.0)(5.2,1.6)
\psline{-}(5.2,1.6)(5.6,1.2) \psline{-}(5.6,1.2)(6.0,0.8)
\psline{-}(6.0,0.8)(6.4,0.4) \psline{-}(6.4,0.4)(6.8,0.8)
\psline{-}(6.8,0.8)(7.2,0.4)

\psset{dotsize=5pt}\psset{dotstyle=o}
\psdots(0.8,0.4)(1.6,1.2)(2.4,2.0)(3.2,2.8)(4.0,2.8)(4.8,2.0)(5.6,1.2)

\psset{dotsize=4pt}\psset{dotstyle=+}
\psset{dotangle=45}\psdots(3.2,2.8)

\psset{dotsize=3pt}\psset{dotstyle=*}
\psdots(1.2,0.8)(2.0,1.6)(2.8,2.4)(3.6,3.2)(4.4,2.4)(5.2,1.6)(6.0,0.8)
(5.6,4.8)(6.4,4.8)

\rput(0.4,3.2){{\small $b)$}}

\psline{-}(7.6,5.2)(8.0,4.8) \psline{-}(8.0,4.8)(8.4,5.2)
\psline{-}(8.4,5.2)(8.8,5.6) \psline{-}(8.8,5.6)(9.2,6.0)
\psline{-}(9.2,6.0)(9.6,6.4) \psline{-}(9.6,6.4)(10.0,6.8)
\psline{-}(10.0,6.8)(10.4,7.2) \psline{-}(10.4,7.2)(10.8,7.6)
\psline{-}(10.8,7.6)(11.2,7.2) \psline{-}(11.2,7.2)(11.6,6.8)
\psline{-}(11.6,6.8)(12.0,6.4) \psline{-}(12.0,6.4)(12.4,6.0)
\psline{-}(12.4,6.0)(12.8,5.6) \psline{-}(12.8,5.6)(13.2,5.2)
\psline{-}(13.2,5.2)(13.6,4.8) \psline{-}(13.6,4.8)(14.0,4.4)
\psline{-}(14.0,4.4)(14.4,4.8) \psline{-}(14.4,4.8)(14.8,5.2)
\psline{-}(14.8,5.2)(15.2,4.8) \psline{-}(15.2,4.8)(15.6,4.4)

\psset{dotsize=5pt}\psset{dotstyle=o}
\psdots(8.4,5.2)(9.2,6.0)(10.0,6.8)(11.6,6.8)(12.4,6.0)(13.2,5.2)

\psset{dotsize=4pt}\psset{dotstyle=+}
\psset{dotangle=45}\psdots(10.0,6.8)

\psset{dotsize=3pt}\psset{dotstyle=*}
\psdots(8.8,5.6)(9.6,6.4)(10.4,7.2)(11.2,7.2)(12.0,6.4)(12.8,5.6)(13.6,4.8)
(14.4,4.8)(15.2,4.8)

\rput(7.6,7.2){{\small $c)$}}

\psline{-}(8.6,0.4)(9.0,0.8) \psline{-}(9.0,0.8)(9.4,1.2)
\psline{-}(9.4,1.2)(9.8,1.6) \psline{-}(9.8,1.6)(10.2,2.0)
\psline{-}(10.2,2.0)(10.6,2.4) \psline{-}(10.6,2.4)(11.0,2.8)
\psline{-}(11.0,2.8)(11.4,2.4) \psline{-}(11.4,2.4)(11.8,2.0)
\psline{-}(11.8,2.0)(12.2,1.6) \psline{-}(12.2,1.6)(12.6,1.2)
\psline{-}(12.6,1.2)(13.0,0.8) \psline{-}(13.0,0.8)(13.4,1.2)
\psline{-}(13.4,1.2)(13.8,0.8) \psline{-}(13.8,0.8)(14.2,0.4)

\psset{dotsize=5pt}\psset{dotstyle=o}
\psdots(9.0,0.8)(9.8,1.6)(10.6,2.4)(11.4,2.4)(12.2,1.6)

\psset{dotsize=4pt}\psset{dotstyle=+}
\psset{dotangle=45}\psdots(10.6,2.4)

\psset{dotsize=3pt}\psset{dotstyle=*}
\psdots(9.4,1.2)(10.2,2.0)(11.8,2.0)(12.6,1.2)(13.8,0.8)

\rput(8.6,3.2){{\small $d)$}}

\end{pspicture}
\end{center}
\end{figure}

The total number of moves of $j$ within $l$ (including the configurations were the particle of charge $j$ ($<l$) is in front) is thus $ 2(l-j)-(2\e_l-1)2\e_j$ .

Summing over all charges larger than $j$ (and recalling that $(2\e_l-1)2\e_j$ is to be counted not more than once and that it does need to be counted once due to (\ref{petcon}))  yields
\begin{equation}\label{defp}
p_j= -2\e_j+\sum_{l= j+\frac12}^k 2(l-j) n_l  .
\end{equation}

The presence of the $-1$ in $p_j$ when $j$ is half-integer necessitates a precision on the definition of the $q$-binomial which must be augmented by the condition:
\begin{equation}
\begin{bmatrix}
-1\\ \phantom{-}0\end{bmatrix}=1\;.\end{equation} 

Finally, we illustrate the construction of a generic path from the corresponding minimal-weight configuration in Fig. \ref{fig9}.


\begin{figure}[ht]
\caption{{\footnotesize Various configurations obtained from the minimal-weight configuration in Fig. \ref{fig4} by successive displacements toward the left of the peaks of charge lower that 2. The successive displacements are as  follows: the peak of charge $\frac32$ is moved by 1 unit in (a); the leftmost peak of charge 1 is moved by 5 units in (b); the other peak of charge 1 is moved by 2 units in (c); the leftmost peak of charge $\frac12$ is moved by 7 units in (d); the last peak is moved by 3 units in (e). The total number of displacements is 18 so that the relative weight of the last  path is $62+18=80$.
 }} \label{fig9}
\begin{center}
\begin{pspicture}(0,0)(14.8,12.4)

\psline{->}(0.4,0.4)(14.4,0.4) \psline{->}(0.4,2.8)(14.4,2.8)
\psline{->}(0.4,5.2)(14.4,5.2) \psline{->}(0.4,7.6)(14.4,7.6)
\psline{->}(0.4,10.0)(14.4,10.0)
\psline{->}(0.4,0.4)(0.4,2.0) \psline{->}(0.4,2.8)(0.4,4.4)
\psline{->}(0.4,5.2)(0.4,6.8) \psline{->}(0.4,7.6)(0.4,9.2)
\psline{->}(0.4,10.0)(0.4,11.6)

\psline{-}(0.4,0.4)(0.4,0.5) \psline{-}(0.8,0.4)(0.8,0.5)
\psline{-}(1.2,0.4)(1.2,0.5) \psline{-}(1.6,0.4)(1.6,0.5)
\psline{-}(2.0,0.4)(2.0,0.5) \psline{-}(2.4,0.4)(2.4,0.5)
\psline{-}(2.8,0.4)(2.8,0.5) \psline{-}(3.2,0.4)(3.2,0.5)
\psline{-}(3.6,0.4)(3.6,0.5) \psline{-}(4.0,0.4)(4.0,0.5)
\psline{-}(4.4,0.4)(4.4,0.5) \psline{-}(4.8,0.4)(4.8,0.5)
\psline{-}(5.2,0.4)(5.2,0.5) \psline{-}(5.6,0.4)(5.6,0.5)
\psline{-}(6.0,0.4)(6.0,0.5) \psline{-}(6.4,0.4)(6.4,0.5)
\psline{-}(6.8,0.4)(6.8,0.5) \psline{-}(7.2,0.4)(7.2,0.5)
\psline{-}(7.6,0.4)(7.6,0.5) \psline{-}(8.0,0.4)(8.0,0.5)
\psline{-}(8.4,0.4)(8.4,0.5) \psline{-}(8.8,0.4)(8.8,0.5)
\psline{-}(9.2,0.4)(9.2,0.5) \psline{-}(9.6,0.4)(9.6,0.5)
\psline{-}(10.0,0.4)(10.0,0.5) \psline{-}(10.4,0.4)(10.4,0.5)
\psline{-}(10.8,0.4)(10.8,0.5) \psline{-}(11.2,0.4)(11.2,0.5)
\psline{-}(11.6,0.4)(11.6,0.5) \psline{-}(12.0,0.4)(12.0,0.5)
\psline{-}(12.4,0.4)(12.4,0.5) \psline{-}(12.8,0.4)(12.8,0.5)
\psline{-}(13.2,0.4)(13.2,0.5) \psline{-}(13.6,0.4)(13.6,0.5)

\psline{-}(0.4,2.8)(0.4,2.9) \psline{-}(0.8,2.8)(0.8,2.9)
\psline{-}(1.2,2.8)(1.2,2.9) \psline{-}(1.6,2.8)(1.6,2.9)
\psline{-}(2.0,2.8)(2.0,2.9) \psline{-}(2.4,2.8)(2.4,2.9)
\psline{-}(2.8,2.8)(2.8,2.9) \psline{-}(3.2,2.8)(3.2,2.9)
\psline{-}(3.6,2.8)(3.6,2.9) \psline{-}(4.0,2.8)(4.0,2.9)
\psline{-}(4.4,2.8)(4.4,2.9) \psline{-}(4.8,2.8)(4.8,2.9)
\psline{-}(5.2,2.8)(5.2,2.9) \psline{-}(5.6,2.8)(5.6,2.9)
\psline{-}(6.0,2.8)(6.0,2.9) \psline{-}(6.4,2.8)(6.4,2.9)
\psline{-}(6.8,2.8)(6.8,2.9) \psline{-}(7.2,2.8)(7.2,2.9)
\psline{-}(7.6,2.8)(7.6,2.9) \psline{-}(8.0,2.8)(8.0,2.9)
\psline{-}(8.4,2.8)(8.4,2.9) \psline{-}(8.8,2.8)(8.8,2.9)
\psline{-}(9.2,2.8)(9.2,2.9) \psline{-}(9.6,2.8)(9.6,2.9)
\psline{-}(10.0,2.8)(10.0,2.9) \psline{-}(10.4,2.8)(10.4,2.9)
\psline{-}(10.8,2.8)(10.8,2.9) \psline{-}(11.2,2.8)(11.2,2.9)
\psline{-}(11.6,2.8)(11.6,2.9) \psline{-}(12.0,2.8)(12.0,2.9)
\psline{-}(12.4,2.8)(12.4,2.9) \psline{-}(12.8,2.8)(12.8,2.9)
\psline{-}(13.2,2.8)(13.2,2.9) \psline{-}(13.6,2.8)(13.6,2.9)

\psline{-}(0.4,5.2)(0.4,5.3) \psline{-}(0.8,5.2)(0.8,5.3)
\psline{-}(1.2,5.2)(1.2,5.3) \psline{-}(1.6,5.2)(1.6,5.3)
\psline{-}(2.0,5.2)(2.0,5.3) \psline{-}(2.4,5.2)(2.4,5.3)
\psline{-}(2.8,5.2)(2.8,5.3) \psline{-}(3.2,5.2)(3.2,5.3)
\psline{-}(3.6,5.2)(3.6,5.3) \psline{-}(4.0,5.2)(4.0,5.3)
\psline{-}(4.4,5.2)(4.4,5.3) \psline{-}(4.8,5.2)(4.8,5.3)
\psline{-}(5.2,5.2)(5.2,5.3) \psline{-}(5.6,5.2)(5.6,5.3)
\psline{-}(6.0,5.2)(6.0,5.3) \psline{-}(6.4,5.2)(6.4,5.3)
\psline{-}(6.8,5.2)(6.8,5.3) \psline{-}(7.2,5.2)(7.2,5.3)
\psline{-}(7.6,5.2)(7.6,5.3) \psline{-}(8.0,5.2)(8.0,5.3)
\psline{-}(8.4,5.2)(8.4,5.3) \psline{-}(8.8,5.2)(8.8,5.3)
\psline{-}(9.2,5.2)(9.2,5.3) \psline{-}(9.6,5.2)(9.6,5.3)
\psline{-}(10.0,5.2)(10.0,5.3) \psline{-}(10.4,5.2)(10.4,5.3)
\psline{-}(10.8,5.2)(10.8,5.3) \psline{-}(11.2,5.2)(11.2,5.3)
\psline{-}(11.6,5.2)(11.6,5.3) \psline{-}(12.0,5.2)(12.0,5.3)
\psline{-}(12.4,5.2)(12.4,5.3) \psline{-}(12.8,5.2)(12.8,5.3)
\psline{-}(13.2,5.2)(13.2,5.3) \psline{-}(13.6,5.2)(13.6,5.3)

\psline{-}(0.4,7.6)(0.4,7.7) \psline{-}(0.8,7.6)(0.8,7.7)
\psline{-}(1.2,7.6)(1.2,7.7) \psline{-}(1.6,7.6)(1.6,7.7)
\psline{-}(2.0,7.6)(2.0,7.7) \psline{-}(2.4,7.6)(2.4,7.7)
\psline{-}(2.8,7.6)(2.8,7.7) \psline{-}(3.2,7.6)(3.2,7.7)
\psline{-}(3.6,7.6)(3.6,7.7) \psline{-}(4.0,7.6)(4.0,7.7)
\psline{-}(4.4,7.6)(4.4,7.7) \psline{-}(4.8,7.6)(4.8,7.7)
\psline{-}(5.2,7.6)(5.2,7.7) \psline{-}(5.6,7.6)(5.6,7.7)
\psline{-}(6.0,7.6)(6.0,7.7) \psline{-}(6.4,7.6)(6.4,7.7)
\psline{-}(6.8,7.6)(6.8,7.7) \psline{-}(7.2,7.6)(7.2,7.7)
\psline{-}(7.6,7.6)(7.6,7.7) \psline{-}(8.0,7.6)(8.0,7.7)
\psline{-}(8.4,7.6)(8.4,7.7) \psline{-}(8.8,7.6)(8.8,7.7)
\psline{-}(9.2,7.6)(9.2,7.7) \psline{-}(9.6,7.6)(9.6,7.7)
\psline{-}(10.0,7.6)(10.0,7.7) \psline{-}(10.4,7.6)(10.4,7.7)
\psline{-}(10.8,7.6)(10.8,7.7) \psline{-}(11.2,7.6)(11.2,7.7)
\psline{-}(11.6,7.6)(11.6,7.7) \psline{-}(12.0,7.6)(12.0,7.7)
\psline{-}(12.4,7.6)(12.4,7.7) \psline{-}(12.8,7.6)(12.8,7.7)
\psline{-}(13.2,7.6)(13.2,7.7) \psline{-}(13.6,7.6)(13.6,7.7)

\psline{-}(0.4,10.0)(0.4,10.1) \psline{-}(0.8,10.0)(0.8,10.1)
\psline{-}(1.2,10.0)(1.2,10.1) \psline{-}(1.6,10.0)(1.6,10.1)
\psline{-}(2.0,10.0)(2.0,10.1) \psline{-}(2.4,10.0)(2.4,10.1)
\psline{-}(2.8,10.0)(2.8,10.1) \psline{-}(3.2,10.0)(3.2,10.1)
\psline{-}(3.6,10.0)(3.6,10.1) \psline{-}(4.0,10.0)(4.0,10.1)
\psline{-}(4.4,10.0)(4.4,10.1) \psline{-}(4.8,10.0)(4.8,10.1)
\psline{-}(5.2,10.0)(5.2,10.1) \psline{-}(5.6,10.0)(5.6,10.1)
\psline{-}(6.0,10.0)(6.0,10.1) \psline{-}(6.4,10.0)(6.4,10.1)
\psline{-}(6.8,10.0)(6.8,10.1) \psline{-}(7.2,10.0)(7.2,10.1)
\psline{-}(7.6,10.0)(7.6,10.1) \psline{-}(8.0,10.0)(8.0,10.1)
\psline{-}(8.4,10.0)(8.4,10.1) \psline{-}(8.8,10.0)(8.8,10.1)
\psline{-}(9.2,10.0)(9.2,10.1) \psline{-}(9.6,10.0)(9.6,10.1)
\psline{-}(10.0,10.0)(10.0,10.1) \psline{-}(10.4,10.0)(10.4,10.1)
\psline{-}(10.8,10.0)(10.8,10.1) \psline{-}(11.2,10.0)(11.2,10.1)
\psline{-}(11.6,10.0)(11.6,10.1) \psline{-}(12.0,10.0)(12.0,10.1)
\psline{-}(12.4,10.0)(12.4,10.1) \psline{-}(12.8,10.0)(12.8,10.1)
\psline{-}(13.2,10.0)(13.2,10.1) \psline{-}(13.6,10.0)(13.6,10.1)

\rput(1.2,0.2){{\small$1$}} \rput(2.0,0.2){{\small$2$}}
\rput(2.8,0.2){{\small$3$}} \rput(3.6,0.2){{\small$4$}}
\rput(4.4,0.2){{\small$5$}} \rput(5.2,0.2){{\small$6$}}
\rput(6.0,0.2){{\small$7$}} \rput(6.8,0.2){{\small$8$}}
\rput(7.6,0.2){{\small$9$}} \rput(8.4,0.2){{\small$10$}}
\rput(9.2,0.2){{\small$11$}} \rput(10.0,0.2){{\small$12$}}
\rput(10.8,0.2){{\small$13$}} \rput(11.6,0.2){{\small$14$}}
\rput(12.4,0.2){{\small$15$}} \rput(13.2,0.2){{\small$16$}}


\psline{-}(0.4,10.0)(0.8,10.4) \psline{-}(0.8,10.4)(1.2,10.8)
\psline{-}(1.2,10.8)(1.6,11.2) \psline{-}(1.6,11.2)(2.0,11.6)
\psline{-}(2.0,11.6)(2.4,11.2) \psline{-}(2.4,11.2)(2.8,10.8)
\psline{-}(2.8,10.8)(3.2,10.4) \psline{-}(3.2,10.4)(3.6,10.8)
\psline{-}(3.6,10.8)(4.0,11.2) \psline{-}(4.0,11.2)(4.4,11.6)
\psline{-}(4.4,11.6)(4.8,11.2) \psline{-}(4.8,11.2)(5.2,10.8)
\psline{-}(5.2,10.8)(5.6,10.4) \psline{-}(5.6,10.4)(6.0,10.0)
\psline{-}(6.0,10.0)(6.4,10.4) \psline{-}(6.4,10.4)(6.8,10.8)
\psline{-}(6.8,10.8)(7.2,11.2) \psline{-}(7.2,11.2)(7.6,11.6)
\psline{-}(7.6,11.6)(8.0,11.2) \psline{-}(8.0,11.2)(8.4,10.8)
\psline{-}(8.4,10.8)(8.8,10.4) \psline{-}(8.8,10.4)(9.2,10.0)
\psline{-}(9.2,10.0)(9.6,10.4) \psline{-}(9.6,10.4)(10.0,10.8)
\psline{-}(10.0,10.8)(10.4,10.4) \psline{-}(10.4,10.4)(10.8,10.0)
\psline{-}(10.8,10.0)(11.2,10.4) \psline{-}(11.2,10.4)(11.6,10.8)
\psline{-}(11.6,10.8)(12.0,10.4) \psline{-}(12.0,10.4)(12.4,10.8)
\psline{-}(12.4,10.8)(12.8,10.4) \psline{-}(12.8,10.4)(13.2,10.8)
\psline{-}(13.2,10.8)(13.6,10.4)

\psline{-}(0.4,7.6) (0.8,8.0) \psline{-}(0.8,8.0)(1.2,8.4)
\psline{-}(1.2,8.4)(1.6,8.0) \psline{-}(1.6,8.0)(2.0,7.6)
\psline{-}(2.0,7.6) (2.4,8.0) \psline{-}(2.4,8.0)(2.8,8.4)
\psline{-}(2.8,8.4)(3.2,8.8) \psline{-}(3.2,8.8)(3.6,9.2)
\psline{-}(3.6,9.2)(4.0,8.8) \psline{-}(4.0,8.8)(4.4,8.4)
\psline{-}(4.4,8.4)(4.8,8.0) \psline{-}(4.8,8.0)(5.2,8.4)
\psline{-}(5.2,8.4)(5.6,8.8) \psline{-}(5.6,8.8)(6.0,9.2)
\psline{-}(6.0,9.2)(6.4,8.8) \psline{-}(6.4,8.8)(6.8,8.4)
\psline{-}(6.8,8.4)(7.2,8.0) \psline{-}(7.2,8.0)(7.6,7.6)
\psline{-}(7.6,7.6) (8.0,8.0) \psline{-}(8.0,8.0)(8.4,8.4)
\psline{-}(8.4,8.4)(8.8,8.8) \psline{-}(8.8,8.8)(9.2,9.2)
\psline{-}(9.2,9.2)(9.6,8.8) \psline{-}(9.6,8.8)(10.0,8.4)
\psline{-}(10.0,8.4)(10.4,8.0) \psline{-}(10.4,8.0)(10.8,7.6)
\psline{-}(10.8,7.6)(11.2,8.0) \psline{-}(11.2,8.0) (11.6,8.4)
\psline{-}(11.6,8.4)(12.0,8.0) \psline{-}(12.0,8.0)(12.4,8.4)
\psline{-}(12.4,8.4)(12.8,8.0) \psline{-}(12.8,8.0)(13.2,8.4)
\psline{-}(13.2,8.4)(13.6,8.0)

\psline{-}(0.4,5.2)(0.8,5.6) \psline{-}(0.8,5.6)(1.2,6.0)
\psline{-}(1.2,6.0)(1.6,5.6) \psline{-}(1.6,5.6)(2.0,5.2)
\psline{-}(2.0,5.2)(2.4,5.6) \psline{-}(2.4,5.6)(2.8,6.0)
\psline{-}(2.8,6.0)(3.2,6.4) \psline{-}(3.2,6.4)(3.6,6.8)
\psline{-}(3.6,6.8)(4.0,6.4) \psline{-}(4.0,6.4)(4.4,6.0)
\psline{-}(4.4,6.0)(4.8,5.6) \psline{-}(4.8,5.6)(5.2,6.0)
\psline{-}(5.2,6.0)(5.6,6.4) \psline{-}(5.6,6.4)(6.0,6.8)
\psline{-}(6.0,6.8)(6.4,6.4) \psline{-}(6.4,6.4)(6.8,6.0)
\psline{-}(6.8,6.0)(7.2,5.6) \psline{-}(7.2,5.6)(7.6,5.2)
\psline{-}(7.6,5.2)(8.0,5.6) \psline{-}(8.0,5.6)(8.4,6.0)
\psline{-}(8.4,6.0)(8.8,5.6) \psline{-}(8.8,5.6)(9.2,5.2)
\psline{-}(9.2,5.2)(9.6,5.6) \psline{-}(9.6,5.6)(10.0,6.0)
\psline{-}(10.0,6.0)(10.4,6.4) \psline{-}(10.4,6.4)(10.8,6.8)
\psline{-}(10.8,6.8)(11.2,6.4) \psline{-}(11.2,6.4)(11.6,6.0)
\psline{-}(11.6,6.0)(12.0,5.6) \psline{-}(12.0,5.6)(12.4,6.0)
\psline{-}(12.4,6.0)(12.8,5.6) \psline{-}(12.8,5.6)(13.2,6.0)
\psline{-}(13.2,6.0)(13.6,5.6)

\psline{-}(0.4,2.8)(0.8,3.2) \psline{-}(0.8,3.2)(1.2,3.6)
\psline{-}(1.2,3.6)(1.6,3.2) \psline{-}(1.6,3.2)(2.0,2.8)
\psline{-}(2.0,2.8)(2.4,3.2) \psline{-}(2.4,3.2)(2.8,3.6)
\psline{-}(2.8,3.6)(3.2,4.0) \psline{-}(3.2,4.0)(3.6,4.4)
\psline{-}(3.6,4.4)(4.0,4.0) \psline{-}(4.0,4.0)(4.4,4.4)
\psline{-}(4.4,4.4)(4.8,4.0) \psline{-}(4.8,4.0)(5.2,3.6)
\psline{-}(5.2,3.6)(5.6,3.2) \psline{-}(5.6,3.2)(6.0,3.6)
\psline{-}(6.0,3.6)(6.4,4.0) \psline{-}(6.4,4.0)(6.8,4.4)
\psline{-}(6.8,4.4)(7.2,4.0) \psline{-}(7.2,4.0)(7.6,3.6)
\psline{-}(7.6,3.6)(8.0,3.2) \psline{-}(8.0,3.2)(8.4,2.8)
\psline{-}(8.4,2.8)(8.8,3.2) \psline{-}(8.8,3.2)(9.2,3.6)
\psline{-}(9.2,3.6)(9.6,3.2) \psline{-}(9.6,3.2)(10.0,2.8)
\psline{-}(10.0,2.8)(10.4,3.2) \psline{-}(10.4,3.2)(10.8,3.6)
\psline{-}(10.8,3.6)(11.2,4.0) \psline{-}(11.2,4.0)(11.6,4.4)
\psline{-}(11.6,4.4)(12.0,4.0) \psline{-}(12.0,4.0)(12.4,3.6)
\psline{-}(12.4,3.6)(12.8,3.2) \psline{-}(12.8,3.2)(13.2,3.6)
\psline{-}(13.2,3.6)(13.6,3.2)

\psline{-}(0.4,0.4)(0.8,0.8) \psline{-}(0.8,0.8)(1.2,1.2)
\psline{-}(1.2,1.2)(1.6,0.8) \psline{-}(1.6,0.8)(2.0,0.4)
\psline{-}(2.0,0.4)(2.4,0.8) \psline{-}(2.4,0.8)(2.8,1.2)
\psline{-}(2.8,1.2)(3.2,1.6) \psline{-}(3.2,1.6)(3.6,2.0)
\psline{-}(3.6,2.0)(4.0,1.6) \psline{-}(4.0,1.6)(4.4,2.0)
\psline{-}(4.4,2.0)(4.8,1.6) \psline{-}(4.8,1.6)(5.2,1.2)
\psline{-}(5.2,1.2)(5.6,0.8) \psline{-}(5.6,0.8)(6.0,1.2)
\psline{-}(6.0,1.2)(6.4,1.6) \psline{-}(6.4,1.6)(6.8,2.0)
\psline{-}(6.8,2.0)(7.2,1.6) \psline{-}(7.2,1.6)(7.6,1.2)
\psline{-}(7.6,1.2)(8.0,0.8) \psline{-}(8.0,0.8)(8.4,0.4)
\psline{-}(8.4,0.4)(8.8,0.8) \psline{-}(8.8,0.8)(9.2,1.2)
\psline{-}(9.2,1.2)(9.6,0.8) \psline{-}(9.6,0.8)(10.0,1.2)
\psline{-}(10.0,1.2)(10.4,0.8) \psline{-}(10.4,0.8)(10.8,0.4)
\psline{-}(10.8,0.4)(11.2,0.8) \psline{-}(11.2,0.8)(11.6,1.2)
\psline{-}(11.6,1.2)(12.0,1.6) \psline{-}(12.0,1.6)(12.4,2.0)
\psline{-}(12.4,2.0)(12.8,1.6) \psline{-}(12.8,1.6)(13.2,1.2)
\psline{-}(13.2,1.2)(13.6,0.8)

\rput(0.2,12.0){{\small$a)$}} \rput(0.2,9.6){{\small$b)$}}
\rput(0.2,7.2){{\small$c)$}} \rput(0.2,4.8){{\small$d)$}}
\rput(0.2,2.4){{\small$e)$}}

\end{pspicture}
\end{center}
\end{figure}
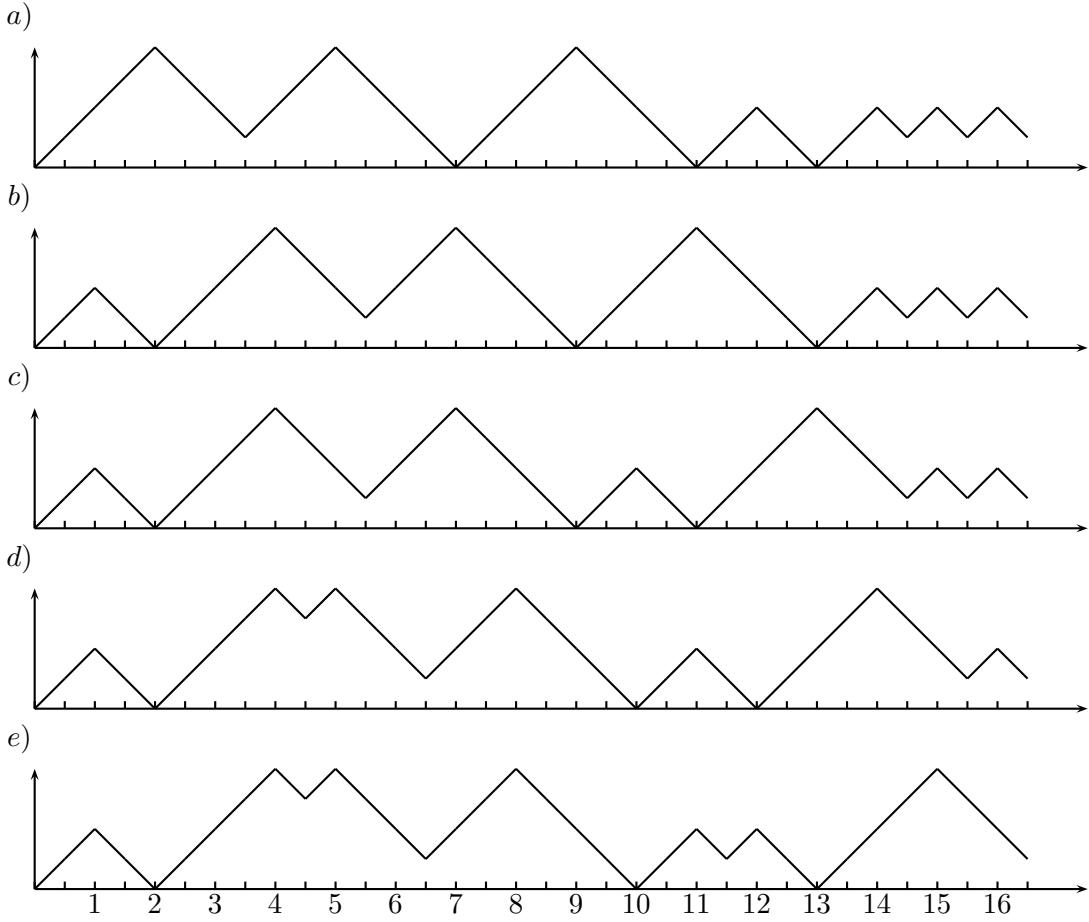

\subsection{Examples: the $\M(2,5)$ and the $\M(3,7)$ models }

The $\M^{[1]}_{(0,\frac12)}$ paths describe the space of states of the finitized version of the vacuum module in the $\M(2,5)$ minimal model. Paths in that case only  have peaks of charge $\frac12 $ or 1. The states in  the finitized module with $L=\frac{11}2$ are displayed in Fig. \ref{fig10}.  The charge content of each path is also given and the number of paths with same charge content  exemplifies  the combinatorial factor 
\begin{equation}
\begin{pmatrix}
p_\frac12+n_\frac12\\ n_\frac12\end{pmatrix}=\begin{pmatrix}
n_\frac12+n_1-1\\ n_\frac12\end{pmatrix}\;.\end{equation} 
For instance, there are three paths with charge content $(n_\frac12,n_1)=(2,2)$.  
The paths of weight 0, 2 and 6 are the minimal-weight configurations with their given charge content and these values of the weight match  the generic expression for ${\cal W}^{(1)} $ obtained from  (\ref{defW}):
\begin{equation}
{\cal W}^{(1)} = n_1^2-n_1.
\end{equation}


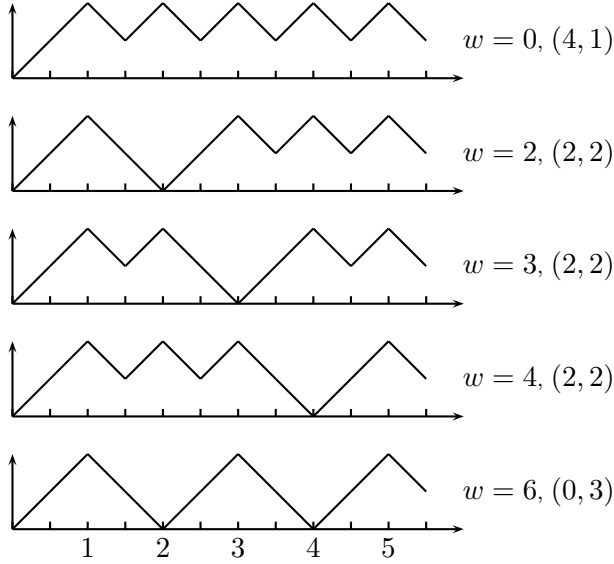
\begin{figure}[ht]
\caption{{\footnotesize The various $\M^{[1]}_{(0,\frac12)}$ paths with $L=\frac{11}2$. For each path, we indicate the relative weight $w$ together with the charge content in the form $(n_\frac12,n_1)$. }} \label{fig10}
\begin{center}
\begin{pspicture}(0,0)(7.0,8.0)
\psline{->}(0.5,0.5)(6.5,0.5) \psline{->}(0.5,2.0)(6.5,2.0)
\psline{->}(0.5,3.5)(6.5,3.5) \psline{->}(0.5,5.0)(6.5,5.0)
\psline{->}(0.5,6.5)(6.5,6.5)
\psline{->}(0.5,0.5)(0.5,1.5)\psline{->}(0.5,2.0)(0.5,3.0)
\psline{->}(0.5,3.5)(0.5,4.5)\psline{->}(0.5,5.0)(0.5,6.0)
\psline{->}(0.5,6.5)(0.5,7.5)

\psset{linestyle=solid}

\psline{-}(0.5,0.5)(0.5,0.6) \psline{-}(1.0,0.5)(1.0,0.6)
\psline{-}(1.5,0.5)(1.5,0.6) \psline{-}(2.0,0.5)(2.0,0.6)
\psline{-}(2.5,0.5)(2.5,0.6) \psline{-}(3.0,0.5)(3.0,0.6)
\psline{-}(3.5,0.5)(3.5,0.6) \psline{-}(4.0,0.5)(4.0,0.6)
\psline{-}(4.5,0.5)(4.5,0.6) \psline{-}(5.0,0.5)(5.0,0.6)
\psline{-}(5.5,0.5)(5.5,0.6) \psline{-}(6.0,0.5)(6.0,0.6)
\rput(2.5,0.25){{\small $2$}}\rput(4.5,0.25){{\small
$4$}}\rput(1.5,0.25){{\small $1$}}\rput(3.5,0.25){{\small $3$}}
\rput(5.5,0.25){{\small $5$}}

\psline{-}(0.5,2.0)(0.5,2.1) \psline{-}(1.0,2.0)(1.0,2.1)
\psline{-}(1.5,2.0)(1.5,2.1) \psline{-}(2.0,2.0)(2.0,2.1)
\psline{-}(2.5,2.0)(2.5,2.1) \psline{-}(3.0,2.0)(3.0,2.1)
\psline{-}(3.5,2.0)(3.5,2.1) \psline{-}(4.0,2.0)(4.0,2.1)
\psline{-}(4.5,2.0)(4.5,2.1) \psline{-}(5.0,2.0)(5.0,2.1)
\psline{-}(5.5,2.0)(5.5,2.1) \psline{-}(6.0,2.0)(6.0,2.1)

\psline{-}(0.5,3.5)(0.5,3.6) \psline{-}(1.0,3.5)(1.0,3.6)
\psline{-}(1.5,3.5)(1.5,3.6) \psline{-}(2.0,3.5)(2.0,3.6)
\psline{-}(2.5,3.5)(2.5,3.6) \psline{-}(3.0,3.5)(3.0,3.6)
\psline{-}(3.5,3.5)(3.5,3.6) \psline{-}(4.0,3.5)(4.0,3.6)
\psline{-}(4.5,3.5)(4.5,3.6) \psline{-}(5.0,3.5)(5.0,3.6)
\psline{-}(5.5,3.5)(5.5,3.6) \psline{-}(6.0,3.5)(6.0,3.6)

\psline{-}(0.5,5.0)(0.5,5.1) \psline{-}(1.0,5.0)(1.0,5.1)
\psline{-}(1.5,5.0)(1.5,5.1) \psline{-}(2.0,5.0)(2.0,5.1)
\psline{-}(2.5,5.0)(2.5,5.1) \psline{-}(3.0,5.0)(3.0,5.1)
\psline{-}(3.5,5.0)(3.5,5.1) \psline{-}(4.0,5.0)(4.0,5.1)
\psline{-}(4.5,5.0)(4.5,5.1) \psline{-}(5.0,5.0)(5.0,5.1)
\psline{-}(5.5,5.0)(5.5,5.1) \psline{-}(6.0,5.0)(6.0,5.1)

\psline{-}(0.5,6.5)(0.5,6.6) \psline{-}(1.0,6.5)(1.0,6.6)
\psline{-}(1.5,6.5)(1.5,6.6) \psline{-}(2.0,6.5)(2.0,6.6)
\psline{-}(2.5,6.5)(2.5,6.6) \psline{-}(3.0,6.5)(3.0,6.6)
\psline{-}(3.5,6.5)(3.5,6.6) \psline{-}(4.0,6.5)(4.0,6.6)
\psline{-}(4.5,6.5)(4.5,6.6) \psline{-}(5.0,6.5)(5.0,6.6)
\psline{-}(5.5,6.5)(5.5,6.6) \psline{-}(6.0,6.5)(6.0,6.6)

\psline{-}(0.5,6.5)(1.0,7.0) \psline{-}(1.0,7.0)(1.5,7.5)
\psline{-}(1.5,7.5)(2.0,7.0) \psline{-}(2.0,7.0)(2.5,7.5)
\psline{-}(2.5,7.5)(3.0,7.0) \psline{-}(3.0,7.0)(3.5,7.5)
\psline{-}(3.5,7.5)(4.0,7.0) \psline{-}(4.0,7.0)(4.5,7.5)
\psline{-}(4.5,7.5)(5.0,7.0) \psline{-}(5.0,7.0)(5.5,7.5)
\psline{-}(5.5,7.5)(6.0,7.0) \rput(7.5,7.0){{\small $w=0$, $(4,1)$}}

\psline{-}(0.5,5.0)(1.0,5.5) \psline{-}(1.0,5.5)(1.5,6.0)
\psline{-}(1.5,6.0)(2.0,5.5) \psline{-}(2.0,5.5)(2.5,5.0)
\psline{-}(2.5,5.0)(3.0,5.5) \psline{-}(3.0,5.5)(3.5,6.0)
\psline{-}(3.5,6.0)(4.0,5.5) \psline{-}(4.0,5.5)(4.5,6.0)
\psline{-}(4.5,6.0)(5.0,5.5) \psline{-}(5.0,5.5)(5.5,6.0)
\psline{-}(5.5,6.0)(6.0,5.5)\rput(7.5,5.5){{\small $w=2$, $(2,2)$}}

\psline{-}(0.5,3.5)(1.0,4.0) \psline{-}(1.0,4.0)(1.5,4.5)
\psline{-}(1.5,4.5)(2.0,4.0) \psline{-}(2.0,4.0)(2.5,4.5)
\psline{-}(2.5,4.5)(3.0,4.0) \psline{-}(3.0,4.0)(3.5,3.5)
\psline{-}(3.5,3.5)(4.0,4.0) \psline{-}(4.0,4.0)(4.5,4.5)
\psline{-}(4.5,4.5)(5.0,4.0) \psline{-}(5.0,4.0)(5.5,4.5)
\psline{-}(5.5,4.5)(6.0,4.0)\rput(7.5,4.0){{\small $w=3$, $(2,2)$}}

\psline{-}(0.5,2.0)(1.0,2.5) \psline{-}(1.0,2.5)(1.5,3.0)
\psline{-}(1.5,3.0)(2.0,2.5) \psline{-}(2.0,2.5)(2.5,3.0)
\psline{-}(2.5,3.0)(3.0,2.5) \psline{-}(3.0,2.5)(3.5,3.0)
\psline{-}(3.5,3.0)(4.0,2.5) \psline{-}(4.0,2.5)(4.5,2.0)
\psline{-}(4.5,2.0)(5.0,2.5) \psline{-}(5.0,2.5)(5.5,3.0)
\psline{-}(5.5,3.0)(6.0,2.5)\rput(7.5,2.5){{\small $w=4$, $(2,2)$}}

\psline{-}(0.5,0.5)(1.0,1.0) \psline{-}(1.0,1.0)(1.5,1.5)
\psline{-}(1.5,1.5)(2.0,1.0) \psline{-}(2.0,1.0)(2.5,0.5)
\psline{-}(2.5,0.5)(3.0,1.0) \psline{-}(3.0,1.0)(3.5,1.5)
\psline{-}(3.5,1.5)(4.0,1.0) \psline{-}(4.0,1.0)(4.5,0.5)
\psline{-}(4.5,0.5)(5.0,1.0) \psline{-}(5.0,1.0)(5.5,1.5)
\psline{-}(5.5,1.5)(6.0,1.0)\rput(7.5,1.0){{\small $w=6$, $(0,3)$}}

\end{pspicture}
\end{center}
\end{figure}

For $k=2$, the corresponding minimal model is $\M(3,7)$. The peaks are now allowed to have charge up to 2. Consider  again $L=\frac{11}2$. It is clear that all the $\M^{[1]}_{(0,\frac12)}$ paths form a subset of the $\M^{[2]}_{(0,\frac12)}$ ones (and this illustrates the natural embedding of states as $k\rw k+1$ in this path  description).  The additional paths pertaining to the case $k=2$ are presented in Fig. \ref{fig11}. The number of configurations with the charge content specified is now
\begin{equation}
\prod_{j=\frac12}^\frac32\begin{pmatrix}
p_j+n_j\\ n_j\end{pmatrix} =\begin{pmatrix}
n_\frac12+n_1+2n_\frac32+3n_2-1\\ n_\frac12\end{pmatrix} 
\begin{pmatrix}
n_1+n_\frac32+2n_2\\ n_1\end{pmatrix}\begin{pmatrix}
n_\frac32+n_2-1\\ n_\frac32\end{pmatrix}
    \;.\end{equation}
For the two minimal-weight configurations appearing in the figure, the values of the  weights agree with  those obtained from following expression read off (\ref{defW}):
\begin{equation}
{\cal W}^{(2)}= 6n_2^2+3n_\frac32^2+n_1^2+8n_2n_\frac32 +4n_2n_1 +3n_\frac32 n_1-2n_2-n_\frac32-n_1.\end{equation}
The other configurations are obtained (successively) by displacing by  one unit toward the left  a peak of charge $<2$.


\begin{figure}[ht]
\caption{{\footnotesize The  $\M^{[2]}_{(0,\frac12)}$ paths with $L=\frac{11}2$ which are not included in the $\M^{[1]}_{(0,\frac12)}$  set  (already displayed in Fig. \ref{fig10}). The charge content is presented in the form $(n_\frac12,n_1,n_\frac32,n_2)$.  }} \label{fig11}
\begin{center}
\begin{pspicture}(0,0)(14.0,13.0)

\psline{->}(0.5,0.5)(6.5,0.5) \psline{->}(7.5,0.5)(13.5,0.5)
\psline{->}(0.5,3.0)(6.5,3.0) \psline{->}(7.5,3.0)(13.5,3.0)
\psline{->}(0.5,5.5)(6.5,5.5) \psline{->}(7.5,5.5)(13.5,5.5)
\psline{->}(0.5,8.0)(6.5,8.0) \psline{->}(7.5,8.0)(13.5,8.0)
\psline{->}(0.5,10.5)(6.5,10.5)
\psline{->}(0.5,0.5)(0.5,2.5) \psline{->}(7.5,0.5)(7.5,2.5)
\psline{->}(0.5,3.0)(0.5,5.0) \psline{->}(7.5,3.0)(7.5,5.0)
\psline{->}(0.5,5.5)(0.5,7.5) \psline{->}(7.5,5.5)(7.5,7.5)
\psline{->}(0.5,8.0)(0.5,10.0) \psline{->}(7.5,8.0)(7.5,10.0)
\psline{->}(0.5,10.5)(0.5,12.5)

\psline{-}(0.5,0.5)(0.5,0.6) \psline{-}(1.0,0.5)(1.0,0.6)
\psline{-}(1.5,0.5)(1.5,0.6) \psline{-}(2.0,0.5)(2.0,0.6)
\psline{-}(2.5,0.5)(2.5,0.6) \psline{-}(3.0,0.5)(3.0,0.6)
\psline{-}(3.5,0.5)(3.5,0.6) \psline{-}(4.0,0.5)(4.0,0.6)
\psline{-}(4.5,0.5)(4.5,0.6) \psline{-}(5.0,0.5)(5.0,0.6)
\psline{-}(5.5,0.5)(5.5,0.6) \psline{-}(6.0,0.5)(6.0,0.6)

\psline{-}(0.5,3.0)(0.5,3.1) \psline{-}(1.0,3.0)(1.0,3.1)
\psline{-}(1.5,3.0)(1.5,3.1) \psline{-}(2.0,3.0)(2.0,3.1)
\psline{-}(2.5,3.0)(2.5,3.1) \psline{-}(3.0,3.0)(3.0,3.1)
\psline{-}(3.5,3.0)(3.5,3.1) \psline{-}(4.0,3.0)(4.0,3.1)
\psline{-}(4.5,3.0)(4.5,3.1) \psline{-}(5.0,3.0)(5.0,3.1)
\psline{-}(5.5,3.0)(5.5,3.1) \psline{-}(6.0,3.0)(6.0,3.1)

\psline{-}(0.5,5.5)(0.5,5.6) \psline{-}(1.0,5.5)(1.0,5.6)
\psline{-}(1.5,5.5)(1.5,5.6) \psline{-}(2.0,5.5)(2.0,5.6)
\psline{-}(2.5,5.5)(2.5,5.6) \psline{-}(3.0,5.5)(3.0,5.6)
\psline{-}(3.5,5.5)(3.5,5.6) \psline{-}(4.0,5.5)(4.0,5.6)
\psline{-}(4.5,5.5)(4.5,5.6) \psline{-}(5.0,5.5)(5.0,5.6)
\psline{-}(5.5,5.5)(5.5,5.6) \psline{-}(6.0,5.5)(6.0,5.6)

\psline{-}(0.5,8.0)(0.5,8.1) \psline{-}(1.0,8.0)(1.0,8.1)
\psline{-}(1.5,8.0)(1.5,8.1) \psline{-}(2.0,8.0)(2.0,8.1)
\psline{-}(2.5,8.0)(2.5,8.1) \psline{-}(3.0,8.0)(3.0,8.1)
\psline{-}(3.5,8.0)(3.5,8.1) \psline{-}(4.0,8.0)(4.0,8.1)
\psline{-}(4.5,8.0)(4.5,8.1) \psline{-}(5.0,8.0)(5.0,8.1)
\psline{-}(5.5,8.0)(5.5,8.1) \psline{-}(6.0,8.0)(6.0,8.1)

\psline{-}(0.5,10.5)(0.5,10.6) \psline{-}(1.0,10.5)(1.0,10.6)
\psline{-}(1.5,10.5)(1.5,10.6) \psline{-}(2.0,10.5)(2.0,10.6)
\psline{-}(2.5,10.5)(2.5,10.6) \psline{-}(3.0,10.5)(3.0,10.6)
\psline{-}(3.5,10.5)(3.5,10.6) \psline{-}(4.0,10.5)(4.0,10.6)
\psline{-}(4.5,10.5)(4.5,10.6) \psline{-}(5.0,10.5)(5.0,10.6)
\psline{-}(5.5,10.5)(5.5,10.6) \psline{-}(6.0,10.5)(6.0,10.6)

\psline{-}(7.5,0.5)(7.5,0.6) \psline{-}(8.0,0.5)(8.0,0.6)
\psline{-}(8.5,0.5)(8.5,0.6) \psline{-}(9.0,0.5)(9.0,0.6)
\psline{-}(9.5,0.5)(9.5,0.6) \psline{-}(10.0,0.5)(10.0,0.6)
\psline{-}(10.5,0.5)(10.5,0.6) \psline{-}(11.0,0.5)(11.0,0.6)
\psline{-}(11.5,0.5)(11.5,0.6) \psline{-}(12.0,0.5)(12.0,0.6)
\psline{-}(12.5,0.5)(12.5,0.6) \psline{-}(13.0,0.5)(13.0,0.6)

\psline{-}(7.5,3.0)(7.5,3.1) \psline{-}(8.0,3.0)(8.0,3.1)
\psline{-}(8.5,3.0)(8.5,3.1) \psline{-}(9.0,3.0)(9.0,3.1)
\psline{-}(9.5,3.0)(9.5,3.1) \psline{-}(10.0,3.0)(10.0,3.1)
\psline{-}(10.5,3.0)(10.5,3.1) \psline{-}(11.0,3.0)(11.0,3.1)
\psline{-}(11.5,3.0)(11.5,3.1) \psline{-}(12.0,3.0)(12.0,3.1)
\psline{-}(12.5,3.0)(12.5,3.1) \psline{-}(13.0,3.0)(13.0,3.1)

\psline{-}(7.5,5.5)(7.5,5.6) \psline{-}(8.0,5.5)(8.0,5.6)
\psline{-}(8.5,5.5)(8.5,5.6) \psline{-}(9.0,5.5)(9.0,5.6)
\psline{-}(9.5,5.5)(9.5,5.6) \psline{-}(10.0,5.5)(10.0,5.6)
\psline{-}(10.5,5.5)(10.5,5.6) \psline{-}(11.0,5.5)(11.0,5.6)
\psline{-}(11.5,5.5)(11.5,5.6) \psline{-}(12.0,5.5)(12.0,5.6)
\psline{-}(12.5,5.5)(12.5,5.6) \psline{-}(13.0,5.5)(13.0,5.6)

\psline{-}(7.5,8.0)(7.5,8.1) \psline{-}(8.0,8.0)(8.0,8.1)
\psline{-}(8.5,8.0)(8.5,8.1) \psline{-}(9.0,8.0)(9.0,8.1)
\psline{-}(9.5,8.0)(9.5,8.1) \psline{-}(10.0,8.0)(10.0,8.1)
\psline{-}(10.5,8.0)(10.5,8.1) \psline{-}(11.0,8.0)(11.0,8.1)
\psline{-}(11.5,8.0)(11.5,8.1) \psline{-}(12.0,8.0)(12.0,8.1)
\psline{-}(12.5,8.0)(12.5,8.1) \psline{-}(13.0,8.0)(13.0,8.1)

\rput(1.5,0.25){{\small $1$}} \rput(2.5,0.25){{\small $2$}}
\rput(3.5,0.25){{\small $3$}} \rput(4.5,0.25){{\small $4$}}
\rput(5.5,0.25){{\small $5$}}

\rput(8.5,0.25){{\small $1$}}  \rput(9.5,0.25){{\small $2$}}
\rput(10.5,0.25){{\small $3$}} \rput(11.5,0.25){{\small $4$}}
\rput(12.5,0.25){{\small $5$}}

\psline{-}(0.5,10.5)(1.0,11.0) \psline{-}(1.0,11.0)(1.5,11.5)
\psline{-}(1.5,11.5)(2.0,12.0) \psline{-}(2.0,12.0)(2.5,12.5)
\psline{-}(2.5,12.5)(3.0,12.0) \psline{-}(3.0,12.0)(3.5,11.5)
\psline{-}(3.5,11.5)(4.0,11.0) \psline{-}(4.0,11.0)(4.5,11.5)
\psline{-}(4.5,11.5)(5.0,11.0) \psline{-}(5.0,11.0)(5.5,11.5)
\psline{-}(5.5,11.5)(6.0,11.0)

\psline{-}(0.5,8.0)(1.0,8.5) \psline{-}(1.0,8.5)(1.5,9.0)
\psline{-}(1.5,9.0)(2.0,9.5) \psline{-}(2.0,9.5)(2.5,10.0)
\psline{-}(2.5,10.0)(3.0,9.5) \psline{-}(3.0,9.5)(3.5,10.0)
\psline{-}(3.5,10.0)(4.0,9.5) \psline{-}(4.0,9.5)(4.5,9.0)
\psline{-}(4.5,9.0)(5.0,8.5) \psline{-}(5.0,8.5)(5.5,9.0)
\psline{-}(5.5,9.0)(6.0,8.5)

\psline{-}(0.5,5.5)(1.0,6.0) \psline{-}(1.0,6.0)(1.5,6.5)
\psline{-}(1.5,6.5)(2.0,6.0) \psline{-}(2.0,6.0)(2.5,6.5)
\psline{-}(2.5,6.5)(3.0,7.0) \psline{-}(3.0,7.0)(3.5,7.5)
\psline{-}(3.5,7.5)(4.0,7.0) \psline{-}(4.0,7.0)(4.5,6.5)
\psline{-}(4.5,6.5)(5.0,6.0) \psline{-}(5.0,6.0)(5.5,6.5)
\psline{-}(5.5,6.5)(6.0,6.0)

\psline{-}(0.5,3.0)(1.0,3.5) \psline{-}(1.0,3.5)(1.5,4.0)
\psline{-}(1.5,4.0)(2.0,4.5) \psline{-}(2.0,4.5)(2.5,5.0)
\psline{-}(2.5,5.0)(3.0,4.5) \psline{-}(3.0,4.5)(3.5,5.0)
\psline{-}(3.5,5.0)(4.0,4.5) \psline{-}(4.0,4.5)(4.5,5.0)
\psline{-}(4.5,5.0)(5.0,4.5) \psline{-}(5.0,4.5)(5.5,4.0)
\psline{-}(5.5,4.0)(6.0,3.5)

\psline{-}(0.5,0.5)(1.0,1.0) \psline{-}(1.0,1.0)(1.5,1.5)
\psline{-}(1.5,1.5)(2.0,1.0) \psline{-}(2.0,1.0)(2.5,1.5)
\psline{-}(2.5,1.5)(3.0,2.0) \psline{-}(3.0,2.0)(3.5,2.5)
\psline{-}(3.5,2.5)(4.0,2.0) \psline{-}(4.0,2.0)(4.5,2.5)
\psline{-}(4.5,2.5)(5.0,2.0) \psline{-}(5.0,2.0)(5.5,1.5)
\psline{-}(5.5,1.5)(6.0,1.0)

\psline{-}(7.5,8.0)(8.0,8.5) \psline{-}(8.0,8.5)(8.5,9.0)
\psline{-}(8.5,9.0)(9.0,8.5) \psline{-}(9.0,8.5)(9.5,9.0)
\psline{-}(9.5,9.0)(10.0,8.5) \psline{-}(10.0,8.5)(10.5,9.0)
\psline{-}(10.5,9.0)(11.0,9.5) \psline{-}(11.0,9.5)(11.5,10.0)
\psline{-}(11.5,10.0)(12.0,9.5) \psline{-}(12.0,9.5)(12.5,9.0)
\psline{-}(12.5,9.0)(13.0,8.5)

\psline{-}(7.5,5.5)(8.0,6.0) \psline{-}(8.0,6.0)(8.5,6.5)
\psline{-}(8.5,6.5)(9.0,7.0) \psline{-}(9.0,7.0)(9.5,7.5)
\psline{-}(9.5,7.5)(10.0,7.0) \psline{-}(10.0,7.0)(10.5,6.5)
\psline{-}(10.5,6.5)(11.0,6.0) \psline{-}(11.0,6.0)(11.5,5.5)
\psline{-}(11.5,5.5)(12.0,6.0) \psline{-}(12.0,6.0)(12.5,6.5)
\psline{-}(12.5,6.5)(13.0,6.0)

\psline{-}(7.5,3.0)(8.0,3.5) \psline{-}(8.0,3.5)(8.5,4.0)
\psline{-}(8.5,4.0)(9.0,4.5) \psline{-}(9.0,4.5)(9.5,5.0)
\psline{-}(9.5,5.0)(10.0,4.5) \psline{-}(10.0,4.5)(10.5,4.0)
\psline{-}(10.5,4.0)(11.0,4.5) \psline{-}(11.0,4.5)(11.5,5.0)
\psline{-}(11.5,5.0)(12.0,4.5) \psline{-}(12.0,4.5)(12.5,4.0)
\psline{-}(12.5,4.0)(13.0,3.5)

\psline{-}(7.5,0.5)(8.0,1.0) \psline{-}(8.0,1.0)(8.5,1.5)
\psline{-}(8.5,1.5)(9.0,1.0) \psline{-}(9.0,1.0)(9.5,0.5)
\psline{-}(9.5,0.5)(10.0,1.0) \psline{-}(10.0,1.0)(10.5,1.5)
\psline{-}(10.5,1.5)(11.0,2.0) \psline{-}(11.0,2.0)(11.5,2.5)
\psline{-}(11.5,2.5)(12.0,2.0) \psline{-}(12.0,2.0)(12.5,1.5)
\psline{-}(12.5,1.5)(13.0,1.0)

\rput(6.0,12.5){{\small$w=4$,$(2,0,0,1)$}}
\rput(6.0,10.0){{\small$w=5$,$(2,0,0,1)$}}
\rput(6.0,7.5){{\small$w=6$,$(2,0,0,1)$}}
\rput(6.0,5.0){{\small$w=6$,$(2,0,0,1)$}}
\rput(6.0,2.5){{\small$w=7$,$(2,0,0,1)$}}
\rput(13.5,10.0){{\small$w=8$,$(2,0,0,1)$}}
\rput(13.5,7.5){{\small$w=8$,$(0,1,0,1)$}}
\rput(13.5,5.0){{\small$w=9$,$(0,1,0,1)$}}
\rput(13.5,2.5){{\small$w=10$,$(0,1,0,1)$}}

\end{pspicture}
\end{center}
\end{figure}
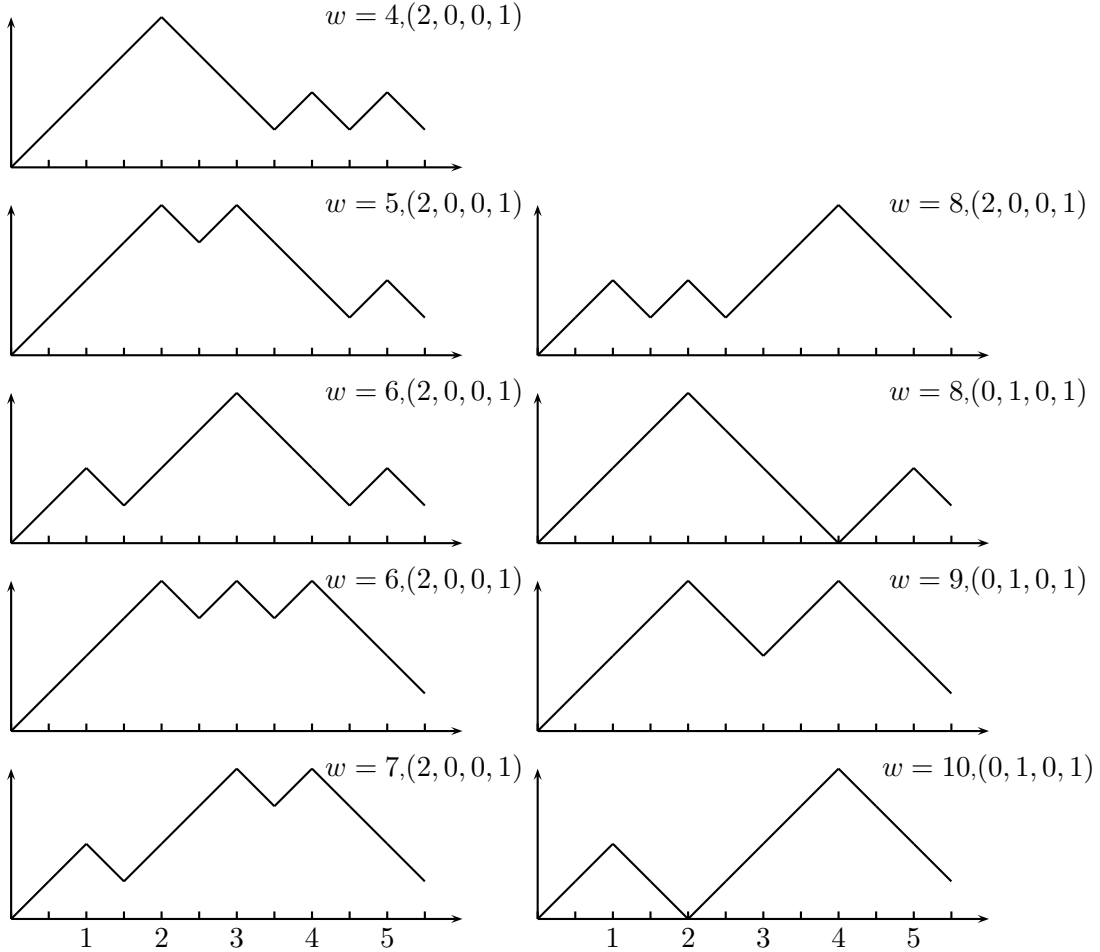

\subsection{The $m$-$n$ system }

It is convenient to attribute  a special name to $p_j+2\e_j$; we call it $m_j$:
\begin{equation}\label{defm}
m_j= \sum_{l= j+\frac12}^k 2(l-j) n_l .
\end{equation}
This definition ensures that $m_j$ is integer.
In \cite{OleJS}, $m_j$ is interpreted as the number of antiparticles with charges $-j$. Note that $m_j$ is independent of $n_\frac12$ and that there are no antiparticles of charge $-k$.
The above relation can  be reverted:
\begin{equation}
n_j= m_{j-\frac12}-2m_j+m_{j+\frac12}+ 2m \delta_{j,\frac12},
\end{equation}
with the understanding that $m_0=m_k=0$.
The expression for ${\cal W}^{(k)}$  in (\ref{defW}) takes the following equivalent form:
\begin{equation}\label{Wenm}
{\cal W}^{(k)}= \frac12\sum_{i,j=\frac12}^{k-\frac12} m_i C_{ij}m_j +\sum_{j=\frac12}^{k-\frac12} (-1)^{2j}m_j\;,
\end{equation}
with 
\begin{equation}\label{Car}
C_{ij}=- \delta_{i,j-\frac12}+2 \delta_{i,j}-  \delta_{i,j+\frac12}\;.
\end{equation}
This is precisely the Cartan matrix of the $A_{2k-1}$ Lie algebra but  with all the root labels divided by 2.
The combinatorial factor can also be expressed in terms of the variables $m_j$ as:
\begin{equation}
m_j+n_j = m_{j-\frac12}-m_j +m_{j+\frac12} +2m \delta_{j,\frac12}
\end{equation}
We can write the  right hand side in matrix form as
\begin{equation}
m_j+n_j =  \sum_{l=\frac12}^{k-\frac12}{\cal J}_{jl}\, m_l +2m \delta_{j,\frac12}\;,
\end{equation}
with 
\begin{equation} {\cal J}_{jl}= \delta_{j,l-\frac12}- \delta_{j,l}+  \delta_{j,l+\frac12}.
\end{equation}
This matrix  ${\cal J}$  is  related to   $C$ as ${\cal J}= I-C$ where $I$ is the identity matrix.


\subsection{Generating function for all paths}

Collecting the results of the previous sections, we have obtained the following generating  function of all paths starting at (0,0) and ending at $(L,\tfrac12)$, with fixed charge content:
\begin{equation} \label{GF}G^{(k)}(n_\frac12, \cdots, n_k ;q)= q^{{\cal W}^{(k)}}\,  \prod_{j=\frac12}^{k-\frac12} \begin{bmatrix}
m_j+n_j-2\e_j\\ n_j\end{bmatrix}\;,
\end{equation}
where 
${\cal W}^{(k)}$ is defined in (\ref{defW}) or in (\ref{Wenm}) and $m_j$ in (\ref{defm}).
Summing over all configurations with fixed total charge $m=\sum jn_j$ yields
\begin{equation} \label{finica}
G^{(k)}_m(q)= \sum_{\substack{ n_\frac12, n_1,\cdots, n_k=0\\ \sum jn_j=m}}^\y q^{{\cal W}^{(k)}} \, \prod_{j=\frac12}^{k-\frac12} \begin{bmatrix}
m_j+n_j-2\e_j\\ n_j\end{bmatrix}\;.
\end{equation}
This is an expression for the  finitized version of the $\M(k+1,2k+3)$ vacuum character.
This can equally be written solely in terms of the $m_j$ variables as 
\begin{equation} \label{forM}
G^{(k)}_m(q)= \sum_{m_\frac12, m_1,\cdots, m_{k-\frac12}=0 }^\y q^{{\cal W}^{(k)}} \, \prod_{j=\frac12}^{k-\frac12}
 \begin{bmatrix}({\cal J} m)_j+2m\delta_{j,\frac12} 
- 2\e_j\\ m_j -2\e_j \end{bmatrix}
\end{equation} In  the second expression, the finitized version is not  manifest from a constraint as an upper bound in the  summation variables but rather by  the presence of the term $2m$ in the $q$-binomial coefficients. Recall in that regard  that: 
\begin{equation} 
\begin{bmatrix}
N\\ n\end{bmatrix}\not=0 \quad \text{only when}\quad 0\leq n\leq N \quad \text{or}\quad n=0,\;N=-1\;.
\end{equation}  This also makes manifest the constraint $m_j\geq 2\e_j$.

The conformal limit is obtained by setting $m\rw\y$ via $n_\frac12\rw \y$. Using 
\begin{equation} 
\lim_{n\rw\y}  \begin{bmatrix}
n\\ p \end{bmatrix}= \frac{1}{(q)_p}\;,
\end{equation} 
we get
\begin{equation} 
\chi_{1,1}^{(k+1,2k+3)}(q)=\lim_{n_\frac12\rw\y} G^{(k)}_m(q) =    \sum_{n_1,n_\frac32,\cdots, n_k=0}^\y \frac{ q^{{\cal W}^{(k)}} } {(q)_{m_\frac12}} \, \prod_{j=\frac12}^{k-\frac12} \begin{bmatrix}
m_j+n_j-2\e_j\\ n_j\end{bmatrix}\;,
\end{equation}
or equivalently, 
\begin{equation} 
\chi_{1,1}^{(k+1,2k+3)}(q) =    \sum_{m_\frac12, m_1,\cdots, m_{k-\frac12}=0 }^\y \frac{ q^{{\cal W}^{(k)}} } {(q)_{m_\frac12}} \, \prod_{j=\frac12}^{k-\frac12} \begin{bmatrix}
({\cal J} m)_j
- 2\e_j  \\ m_j-2\e_j\end{bmatrix}\; .
\end{equation}

For instance, for $k=1$,  with $m_\frac12 = n$, we have 
\begin{equation}  \chi_{1,1}^{(2,5)}(q)= \sum_{ n=1}^\y \frac{q^{n^2-n} }{(q)_{n-1}} = 
 \sum_{ n=0}^\y \frac{q^{n^2+n} }{(q)_{n}}\end{equation}
 where   in the last step we have replaced $n\rw n+1$. This agrees with the known  vacuum character of the $\M(2,5)$ model \cite{FNO}. For $k=2$, this yields 
\begin{equation}  \chi_{1,1}^{(3,7)}(q)= \sum_{m_\frac12, m_1,m_\frac32\geq 0} \frac{q^{{\cal W}^{(2)} } }{(q)_{m_\frac12-1}} 
\begin{bmatrix}
m_\frac12-m_1+m_\frac32 \\ m_1\end{bmatrix} 
\begin{bmatrix}
m_1-m_\frac32-1\\m_\frac32-1\end{bmatrix} 
\end{equation}
This fermionic form is quite different form the specialization to $p=7$ of the general expression  obtained in \cite{Byt,Jaca,Jacb} for the vacuum $\M(3,p)$ characters (which have no $q$-binomial factors and a different set of quasi-particles). 
Similarly, this expression for the vacuum character of the  $\M(k+1,2k+3)$ models for $k\geq 2$ differs structurally (by the number of modes  that are summed and the form of the $q$-binomial factors) from the previously found expressions in \cite{KKMMb,FQ}.

\subsection{Computation of the central charge} 

The identification of the above  generating function with the vacuum  character of the $\M(k+1,2k+3)$ minimal model relies, for $k>1$, on  the comparison of its $q$-expansion to high order with a standard form of the character.  It is thus of interest to provide an asymptotic confirmation of this correspondence and this  is the subject of this subsection.

 As shown in \cite{Nahm}  (further illustrated in \cite{KKMMa,KKMMb} and reviewed in \cite{Kir}), the fermionic expression of the characters codes the information concerning the central  charge. This connection comes from the  comparison of  (1): the asymptotic expression of the  character (of any irreducible module) in the limit $q\rw 1^-$  obtained by a  saddle-point analysis of the positive multiple sum, with  (2): the leading term of the character obtained by a modular transformation of the expression in the limit $q\rw 0^-$.  For the  type of fermionic character obtained here, in particular, in  the form (\ref{forM}), the appropriate saddle-point analysis is performed in \cite{KKMMb}. We thus only quote the dilogarithmic summation formula that needs  to be checked. 
 
 Let 
 \begin{equation}  {\cal L}(z) = -\frac12\int_0^z dt\,  \frac{ \ln(1-t)}{t} +\frac12 \ln(z)\ln(1-z)\;,
 \end{equation}
 and  $x_i,\, y_i$ be  defined by the relations (cf. (eqs (5.5) in \cite{KKMMb}):
 \begin{equation}\label{xy}
 1-x_i= \prod_{j=\frac12}^{k-\frac12} x_j^{C_{ij}} \qquad {\rm and} \qquad 1-y_i= \prod_{j=1}^{k-\frac12} y_j^{C_{ij}}\;.
 \end{equation}
Note that the two set of equations are identical  except that the second one does not involve $y_{\frac12}$, which  is forced to be $1$. We recall that $C$ stands for the $A_{2k-1}$-Cartan matrix  with all  the node labels divided by 2. 
The general solution  of these equations is known to be\footnote{This solution was pointed out to us by O. Warnaar.}
 \begin{equation}  x_i = S[i;2k+3]\ \qquad i=\frac12,1,\cdots,k-\frac12 \;,
  \end{equation}
 and \
 \begin{equation} y_{i+\frac12}=S[i;2k+2]\qquad i=\frac12,1,\cdots,k-1\;,
  \end{equation}
   where 
 \begin{equation}    S[i;\kappa]\equiv  \frac{\sin \frac{\pi}{ \kappa} } {  \sin\frac{(2i+1)\pi }{ \kappa} }   \frac{ \sin\frac{2\pi }{ \kappa} }{ \sin\frac{(2i+2) \pi }{ \kappa} }\;.
   \end{equation}

The $\M(k+1,2k+3)$ effective central charge 
 \begin{equation} c_{{\rm eff}}=c-24 h_{{\rm min}}= 1-\frac6{(k+1)(2k+3)}\;,
 \end{equation}
 must be related to the following sums of dilogarithms \cite{KKMMb}:
 \begin{equation}\label{diLOG}
 \sum_{i=\frac12}^{k-\frac12} \left[ {\cal L}(1-x_i)-  {\cal L}(1-y_i)\right] = \frac{\pi^2}6 \,c_{{\rm eff}} = \frac{\pi^2}6 \frac{ (3+k)(2k-1)}{(k+1)(2k+3)}\;.
 \end{equation}
 For $k=1$, this reduces to the well-known identity  (cf. \cite{Kir} Eq 1.10):
   \begin{equation}
 {\cal L}\left(\frac12(\sqrt{5}-1)\right)= \frac{\pi^2}{10} \;.\end{equation}
 (since the relations  (\ref{xy}) become simply $1-x=x^2$ with $x\equiv x_\frac12$).

The relation (\ref{diLOG}) has been checked explicitly for very high values of $k$. This  provides an indirect test of the new character expression of the $\M(k+1,2k+3))$ models.  Note that the dilogarithmic identity (\ref{diLOG}) reduces to (expressing the right-hand side as $f(k)-f(k-\frac12)$): 
 \begin{equation}
 \sum_{i=\frac12}^{k-\frac12}  {\cal L}(1-x_i)=  \frac{\pi^2}6  \frac{ 2 k\,  (2k-1)} {(2k+3)}\;.
 \end{equation}

\def\W{{\cal W}}
\section{Other modules}

\subsection{Characterization of the irreducible  modules  in terms of boundary conditions}

The irreducible  character of the highest-weight module $|\phi_{r,s}\R$, with $1\leq r\leq k$ and $1\leq s\leq 2k+1$, with $s$ odd,  is the generating function of the paths with boundary conditions prescribed by $r$ and $s$:  
 \begin{equation}
 r=b+\frac12\; , \qquad s=2a+1\;.
 \end{equation}
 The path is  still required to terminate with a SE edge. 
There are thus $k+1$ values of $s$ and $k$ values of $r$, making a total of $k(k+1)$ distinct fields, as it should.

Note that the path description does not generate two different descriptions of the same module, one for $\phi_{r,s}$ and the other for $\phi_{k+1-r,2k+3-s}$ (and recall that $\phi_{r,s}= \phi_{p'-r,p-s}$). 

Modifying the boundary conditions affects the path generating function in two ways: the minimal weight configuration is modified (which thereby  affects its weight) and the combinatorial factor is modified. The analysis of these separate effects is  considered  in the following two subsections.

\subsection{Dependence of the minimal-weight configuration upon the boundary conditions}

Our first goal is to determine the modification in the expression of  the minimal-weight configuration   when the boundary conditions are changed from $(0,\frac12)$ to $(a,b)$. Let us denote this modified expression as $\W^{(k)}_{(a,b)}$ (with  $\W^{(k)}_{(0,\frac12)}\equiv\W^{(k)}$).  

Let us first consider the effect of modifying the vertical position of the initial vertex  from 0 to $a\not=0$. Figure  \ref{fig12} (a) displays the minimal-weight configuration for $a=2$ (and here we can take $k=2)$. It is clear from the figure  that increasing $y_0$ from 0 to $a$  induces a displacement of all the peaks by $a$. (Of course, here we suppose that $L\rw L+a)$. Since a peak of charge $j$  centered at $x$ contributes $(2j-1)x$ to the weight, one has
\begin{equation}  
\W^{(k)}_{(a,\frac12)} = \W^{(k)}_{(0,\frac12)}+ a\sum_{j=1}^k (2j-1)n_j \;.
\end{equation}

Let us next  return to the case where $a=0$ and consider next the effect resulting from changing the final height from $\frac12$ to $b$.  Take for instance $b=\frac32$. We see from  Fig. \ref{fig12} (b) that the 
minimal-weight configuration has some edges removed (here, the two ones that previously connected the last peak of charge $\frac32$ to the $x$-axis).  As a result, the peaks of charge 1 and $\frac12$ are displaced toward the left by one unit each. This illustrates the general pattern: increasing $b$  by one unit, two edges need to be removed at the end of the sequence of peaks of charge $b$ and all the peaks with charge lower than $b$  get displaced by one unit. These two effects are captured in the expression:
\begin{equation}  
\W^{(k)}_{(0,b)} = \W^{(k)}_{(0,\frac12)} -\l(b-\frac12\r)\sum_{j=b}^k 2j n_j + \sum_{\substack{b'=\frac12\\ b'\in\NN+\frac12}}^b \sum_{j=1}^{b'-\frac12}(2j-1)  n_j\;.
\end{equation} 
The first correction term is related to the removing of two edges each times $b$ increases by 1 and the double sum describes the effect of moving the peaks of charge $< b$  toward the left. (Here we assume that  $L\rw L-(b-\frac12)$.)



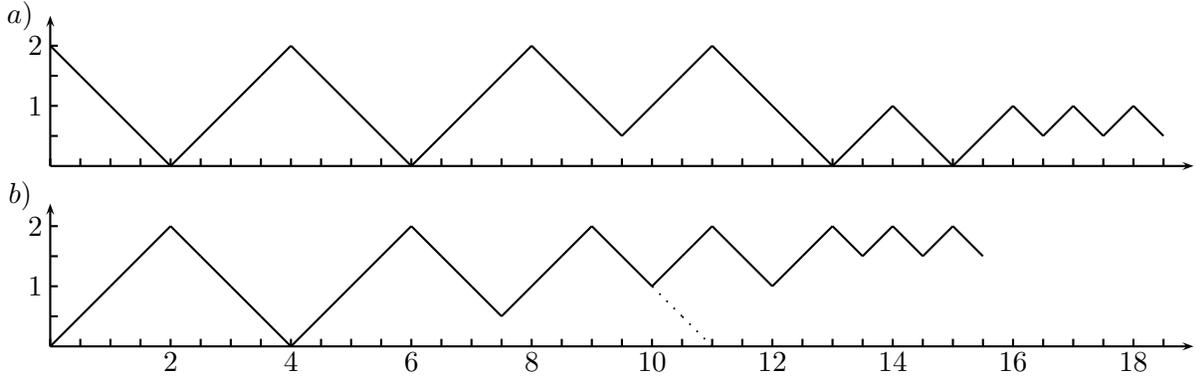
\begin{figure}[ht]
\caption{{\footnotesize The minimal-weight configurations (as modified from Fig. \ref{fig4}) in  the case where $a$ is changed from 0 to 2 in (a) and when $b$ is changed from $\frac12$ to $\frac32$ in  (b). In the second case, the two edges that are removed are indicated by dotted lines.}} \label{fig12}
\begin{center}
\begin{pspicture}(0,0)(26.0,4.8)

\psline{->}(0.4,0.4)(15.6,0.4) \psline{->}(0.4,2.8)(15.6,2.8)
\psline{->}(0.4,0.4)(0.4,2.3) \psline{->}(0.4,2.8)(0.4,4.8)

\psline{-}(0.4,0.4)(0.4,0.5) \psline{-}(0.8,0.4)(0.8,0.5)
\psline{-}(1.2,0.4)(1.2,0.5) \psline{-}(1.6,0.4)(1.6,0.5)
\psline{-}(2.0,0.4)(2.0,0.5) \psline{-}(2.4,0.4)(2.4,0.5)
\psline{-}(2.8,0.4)(2.8,0.5) \psline{-}(3.2,0.4)(3.2,0.5)
\psline{-}(3.6,0.4)(3.6,0.5) \psline{-}(4.0,0.4)(4.0,0.5)
\psline{-}(4.4,0.4)(4.4,0.5) \psline{-}(4.8,0.4)(4.8,0.5)
\psline{-}(5.2,0.4)(5.2,0.5) \psline{-}(5.6,0.4)(5.6,0.5)
\psline{-}(6.0,0.4)(6.0,0.5) \psline{-}(6.4,0.4)(6.4,0.5)
\psline{-}(6.8,0.4)(6.8,0.5) \psline{-}(7.2,0.4)(7.2,0.5)
\psline{-}(7.6,0.4)(7.6,0.5) \psline{-}(8.0,0.4)(8.0,0.5)
\psline{-}(8.4,0.4)(8.4,0.5) \psline{-}(8.8,0.4)(8.8,0.5)
\psline{-}(9.2,0.4)(9.2,0.5) \psline{-}(9.6,0.4)(9.6,0.5)
\psline{-}(10.0,0.4)(10.0,0.5) \psline{-}(10.4,0.4)(10.4,0.5)
\psline{-}(10.8,0.4)(10.8,0.5) \psline{-}(11.2,0.4)(11.2,0.5)
\psline{-}(11.6,0.4)(11.6,0.5) \psline{-}(12.0,0.4)(12.0,0.5)
\psline{-}(12.4,0.4)(12.4,0.5) \psline{-}(12.8,0.4)(12.8,0.5)
\psline{-}(13.2,0.4)(13.2,0.5) \psline{-}(13.6,0.4)(13.6,0.5)
\psline{-}(14.0,0.4)(14.0,0.5) \psline{-}(14.4,0.4)(14.4,0.5)
\psline{-}(14.8,0.4)(14.8,0.5) \psline{-}(15.2,0.4)(15.2,0.5)

\psline{-}(0.4,2.8)(0.4,2.9) \psline{-}(0.8,2.8)(0.8,2.9)
\psline{-}(1.2,2.8)(1.2,2.9) \psline{-}(1.6,2.8)(1.6,2.9)
\psline{-}(2.0,2.8)(2.0,2.9) \psline{-}(2.4,2.8)(2.4,2.9)
\psline{-}(2.8,2.8)(2.8,2.9) \psline{-}(3.2,2.8)(3.2,2.9)
\psline{-}(3.6,2.8)(3.6,2.9) \psline{-}(4.0,2.8)(4.0,2.9)
\psline{-}(4.4,2.8)(4.4,2.9) \psline{-}(4.8,2.8)(4.8,2.9)
\psline{-}(5.2,2.8)(5.2,2.9) \psline{-}(5.6,2.8)(5.6,2.9)
\psline{-}(6.0,2.8)(6.0,2.9) \psline{-}(6.4,2.8)(6.4,2.9)
\psline{-}(6.8,2.8)(6.8,2.9) \psline{-}(7.2,2.8)(7.2,2.9)
\psline{-}(7.6,2.8)(7.6,2.9) \psline{-}(8.0,2.8)(8.0,2.9)
\psline{-}(8.4,2.8)(8.4,2.9) \psline{-}(8.8,2.8)(8.8,2.9)
\psline{-}(9.2,2.8)(9.2,2.9) \psline{-}(9.6,2.8)(9.6,2.9)
\psline{-}(10.0,2.8)(10.0,2.9) \psline{-}(10.4,2.8)(10.4,2.9)
\psline{-}(10.8,2.8)(10.8,2.9) \psline{-}(11.2,2.8)(11.2,2.9)
\psline{-}(11.6,2.8)(11.6,2.9) \psline{-}(12.0,2.8)(12.0,2.9)
\psline{-}(12.4,2.8)(12.4,2.9) \psline{-}(12.8,2.8)(12.8,2.9)
\psline{-}(13.2,2.8)(13.2,2.9) \psline{-}(13.6,2.8)(13.6,2.9)
\psline{-}(14.0,2.8)(14.0,2.9) \psline{-}(14.4,2.8)(14.4,2.9)
\psline{-}(14.8,2.8)(14.8,2.9) \psline{-}(15.2,2.8)(15.2,2.9)

\rput(2.0,0.2){{\small$2$}} \rput(3.6,0.2){{\small$4$}}
\rput(5.2,0.2){{\small$6$}} \rput(6.8,0.2){{\small$8$}}
\rput(8.4,0.2){{\small$10$}} \rput(10.0,0.2){{\small$12$}}
\rput(11.6,0.2){{\small$14$}} \rput(13.2,0.2){{\small $16$}}
\rput(14.8,0.2){{\small$18$}}

\psline{-}(0.4,0.8)(0.5,0.8) \psline{-}(0.4,1.2)(0.5,1.2)
\psline{-}(0.4,1.6)(0.5,1.6) \psline{-}(0.4,2.0)(0.5,2.0)

\psline{-}(0.4,3.2)(0.5,3.2) \psline{-}(0.4,3.6)(0.5,3.6)
\psline{-}(0.4,4.0)(0.5,4.0) \psline{-}(0.4,4.4)(0.5,4.4)

\rput(0.2,3.6){{\small $1$}}\rput(0.2,4.4){{\small $2$}}

\rput(0.2,1.2){{\small $1$}}\rput(0.2,2.0){{\small $2$}}

\psline{-}(0.4,4.4)(0.8,4.0) \psline{-}(0.8,4.0)(1.2,3.6)
\psline{-}(1.2,3.6)(1.6,3.2) \psline{-}(1.6,3.2)(2.0,2.8)
\psline{-}(2.0,2.8)(2.4,3.2) \psline{-}(2.4,3.2)(2.8,3.6)
\psline{-}(2.8,3.6)(3.2,4.0) \psline{-}(3.2,4.0)(3.6,4.4)
\psline{-}(3.6,4.4)(4.0,4.0) \psline{-}(4.0,4.0)(4.4,3.6)
\psline{-}(4.4,3.6)(4.8,3.2) \psline{-}(4.8,3.2)(5.2,2.8)
\psline{-}(5.2,2.8)(5.6,3.2) \psline{-}(5.6,3.2)(6.0,3.6)
\psline{-}(6.0,3.6)(6.4,4.0) \psline{-}(6.4,4.0)(6.8,4.4)
\psline{-}(6.8,4.4)(7.2,4.0) \psline{-}(7.2,4.0)(7.6,3.6)
\psline{-}(7.6,3.6)(8.0,3.2) \psline{-}(8.0,3.2)(8.4,3.6)
\psline{-}(8.4,3.6)(8.8,4.0) \psline{-}(8.8,4.0)(9.2,4.4)
\psline{-}(9.2,4.4)(9.6,4.0) \psline{-}(9.6,4.0)(10.0,3.6)
\psline{-}(10.0,3.6)(10.4,3.2) \psline{-}(10.4,3.2)(10.8,2.8)
\psline{-}(10.8,2.8)(11.2,3.2) \psline{-}(11.2,3.2)(11.6,3.6)
\psline{-}(11.6,3.6)(12.0,3.2) \psline{-}(12.0,3.2)(12.4,2.8)
\psline{-}(12.4,2.8)(12.8,3.2) \psline{-}(12.8,3.2)(13.2,3.6)
\psline{-}(13.2,3.6)(13.6,3.2) \psline{-}(13.6,3.2)(14.0,3.6)
\psline{-}(14.0,3.6)(14.4,3.2) \psline{-}(14.4,3.2)(14.8,3.6)
\psline{-}(14.8,3.6)(15.2,3.2)

\psline{-}(0.4,0.4)(0.8,0.8) \psline{-}(0.8,0.8)(1.2,1.2)
\psline{-}(1.2,1.2)(1.6,1.6) \psline{-}(1.6,1.6)(2.0,2.0)
\psline{-}(2.0,2.0)(2.4,1.6) \psline{-}(2.4,1.6)(2.8,1.2)
\psline{-}(2.8,1.2)(3.2,0.8) \psline{-}(3.2,0.8)(3.6,0.4)
\psline{-}(3.6,0.4)(4.0,0.8) \psline{-}(4.0,0.8)(4.4,1.2)
\psline{-}(4.4,1.2)(4.8,1.6) \psline{-}(4.8,1.6)(5.2,2.0)
\psline{-}(5.2,2.0)(5.6,1.6) \psline{-}(5.6,1.6)(6.0,1.2)
\psline{-}(6.0,1.2)(6.4,0.8) \psline{-}(6.4,0.8)(6.8,1.2)
\psline{-}(6.8,1.2)(7.2,1.6) \psline{-}(7.2,1.6)(7.6,2.0)
\psline{-}(7.6,2.0)(8.0,1.6) \psline{-}(8.0,1.6)(8.4,1.2)
\psline{-}(8.4,1.2)(8.8,1.6) \psline{-}(8.8,1.6)(9.2,2.0)
\psline{-}(9.2,2.0)(9.6,1.6) \psline{-}(9.6,1.6)(10.0,1.2)
\psline{-}(10.0,1.2)(10.4,1.6) \psline{-}(10.4,1.6)(10.8,2.0)
\psline{-}(10.8,2.0)(11.2,1.6) \psline{-}(11.2,1.6)(11.6,2.0)
\psline{-}(11.6,2.0)(12.0,1.6) \psline{-}(12.0,1.6)(12.4,2.0)
\psline{-}(12.4,2.0)(12.8,1.6)
\psset{linestyle=dotted} \psline{-}(8.4,1.2)(8.8,0.8)
\psline{-}(8.8,0.8)(9.2,0.4)

\rput(0,4.8){{\small $a)$}} \rput(0,2.4){{\small $b)$}}

\end{pspicture}
\end{center}
\end{figure}

Combining these two  modifications (which do not influence each other), we arrive at the following expression:
 \begin{align}\label{Wmod}
 \W^{(k)}_{(a,b)}& = \W^{(k)}_{(0,\frac12)} + \sum_{j=1}^k \big(j(2a-2b+1) -a+{\rm max}\; (\lceil b-j\rceil,0)\big) n_j 
 \nonumber \\& =\sum_{i,j=1}^k n_iB_{ij}n_j + \sum_{j=1}^k \big( j(2a-2b+1) -a-\lf j\rf +{\rm max}\; (\lceil b-j\rceil,0)\big) n_j \;, 
 \end{align}
where $B_{ij}$ is defined in (\ref{defB}).

As usual with fermionic expressions, we see that the boundary conditions only affect the linear terms in the expression of the weight.

\subsection{Dependence of the number of  configuration upon the boundary conditions}

The combinatorial factor is also modified by a constant term that depends upon the values of $a$ and $b$. Let us write the modified $q$-combinatorial factor as
\begin{equation}
\begin{bmatrix} n_j+p_j\\ n_j\end{bmatrix} \equiv 
\begin{bmatrix} n_j+m_j+u_j^{(0,\frac12)} \\ n_j\end{bmatrix}\rw \begin{bmatrix} n_j+m_j+u_j^{(a,b)} \\ n_j\end{bmatrix}\;, 
\end{equation} 
with $u_j^{(0,\frac12)} = -2\e_j$ and $m_j$ is defined in (\ref{defm}).

Again the effect of the two boundary terms can be studied independently and an analysis similar to the preceding one yields:
\begin{equation}\label{Dmod}
u_j^{(a,b)}= -2\e_j -{\rm max}\, (\lceil b-j\rceil,0)+{\rm min}\, (\lfloor  k+\frac12-j\rfloor,a).
\end{equation}
 
\subsection{Examples}

To illustrate the two key formulae (\ref{Wmod}) and (\ref{Dmod}), we have displayed in three tables, for $k=1,2$ and $3$, the coefficients of the linear terms (identified by the  mode $n_j$ they multiply) and the  correcting combinatorial factors $u_j$,  for the different boundary conditions $(a,b)$.

\begin{table}[htp]
  \centering
  \caption{$\M(2,5)$ data}\label{table1}
\begin{tabular}{c|cc||c||cc}
  \hline
  \hline
  $\phi_{rs}$ & $a$ & $b$ & $n_1$ & $u_{\frac{1}{2}}$ & \\
 \hline
  $\phi_{11}$ & 0 & $1/2$ & -1 & -1  \\
  $\phi_{13}$ & $1$ & 1/2 & 0 & 0  \\
  \hline
\end{tabular}
\end{table}

\begin{table}[htp]
  \centering
  \caption{$\M(3,7)$ data}\label{table2}
\begin{tabular}{c|cc||ccc||cccc}
  \hline
  \hline
  $\phi_{rs}$ & $a$ & $b$ & $n_1$ & $n_{\frac{3}{2}}$ & $n_2$ & $u_{\frac{1}{2}}$ & $u_1$ & $u_{\frac{3}{2}}$ \\
  \hline
  $\phi_{11}$ & 0 & $1/2$ & -1 & -1 & -2 & -1 & 0 & -1  \\
  $\phi_{21}$ & 0 & $3/2$ & -2 & -4 & -6 & -2 & -1 & -1  \\
  $\phi_{13}$ & 1 & $1/2$ & 0 & 1 & 1 & 0 & 1 & 0  \\
  $\phi_{23}$ & 1 & $3/2$ & -1 & -2 & -3 & -1 & 0 & 0  \\
  $\phi_{15}$ & 2 & $1/2$ & 1 & 3 & 4 & 1 & 1 & 0  \\
  $\phi_{25}$ & 2 & $3/2$ & 0 & 0 & 0 & 0 & 0 & 0  \\
  \hline
\end{tabular}
\end{table}

\begin{table}[htp]
  \centering
  \caption{$\M(4,9)$ data}\label{table3}
\begin{tabular}{c|cc||ccccc||ccccc}
  \hline
  \hline
   $\phi_{rs}$ & $a$ & $b$ & $n_1$ & $n_{\frac{3}{2}}$ & $n_2$ & $n_{\frac{5}{2}}$ & $n_5$ & $u_{\frac{1}{2}}$ & $u_1$ & $u_{\frac{3}{2}}$ & $u_2$& $u_{\frac{5}{2}}$\\
  \hline
   $\phi_{11}$ & 0 & $1/2$ & -1 & -1 & -2 & -2 & -3 & -1 & 0 & -1 & 0 & -1 \\
   $\phi_{21}$ & 0 & $3/2$ & -2 & -4 & -6 & -7 & -9 & -2 & -1 & -1 & 0 & -1 \\
   $\phi_{31}$ & 0 & $5/2$ & -3 & -6 & -9 & -12 & -15 & -3 & -2 & -2 & -1 & -1 \\
   $\phi_{13}$ & 1 & $1/2$ & 0 & 1 & 1 & 2 & 2 & 0 & 1 & 0 & 1 & 0\\
   $\phi_{23}$ & 1 & $3/2$ & -1 & -2 & -3 & -3 & -4 & -1 & 0 & 0 & 1 & 0\\
   $\phi_{33}$ & 1 & $5/2$ & -2 & -4 & -6 & -8 & -10 & -2 & -1 & -1 & 0 & 0 \\
   $\phi_{15}$ & 2 & $1/2$ & 1 & 3 & 4 & 6 & 7 & 1 & 2 & 1 & 1 & 0 \\
   $\phi_{25}$ & 2 & $3/2$ & 0 & 0 & 0 & 1 & 1 & 0 & 1 & 1 & 1 & 0 \\
   $\phi_{35}$ & 2 & $5/2$ & -1 & -2 & -3 & -4 & -5 & -1 & 0 & 0 & 0 & 0 \\
   $\phi_{17}$ & 3 & $1/2$ & 2 & 5 & 7 & 10 & 12 & 2 & 2 & 1 & 1 & 0 \\
   $\phi_{27}$ & 3 & $3/2$ & 1 & 2 & 3 & 5 & 6 & 1 & 1 & 1 & 1 & 0 \\
   $\phi_{37}$ & 3 & $5/2$ & 0 & 0 & 0 & 0 & 0 & 0 & 0 & 0 & 0 & 0 \\
  \hline
\end{tabular}
\end{table}

\let\a\alpha

These tables indicate that the path characteristics for the field of lowest conformal dimension, i.e., $\phi_{k,k+1}$,
have the simplest possible form in  that there are no linear terms in $\W^{(k)}$ and all the $u_j$ factors are zero. Let us check that this is generic and not simply a special feature of the lowest values of $k$. Set then $a=k$ and $b=k-\frac12$, with $j\leq k$. Denote by $\alpha_j$ the coefficient of $n_j$. We have, using 
\begin{equation}
{\rm max}\; (\lceil k-\frac12-j\rceil,0)= k-\lc j \rc\qquad {\rm and}\qquad 2j= \lf j \rf + \lc j \rc\;, 
\end{equation}
that
\begin{equation}
\alpha_j= 2j-k+{\rm max}\; (\lceil k-\frac12-j\rceil,0)- \lf j\rf = \lc j\rc +\lf j\rf -k+k-\lc j\rc-\lf j\rf=0\;.
\end{equation}
Similarly, using
\begin{equation}
{\rm min}\; (\lfloor k+\frac12-j\rfloor,k)= \lfloor k+\frac12-j\rfloor = k-\lf j \rf\qquad {\rm and}\qquad 2\e_j= \lc j \rc- \lf j \rf\;, 
\end{equation}
we have that
\begin{equation}
u_j^{(k,k-\frac12)}= \lc j \rc- \lf j \rf\ -k+ \lc j \rc +k-\lf j \rf=0\;. \end{equation}

To complete this discussion, let us show how these boundary-dependent  data can be very easily  obtained  by a simple case-by-case method, for a fixed value of $k$. Consider, for an illustrative purpose, the case $(a,b)=(2,\frac12)$, with $k=3$.  The ground state is as follows: $---+-+-\ldots +-$ (with $+$ and $-$ denoting NE and SE edges respectively).
Consider  the case of a single peak of charge $\frac12$. There are two allowed configurations:
$---+-$ and $-+---$. 
This readily fixes $u_\frac12^{(2,\frac12)}$:
\begin{equation}
\begin{pmatrix}
n_\frac12+u_\frac12^{(2,\frac12)} \\ n_\frac12\end{pmatrix}=\begin{pmatrix}
1+u_\frac12^{(2,\frac12)}\\ 1\end{pmatrix}= 2\qquad\Rightarrow  \qquad u_\frac12^{(2,\frac12)} = 1\;.\end{equation}
 Consider now the following configuration with $n_\frac32=1$: $---+++---$. This configuration is the one of minimal weight among the two possible configurations, the other one being $-+++-----$. That there are two allowed configurations shows  that $u_\frac32^{(2,\frac12)}=0$. With respect to the ground state, the  weight of the minimal-weight configuration is 6. The only non-zero $n_j$ being $n_\frac32$, the sole contributing quadratic term in  $\W^{(3)}_{(2,\frac32)}$ is $3n_\frac32^2$. Writing the (sought for) coefficient of the linear term as $\a_\frac32 $, we have $3n_\frac32^2+\a_\frac32 n_\frac32= 3+\a_\frac32= 6 $, so that $\a_\frac32=3$.

\subsection{Character formulae for all the irreducible modules}

The expressions of $\W^{(k)}_{(a,b)}$ and $u_j^{(a,b)} $, given in (\ref{Wmod}) and (\ref{Dmod}) respectively, are all what is required in order to write the expression of the finitized character for any module. The result is:
\begin{equation}
\chi_{r,s}^{(m)} (q)= \chi_{b+\frac12,2a+1}^{(m)} (q)=\sum_{\substack{n_\frac12, \ldots,n_k=0\\ \sum jn_j = m}}^\y q^{ \W^{(k)}_{(a,b)}}  \prod_{j=\frac12}^{k-\frac12}\begin{bmatrix}
n_j+m_j+u_j^{(a,b)} \\ n_j\end{bmatrix},
\end{equation} 
As stressed previously, for $k>1$, these are completely new expressions for the finitized characters of the $\M(k+1,2k+3)$ minimal models.
As for the vacuum module, the conformal characters are obtained by setting $n_\frac12\rw \y$.

\section{Duality transformation: recovering the characters of graded parafermions}

\subsection{Introducing the duality transformation}

Let us now consider the formal analogue of the path duality transformation that relates the $\M(k+1,k+2)$ and the usual $\z_k$ parafermionic   models \cite{ABF,BMlmp,OleJS, Kyoto,FWa}. The duality is actually  formulated at the level of the respective weight functions: the dual paths are the very same paths considered so far  but weighted differently. The weight of a path appropriate to the description of the  $\M(k+1,2k+3)$ models is   given by  (\ref{weig}). In the dual case, its expression is
\begin{equation}\label{weigdu}
\w= \sum_{x=\frac12}^{L-\frac12} \w(x)\qquad \text{where} \qquad \w(x)= \frac{x}2\left[1-\left|\, y_{x+\frac12}-y_{x-\frac12}\, \right| \right]\;.
\end{equation}
For paths weighted by the  function $\w$, only the local extrema are found to contribute, and each contributes $x/2$. 

The duality relation between the weight functions $w(x) $ and $\w(x)$ can equally be formulated as  the statement that that every vertex of the path contributes $x/2$ to either function:
\begin{equation} w(x)+\w(x)= \frac{x}2.
\end{equation}
For the corresponding  generating functions of paths of lenght $L$
\begin{equation}\label{cfig}
G(q)= \sum_{{\rm paths}}q^{w} \qquad {\rm and }\qquad  {G}^*(q)=\sum_{{\rm paths}}q^\w,
\end{equation}
 the duality transformation reduces to  the interchange $q\lrw 1/q$, up to an $L$-dependent overall factor:
 \begin{equation}\label{cfig}
G(q)= q^{L(L-1)/4} { G}^*(q^{-1})\;.
\end{equation}
The bottom line is that, up to a $L$-dependent correcting factor, the dual characters are obtained by the simple transformation $q\rw 1/q$. 

\subsection{The dual characters}

The question we want to tackle is that of determining which  finitized characters do correspond to the set of states associated to paths of the $\M(k+1,2k+3)$-type but weighted by $\w$. 
With this goal in mind, let us first notice that the resulting  finitized characters should have a smooth infinite length limit. And if for the minimal models the $m\rw \y$ limit is done by taking $n_\frac12\rw\y$, for paths weighted by the function $\w(x)$, the limit is rather achieved by setting $n_k\rw\y$. Indeed, the ground state configuration (as it will be detailed below) is essentially obtained for paths with as many peaks of charge $k$ as possible. And a tail of peaks of charge $k$ should not contribute to the weight of a path relative to the ground state. In other words, the correcting factor should eliminate all  dependence upon the mode $n_k$ in the $q$ exponent.  \footnote{We stress that the necessity of defining the weight relative to the ground state  and  the $q^{L(L-1)/4}$ factor identified before are the two sources of the correcting $L$ factor needed to ensure a well-defined $m\rw\y$  limit.}

To analyze this more precisely and within the simplest possible context, let us again  focus on the vacuum case and consider the duality transformation of the expression (\ref{finica}), where we replace $m_j-2\e_j$ by $p_j$. This yields:
\begin{equation} \label{finicat}
G^{(k)}_m(q^{-1})= \sum_{\substack{ n_\frac12, n_1,\cdots, n_k=0\\ \sum jn_j=m}}^\y q^{-{\cal W}^{(k)}-\sum_{j=\frac12}^{k-\frac12} n_j p_j} \, \prod_{j=\frac12}^{k-\frac12} \begin{bmatrix}
n_j+p_j\\ n_j\end{bmatrix}
\end{equation}
where we used the simply derived transformation:
\begin{equation}
\begin{bmatrix}
n+p\\ n\end{bmatrix}_{q^{-1}}= q^{-np} \begin{bmatrix}
n+p\\ n \end{bmatrix}_q\equiv q^{-np}  \begin{bmatrix}
n+p\\ n \end{bmatrix}
\end{equation}
Let us now introduce a $L$- or $m$-dependent multiplying correction factor of the form $q^{a_1 m^2+a_2m}$ and fix the value of the constants $a_1$ and $a_2$ to be such that the exponent of $q$ no longer depends upon $n_k$. Making explicit the dependence on $n_k$ of ${\cal W}^{(k)}+\sum_{j=\frac12}^{k-\frac12} n_j p_j $, we have
\begin{align}
 {\cal W}^{(k)}+\sum_{j=\frac12}^{k-\frac12} n_j p_j& =k(2k-1)n_k^2+2kn_k\sum_{j=\frac12}^{k-\frac12} (2j-1)n_j  -kn_k+2n_k \sum_{j=\frac12}^{k-\frac12} (k-j)n_j+\ldots\nonumber \\  
&= \frac{2k-1}{k}\l [k^2n_k^2+2k n_k \sum_{j=\frac12}^{k-\frac12} jn_j \r] -kn_k+\ldots
\end{align}
This is to be compared with $a_1m^2+a_2m$, with $m=\sum_{j=\frac12}^{k} jn_j$,  which fixes the value of the two constants to
\begin{equation}
a_1 = \frac{2k-1}{k} \qquad{\rm and}\qquad  a_2 = -1\;. 
\end{equation}
The resulting $q$-exponent can then be reexpressed as $q^{h^{(k)}_0}$ (where the 0 reminds that this holds for the dual vacuum), that is
\begin{align}
h^{(k)}_0& =  2m^2-m-\frac{m^2}{k}- {\cal W}^{(k)}-\sum_{j=\frac12}^{k-\frac12} n_j p_j\nonumber \\ &
=\frac12 \sum_{i,j=\frac12}^{k-\frac12} r_{ij} n_in_j+\frac12\sum_{\substack{j=\frac12\\ j\in\NN+\frac12}}^{k-\frac12} n_j- \frac{\mb^2}{k}\;,
\end{align} 
with 
\begin{equation}\label{defmb}
\mb= \sum_{j=\frac12}^{k-\frac12} jn_j = m-kn_k \;. 
\end{equation} 
An even more concise expression uses the inverse of the Cartan matrix $C$ defined in (\ref{Car}):
\begin{equation}\label{henC}
h^{(k)}_{0}= \sum_{i,j=\frac12}^{k-\frac12} n_iC^{-1}_{ij}n_j+\frac12\sum_{\substack{j=\frac12\\ j\in\NN+\frac12}}^{k-\frac12} n_j\;,
\end{equation} 
with
\begin{equation}
C^{-1}_{ij}= {\rm min}\; (i,j) -\frac{ij}k\;.
\end{equation} 
For instance, we have:
\begin{align}
& h^{(1)}_{0}=\frac14( n_\frac12^2+ 2 n_\frac12)\nonumber\\
& h^{(2)}_{0}=\frac18 (3n_\frac12^2+4 n_1^2+3n_\frac32^2+4n_\frac12 n_1 +2n_\frac12 n_\frac32 +4n_1n_\frac32 +4 n_\frac12+  4n_\frac32)
\end{align}
The renormalized dual form of the vacuum character reads thus
\begin{equation} \label{finicatp}
q^{2m^2-m-m^2/k}\, G^{(k)}_m(q^{-1})= \sum_{\substack{ n_\frac12, n_1,\cdots, n_k=0\\ \sum jn_j=m}}^\y q^{h_0^{(k)}} \, \prod_{j=\frac12}^{k-\frac12} \begin{bmatrix}
n_j+p_j\\ n_j\end{bmatrix}
\end{equation}
Its conformal limit is obtained, as said before, by setting $n_k\rw\y$. This implies that $p_j\rw\y$ for all $j$. Using
\begin{equation}\lim_{p\rw\y} \begin{bmatrix}
n+p\\ n \end{bmatrix}= \frac{1}{(q)_n}\;,
\end{equation}  
one gets
\begin{equation} \label{finicatp}
\lim_{m\rw\y} q^{2m^2-m-m^2/k}G^{(k)}_m(q^{-1})= \sum_{ n_\frac12, n_1,\cdots, n_{k-\frac12}=0}^\y \frac{ q^{h_0^{(k)}} }{ (q)_{n_{\frac12}}\cdots (q)_{n_{k-\frac12}} } \end{equation}
The right-hand side  is precisely the expression for the vacuum character for the  graded $\z_k$ parafermionic models \cite{CRS} in a form involving $2k-1$ quasi-particles \cite{JM.A}.

Even if we did not know this particular expression of the graded parafermionic vacuum character,  there are notable features of the resulting multiple sum which readily indicated a connection with parafermionic theories. A first one is  the symmetry of the $q$-exponent,  $h^{(k)}_0$, with respect to the interchange of $n_j$ and $n_{k-j}$ -- which is manifest when this factor is expressed in terms of the inverse Cartan matrix in (\ref{henC}) (see also the above expression for $h^{(2)}_0$). Another clear parafermionic signature is the presence in the correcting factor of the term  $m^2/k$, which is related to the `fractional dimension' of the parafermionic modes (and this will be made explicit below).\footnote{On second thoughts, this latter point raises a delicate issue. As said above, this character expression reflects a quasi-particle formulation in terms of $2k-1$ quasi-particle  types  (cf. the different  factors $(q)_{n_j}$ in the denominator of (\ref{finicatp})). But the correcting factor $m^2/k$ also involves the $n_k$ modes (as it should: it is intended to cancel undesired terms of that sort). This means that the dual paths involve (in a sense to be made precise) $2k$ parafermionic modes (since there is the extra $n_k$ mode in addition to the expected $2k-1$ modes). In other words, the dual paths  do not have an immediate interpretation in terms of the parafermionic quasi-particle basis states. This subtle issue will be fully clarified in the sequel.}

Summarizing the result of this brief analysis, we  have thus shown, at the level of the (vacuum) character that the  $\M(k+1,2k+3)$ and the  graded $\z_k$ parafermions are dual to each other!
In the remaining of this section, this conclusion will be much strengthened. It will indeed be shown that the paths dual to the $\M^{[k]}$ ones are in one-to-one correspondence with  the graded parafermionic basis states. 
The outline of the argument is presented in the next subsection.

 \subsection{Statement of the results: dual  paths vs  parafermionic states }
Let us first define the dual paths in a precise way. 
In the dual case, it will be  understood that all paths terminate on the $x$-axis (the sequence of SE edges added to a path to reach the $x$-axis from $y_L$ do not contribute to the weight). The different classes of paths are thus completely characterized by the initial (integer) 
vertical position $\ell$, with $0\leq \ell\leq k$. 

From now on, we will refer to a $\LP$ path as a path with the following characteristics: 

\n {\bf $\LP$ paths} {\it  are defined in the strip $x\geq 0$ and $0\leq y\leq k$ in the $(x,y)$  half-integer lattice, with  their peaks at integer positions. They terminate on the $x$-axis and  are weighted by the functions $\w$ defined in (\ref{weigdu}). A path with specified initial point will be denoted $\LP_\ell$.}


Our main result in relation to the dual path description can thus be restated more precisely as follows. 
We provide a one-to-one correspondence between  a $\LP_\ell$ path and a basis state in the graded parafermionic module headed by the highest-weight state of parafermionic charge $\ell$. This is done in two steps.


\begin{enumerate}
\item  We first provide a bijection between the $\LP$ paths to a different type  of paths, namely the $(k-\tfrac12$)-restricted (Bressoud-type \cite{BreL}) paths defined  in the reduced strip  $0\leq y\leq k-\frac12$, but for which horizontal edges  are allowed if they lie on the $x$-axis and whose weight is the sum of the $x$-position of the peaks.  

\item These new lattice paths are shown to be in one-to-one correspondence with the states of the graded $\z_k$ parafermionic models.
\end{enumerate}

The first step is a natural extension of the bijection presented in \cite{Path} for the two known path descriptions of the usual $\z_k$ parafermions. The second one puts together results from \cite{JPath} and \cite{JM.A}.
 
 An immediate corollary  of this result is that, 
after a simple adjustment of the weight (to be explained below),  the generating function for the
$\LP_\ell$ paths with a fixed total charge is equivalent to a  finitized version of the character of the $\z_k$  graded parafermionic irreducible module parametrized by $\ell$.

\subsection{Paths and multiple partitions}

We start by describing a canonical rewriting of the $\LP$ paths \cite{Mult}.  Note at first that for fixed boundary conditions (that is, the specification of $\ell$ in the present context), a path is fully determined by the sequence of its peak positions   and their charge. It will be understood that the sequence of peaks is to be read from right to left (that is, in decreasing values of $x$). Moreover, a peak of charge $i$ at position $x$ will be denoted as $x^{(i)}$ (and recall that $x$ is integer but $i$ can be half-integer and, for $\LP$ paths,  it is bounded by $\frac12\leq i\leq k$).

We next introduce a formal operation which interchanges the ordering  of two peaks \cite{Mult}:
\begin{equation} \label{com}
 x^{(i)} {x'}^{(j)} \quad \rw \quad ({x'}+r_{ij}) ^{(j)} (x-r_{ij})^{(i)} \;, 
 \end{equation}
 where $r_{ij}$ is defined as
 \begin{equation} \label{rij}
 r_{ij} = 2\, {\rm min} \; (i,j)\;, 
 \end{equation}
 After an interchange, the peak interpretation is lost and it is convenient to refer to $x^{(i)}$ as a cluster of charge $i$. This interchange  operation preserves the charges and the integrality of  the new `$x$-coordinates'.
 
 The announced canonical rewriting of the path is the following: using (\ref{com}), we reorder the path clusters (the original sequence of peaks) in increasing values of the charge and, within each subsequence of fixed-charge clusters, with decreasing value of $x$. The result is essentially a sequence of the form
 \begin{equation}\label{Gkmp}
\g_1^{(\frac12)}\, \g_2^{(\frac12)}\cdots   \g^{(\frac12)}_{n_\frac12}\,  \g^{(1)}_{1} \cdots\g^{(1)}_{n_1} \cdots  \g_1^{(k)} \cdots \g^{(k)}_{n_k} \;, 
\end{equation}
with the ordering $\g_l^{(i)}\geq \g_{l+1}^{(i)}$.
The label $n_i$ indicates the number of peaks of charge $i$ in the original path. This sequence is equivalent to the following set of $2k$ ordered partitions:
  \begin{equation}\label{Gk}\Gamma^{(2k)}\equiv(\g^{(\frac12)},\cdots ,  \g^{(k)}), \qquad {\rm with}\qquad 
\g^{(j)}= (\g^{(j)}_1, \cdots , \g^{(j)}_{n_j}).
\end{equation}

The basic characteristics of a path  imply that  the parts within each partition satisfy stronger  conditions than the stated one $\g_l^{(i)}\geq \g_{l+1}^{(i)}$. The characteristics referred to are essentially that there must be  a minimal distance between two peaks of given charge, a distance that depends upon their individual charge as well as the charge content of the peaks in-between.
Indeed, if between two  peaks $ x^{(i)}$ and $ {x'}^{(j)}$ there are peaks all with charge lower than ${\rm min}\, (i,j)$ and whose total charge sums to $c$, then \cite{Mult}:
\begin{equation}\label{dist} x- x' \geq r_{ij}+\chi_{i>j}+2c\;, \end{equation} 
where $ r_{ij} $ is given in ({\ref{rij}) 
 and $\chi_{b} =1$ if $b$ is true and 0 otherwise. 
To be explicit, let us stress that if there are peaks of charge larger that ${\rm min}\, (i,j)$ between a given pair $ x^{(i)}$ and $ {x'}^{(j)}$, the condition does not apply  to this particular pair. However, such a minimal distance condition is applicable  at least to  all pairs of adjacent peaks, where the adjacency condition corresponds to the case where $c=0$.\footnote{This minimal distance condition between adjacent peaks has already been encountered in Sect 3.3. It can also  be illustrated with some of the previous figures. For instance, in Fig. \ref{fig8}, the open circles closest to the center of the largest charged particle from either side, corresponds to the closest possible initial or final vertex of the lower charged particle, from which the minimal distance -- measured  from one peak to the other -- is easily found to match the above result.  Also, in the configuration of  Fig. \ref{fig4}, the distance between adjacent peaks is everywhere minimal except between the third and the fourth peak (at respective position 9 and 12).} In addition, the initial condition $(y_0=\ell$) induces the following bound on the left-most peak, say of charge $j$ and position $x_0$:
\begin{equation}\label{loBg}
   x_{0} \geq \lc  j\rc 
   +{\rm max}(\lf j\rf -k+\ell,0)\;. 
\end{equation}
The  conditions (\ref{dist}) and (\ref{loBg}), that are necessary satisfied by a path,  imply that \cite{Mult}:
\begin{equation} \label{diCg}
\g^{(j)}_l \geq \g^{(j)}_{l+1} + 2j\;, 
\end{equation} and 
\begin{equation}\label{loBg}
   \g^{(j)}_{n_j} \geq  \lc  j\rc 
   +{\rm max}(\lf j\rf -k+\ell,0) + 
2j (n_{j+1}+\cdots +  n_{k}).
\end{equation}

 As an example, the path of Fig. \ref{fig1}
 corresponds to the following sequence
\begin{equation} 
16^{(2)} \,14^{(\frac12)} \,12^{(1)}\,9^{(\frac32)}\,7^{(\frac12)}\,5^{(1)}\,3^{(2)}\,1^{(\frac12)} 
\end{equation}
whose reordering yields 
\begin{equation} \,15^{(\frac12)} \,10^{(\frac12)}\,6^{(\frac12)}\,12^{(1)}\,8^{(1)}\,8^{(\frac32)}\,6^{(2)}\,2^{(2)}\;. 
\end{equation}
The conditions (\ref{diCg}) and (\ref{loBg}) are easily verified for all  $\g^{(j)}_l$. The correponding multiple partition is $( \g^{(\frac12)},\,  \g^{(1)},  \g^{(\frac32)}\,  \g^{(2)})$, with
\begin{equation}  \g^{(\frac12)} = ( 15, 10,6),\qquad \g^{(1)} = ( 12,8) ,\qquad \g^{(\frac32)} = (8), \qquad  \g^{(2)} = (6,2)
.\end{equation}

 After having reordered the peaks of the path in increasing values of the charge, the next step amounts to displacing  all clusters of charge $k$ to the  left using the exchange relation (\ref{com}). This modifies the parts of all the partitions, which are then denoted by $\la_l^{(j)}$. Next, the  charge-$k$ clusters are removed. This transforms the set of $2k$ ordered partitions $\Gamma^{(2k)}$ into a set of $(2k-1)$ ordered partitions  as follows:
  \begin{align}
  \Gamma^{(2k)}= (\g^{(\frac12)},\cdots ,  \g^{(k)}) & \rw (\la^{(k)},\, \la^{(\frac12)},\cdots ,  \la^{(k-\frac12)}) \nonumber \\&\rw 
(\la^{(\frac12)},\cdots ,  \la^{(k-\frac12)}) \equiv \Lambda^{(2k-1)}\;. \end{align}
The modified expression of the parts is 
\begin{equation}\label{glal}
\la^{(k)}_l= \g^{(k)}_l+2\sum_{j=\frac12}^{k-\frac12} jn_j\,\qquad  {\rm and}\qquad \la_l^{(j)} = \g^{(j)}_l-2jn_k \quad (j<k)\;.
\end{equation}

Now, the multiple partition $\Lambda^{(2k-1)} $ is the canonical rewriting of the $k-\frac12$ Bressoud lattice path \cite{BreL}, denoted $\lp$, and defined as follows:

\n {\bf $\lp$ paths} {\it are  defined on a half-integer lattice, within the strip $x\geq 0$ and $0\leq y\leq k-\frac12$, with peaks at integer positions; a path is composed of NE, SE and WE edges, with the WE edges  allowed only if restricted to the $x$-axis.\footnote{The bar in $\lp$ is intended to remind of the possibility of horizontal edges. We also stress that this actually describes a deformation of the Bressoud paths, originally defined for an integer lattice \cite{BreL}.}  They are weighted by \cite{BreL,JPath}:
\begin{equation}\label{weB}
\wb= \sum_{x=\frac12}^{L-\frac12} \wb(x)\qquad \text{where} \qquad \wb(x)= x \delta_{x,{\rm peak}}\;,
\end{equation}
that is, $\wb(x)$ is zero unless $x$ is a peak position, where the weight  is $x$.}

A $\lp$ path is reconstructed from the multiple partition $\Lambda^{(2k-1)}$ by  means of the exchange relations (\ref{com}). The  ordering of the clusters that corresponds to the actual path is uniquely fixed by ensuring the conditions (\ref{dist}) to be everywhere satisfied \cite{Mult}. For instance, the path of Fig. \ref{fig1} becomes the $\lp$  one in Fig. \ref{fig1a}. A more algorithmic description for reconstructing the $\lp$  path  is described in Appendix A.


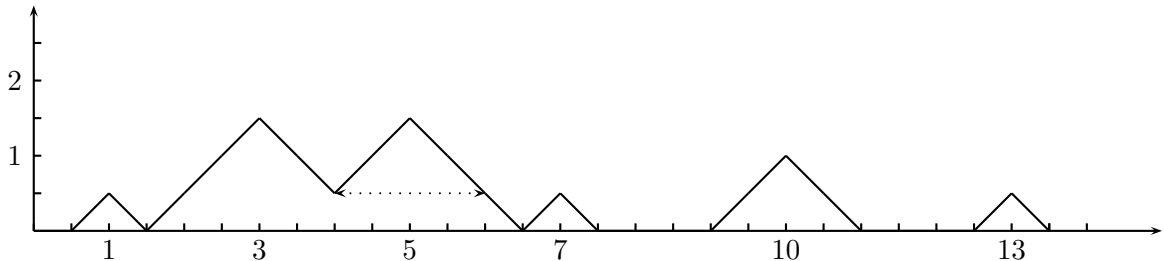
\begin{figure}[ht]
\caption{{\footnotesize The Bressoud path ${{\bar {{\rm P}}^{[\frac32]}}}$ that corresponds to the $\M^{[2]}$ path of Fig. \ref{fig1}.}}
\label{fig1a}
\begin{center}
\begin{pspicture}(0,0)(15.5,4)
\psline{->}(0.5,0.5)(0.5,3.5) \psline{->}(0.5,0.5)(15.5,0.5)
\psset{linestyle=dotted}
\psline{<->}(4.5,1.0)(6.5,1.0)
\psset{linestyle=solid}
\psline{-}(0.5,0.5)(0.5,0.6) \psline{-}(1.0,0.5)(1.0,0.6)
\psline{-}(1.5,0.5)(1.5,0.6) \psline{-}(2.0,0.5)(2.0,0.6)
\psline{-}(2.5,0.5)(2.5,0.6) \psline{-}(3.0,0.5)(3.0,0.6)
\psline{-}(3.5,0.5)(3.5,0.6) \psline{-}(4.0,0.5)(4.0,0.6)
\psline{-}(4.5,0.5)(4.5,0.6) \psline{-}(5.0,0.5)(5.0,0.6)
\psline{-}(5.5,0.5)(5.5,0.6) \psline{-}(6.0,0.5)(6.0,0.6)
\psline{-}(6.5,0.5)(6.5,0.6) \psline{-}(7.0,0.5)(7.0,0.6)
\psline{-}(7.5,0.5)(7.5,0.6) \psline{-}(8.0,0.5)(8.0,0.6)
\psline{-}(8.5,0.5)(8.5,0.6) \psline{-}(9.0,0.5)(9.0,0.6)
\psline{-}(9.5,0.5)(9.5,0.6) \psline{-}(10.0,0.5)(10.0,0.6)
\psline{-}(10.5,0.5)(10.5,0.6) \psline{-}(11.0,0.5)(11.0,0.6)
\psline{-}(11.5,0.5)(11.5,0.6) \psline{-}(12.0,0.5)(12.0,0.6)
\psline{-}(12.5,0.5)(12.5,0.6) \psline{-}(13.0,0.5)(13.0,0.6)
\psline{-}(13.5,0.5)(13.5,0.6) \psline{-}(14.0,0.5)(14.0,0.6)
\psline{-}(14.0,0.5)(14.0,0.6) \psline{-}(14.5,0.5)(14.5,0.6)
\rput(1.5,0.25){{\small $1$}} \rput(3.5,0.25){{\small $3$}}
\rput(5.5,0.25){{\small $5$}} \rput(7.5,0.25){{\small $7$}}
\rput(10.5,0.25){{\small $10$}} \rput(13.5,0.25){{\small $13$}}
 \psline{-}(0.5,1.0)(0.6,1.0)
\psline{-}(0.5,1.5)(0.6,1.5) \psline{-}(0.5,2.0)(0.6,2.0)
\psline{-}(0.5,2.5)(0.6,2.5) \psline{-}(0.5,3.0)(0.6,3.0)
\rput(0.25,1.5){{\small $1$}} \rput(0.25,2.5){{\small $2$}}
\psline{-}(0.5,0.5)(1.0,0.5) \psline{-}(1.0,0.5)(1.5,1.0)
\psline{-}(1.5,1.0)(2.0,0.5) \psline{-}(2.0,0.5)(2.5,1.0)
\psline{-}(2.5,1.0)(3.0,1.5) \psline{-}(3.0,1.5)(3.5,2.0)
\psline{-}(3.5,2.0)(4.0,1.5) \psline{-}(4.0,1.5)(4.5,1.0)
\psline{-}(4.5,1.0)(5.0,1.5) \psline{-}(5.0,1.5)(5.5,2.0)
\psline{-}(5.5,2.0)(6.0,1.5) \psline{-}(6.0,1.5)(6.5,1.0)
\psline{-}(6.5,1.0)(7.0,0.5) \psline{-}(7.0,0.5)(7.5,1.0)
\psline{-}(7.5,1.0)(8.0,0.5) \psline{-}(8.0,0.5)(8.5,0.5)
\psline{-}(8.5,0.5)(9.0,0.5) \psline{-}(9.0,0.5)(9.5,0.5)
\psline{-}(9.5,0.5)(10.0,1.0)
 \psline{-}(10.0,1.0)(10.5,1.5)\psline{-}(10.5,1.5)(11.0,1.0)
\psline{-}(11.0,1.0)(11.5,0.5) \psline{-}(11.5,0.5)(12.0,0.5)
\psline{-}(12.0,0.5)(12.5,0.5) \psline{-}(12.5,0.5)(13.0,0.5)
\psline{-}(13.0,0.5)(13.5,1.0) \psline{-}(13.5,1.0)(14.0,0.5)


\end{pspicture}
\end{center}
\end{figure}



We have thus shown how to relate a $\LP$ path to a $\lp$ one. The inverse procedure is also well defined \cite{Path}: we reorder the peaks of the path $\lp$ to get $\La^{(2k-1)}$ and add to its right a sequence of $n_k$ cluster of charge $k$ all with the minimal values of $\la_l^{(k)} $ allowed and reorder the sequence to reconstruct a $\LP$ path. The value of $n_k$ is fixed as follows: this is the minimal number of clusters of charge $k$ that needs to be introduced for the resulting path to be  free of horizontal edges. 

We have thus found the following sequence of one-to-one correspondences:
\begin{equation}
\LP\leftrightarrow  \Gamma^{(2k)} \leftrightarrow  \La^{(2k-1)} \leftrightarrow \lp\; .
\end{equation} 
The only point that needs to be clarified is the relation between the weight functions.  This is addressed in the next subsection.

Actually, in order to make contact between the paths $\LP$ and the parafermionic  basis of states, we only require the connection  $\LP\leftrightarrow    \La^{(2k-1)}$. However, extending this connection to the Bressoud paths clarifies the relation between the weigh of a $\LP$ path and that of a parafermionic state.

Let us point out that the original basis of states for graded parafermions has been formulated in terms of jagged partitions \cite{JM}. The various links between jagged partitions, $\lp$ paths and multiple partitions are presented in Appendix A. This actually   provides a combinatorial proof of the equivalence of  two bases of states for graded parafermions (jagged vs multiple partitions) via a bijective relation between the states. In  contrast, the proof in \cite{JM.A} relies on a conformal-field-theoretical argument, supported there by the demonstration the equivalence of the generating functions.

 \subsection{Basis for graded parafermions} 

Recall that the graded parafermionic conformal theory is defined by an extension of the usual $\z_k$ parafermionic algebra  by  a $\z_2$ grading, that is, by adding a  new  parafermion $\psi_{1/2}$  of conformal dimension
$1-1/4k$ and such that $\psi_{1/2} \times \psi_{1/2}  \sim \psi_1$, where $(\psi_{1})^{k}\sim I$ \cite{CRS}. The corresponding central charge is
$-3/(2k+3)$.
The parafermionic primary fields
$\phi_\ell$  are  labeled by an integer $\ell$ with
$0\leq\ell\leq k$.
The corresponding highest-weight states are written  $|\phi_\ell\R$ and their conformal dimension 
is
\begin{equation}
h_{\ell}=\frac{\ell(2k-3\ell)}{4k(2k+3)}.
\end{equation}
 The normalization for the parafermionic 
charge is fixed by setting that of 
$\psi_{1/2}$ to be  1 and the charge is defined modulo $2k$ (since $(\psi_{1/2})^{2k}\sim I$).


\def\A{{\cal A}}
The mode decomposition is defined on a generic  field of charge $q$ as follows:
\begin{equation}
{ \psi}_j(z)\varphi_{q}(0)  = \sum_{m=-\y}^\y
z^{-jq/k-m-2j+\lf j\rf }{A}^{(j)}_{j(j+q)/k+m}\, \varphi_{q}(0)  \;,
\end{equation}
Observe that the charge of the field (or the state) on which the mode acts affects its conformal dimension, which is minus its mode index.
In  the following, we omit the fractional part and write
\begin{equation}
{\A}^{(j)}_{m}\equiv  {A}^{(j)}_{j(j+q)/k+m} \;.
\end{equation}


The following basis of states has been obtained in \cite{JM.A}:
\begin{equation}\label{parast}
{\A}^{(\frac12)}_{-\la^{(\frac12)}_1}\cdots {\A}^{(\frac12)}_{-\la^{(\frac12)}_{n_{\frac12}}}{\A}^{(1)}_{-\la^{(1)}_1}\cdots {\A}^{(1)}_{-\la^{(1)}_{\la_{n_1}}}
\cdots {\A}^{(k-\frac12)}_{-\la^{(k-\frac12)}_1} \cdots{\A}^{(k-\frac12)}_{-\la^{(k-\frac12)}_{n_{k-\frac12}}}\; | \phi_\ell\R \;,
\end{equation}
with
\begin{equation}
\la^{(j)}_l \geq \la^{(j)}_{l+1} + 2j\; , \qquad (\text{with} \quad \frac12 \leq j \leq k-\frac12)\;, \end{equation} and the boundary conditions:
\begin{equation}
    \la^{(j)}_{n_j} \geq \lc j\rc+ {\rm max}\, \left(\lf j\rf  + k -\ell,0\right)+
2j (n_{j+\frac12}+n_{j+1}+\cdots +  n_{k-\frac12})\;,
 \end{equation}
  These are precisely the conditions satisfied by the parts $\la^{(j)}_l$ of $\La^{(2k-1)}$ as obtained from the relations (\ref{diCg}), (\ref{loBg}) and (\ref{glal}).

 The weight of the parafermionic state (\ref{parast}) (relative to its highest-weight state) is 
\begin{equation}
h= |\La| -h_{{\rm frac}}^{(\mb)} \;, 
\end{equation} 
where  
\begin{equation}
 |\La| = \sum_{j=1}^{k-\frac12} \sum_{l=1}^{m_j} \la_l^{(j)} \;, \end{equation} 
 while $  h_{{\rm frac}}^{(\mb)}$ stands for the contribution to the conformal dimension that comes from the removed fractional part. This depends solely upon the total charge  $\mb$ (defined in (\ref{defmb})) of the descendant  state.
 This fractional dimension
is easily found to be 
\begin{equation}
h_{{\rm frac}}^{(\mb)}
=\frac{\mb(\mb+\ell)}{k}\;.
\end{equation}
The  peak  described by the cluster $x^{(i)}$ is thus associated to a parafermionic mode $\A_{-x}^{(i)}$.
Note that the charge of the peak $(i)$ is half the parafermionic charge of the corresponding mode $\A^{(i)}$  (which is $2i$).

The rearrangement of the multiple partition $\La^{(2k-1)}$
into a $\lp$ path is done by means of the exchange relation (\ref{com}) and this operation preserves the value of $|\La |$. In other words the weight of the corresponding $\lp$ path is 
\begin{equation}
 \wb = |\La| \;.\end{equation}
 Therefore, to a $\lp$ path of charge $\mb$, one associates a parafermionic state of parafermionic charge $2\mb$ and  conformal dimension
\begin{equation}
h= \wb-h_{{\rm frac}}^{(\mb)} \;.
\end{equation} 

The bijection between $\LP$ and $\lp$ paths entails a natural correspondence between a $\LP$ path and a parafermionic state, or equivalently, between a multiple partition $\Gamma^{(2k)}$ and a parafermionic state. But the resulting  parafermionic state appears in an  unusual form in that it contains a sequence of $\A^{(k)}$ modes at the right  (which are superfluous in the parafermionic context as $\A^{(k)}\sim I$). More explicitly, $\Gamma^{(2k)}$ is associated to the state:
\begin{equation}\label{parask}
{\A}^{(\frac12)}_{-\gamma^{(\frac12)}_1}\cdots {\A}^{(k-\frac12)}_{-\gamma^{(k-\frac12)}_{n_{k-\frac12}}}\, 
{\A}^{(k)}_{-\gamma^{(k)}_1}\cdots {\A}^{(k)}_{-\gamma^{(k)}_{n_{k}}}\; | \phi_\ell\R \;,
\end{equation}
where the $\gamma^{(j)}_l$ satisfy 
 (\ref{loBg}) and (\ref{diCg}).
The relative conformal dimension of this state (whose total parafermionic charge is $2m$) is
\begin{equation}
h= |\Gamma|-h_{{\rm frac}}^{(m)} \;.
\end{equation} 
But $ |\Gamma|$ (the sum of the parts of the multiple partition $\Gamma$) is simply the weight $\wb$ of the $\LP$ path, so that the corresponding parafermionic conformal dimension is 
\begin{equation}
h= \wb-\frac{m(m+\ell)}{k} \;. 
\end{equation} 
We stress that this is $\wb$ and not $\w$ that appears in this expression: the weight of a parafermionic state associated to a  $\LP$ path is now directly expressed in terms of the sum of the peak $x$-positions.

 \subsection{The generating function of  $\LP$ paths}
 
 We are now in position to construct the  generating function of  $\LP$ paths starting at $y_0=\ell$ and  weighted by the relative parafermionic conformal  dimension of the corresponding state. This expression will be  interpreted as a finitized   version of the graded parafermionic character of the irreducible module headed by $|\phi_\ell\R$.
 
The construction of the  generating function will be done in two steps. We first   identify the minimal-weight configuration for a given charge content and calculate its weight. The second step amounts  to determining all the configurations that contribute for a fixed charge content and evaluating their relative weight.

The minimal-weight configuration for $y_0=\ell$ for a given charge content has already been identified;  its weight has also been determined  in \cite{JPath}.\footnote{The result of Sect. 4 of \cite{JPath} is expressed in terms of an integer lattice with peaks at even $x$-positions. Also, the  paths considered there are  Bressoud-type paths. However, because the minimal-weight configuration has no WE edges, the result can be directly lifted to the present context. We simply let $K\rw K+1$, with $K=2k$ and divide all mode labels by 2. The weight $\wb$ of the minimal-weight configuration for a fixed charge content is read off the exponent of $q$ in the numerator of the right-hand side  of the multiple sum in Prop. 4 there, with the adjustment indicated.} It corresponds to  the configuration with peaks ordered in decreasing values of the charge (although a peak of charge 1 has to precedes the peaks of charge $\frac12$  to respect the integer $x$-position for the peaks). Its weight is 
\begin{equation}
w_{{\rm mwc}(\ell)}= \frac12\sum_{i,j=\frac12}^{k} r_{ij} n_in_j+\frac12\sum_{\substack{j=\frac12\\ j\in\NN+\frac12}}^{k-\frac12} n_j+ \frac12\sum_{\substack{j=1\\ j\in\NN}}^k {\rm max}(j-k+\ell,0) n_j\;.
\end{equation} 
The corresponding parafermionic (relative)  conformal dimension is thus 
 \begin{equation}
h_{{\rm mwc}(\ell)} = w_{{\rm mwc}(\ell)}-\frac{m(m+\ell)}{k} \;.\end{equation}
With $m=\mb+kn_k$, we have
\begin{equation}
h_{{\rm mwc}(\ell)}=  \frac12\sum_{i,j=\frac12}^{k-\frac12} r_{ij} n_in_j+\frac12\sum_{\substack{j=\frac12\\ j\in\NN+\frac12}}^{k-\frac12} n_j+ \frac12\sum_{\substack{j=1\\ j\in\NN}}^{k-1} {\rm max}(j-k+\ell,0) n_j- \frac{\mb(\mb+\ell)}{k}\;.
\end{equation} 
We note that this expression no longer depends upon $n_k$.

Consider then the enumeration of all possible configurations and the determination of their weight relative to minimal-weight configuration just identified. This number  is the very same number of configurations identified for the $\M(k+1,2k+3) $ paths.  The only difference is that these are now obtained from the minimal-weight configuration described above by displacing the peaks toward the right. But the net effect is the same: all these configurations contribute to a factor
\begin{equation}\prod_{j=\frac12}^{k-\frac12}\begin{bmatrix}
p_j(\ell)+n_j\\ n_j\end{bmatrix}
\end{equation} 
where $p_j(\ell)= u_j^{(\ell,\frac12)}$ defined in (\ref{Dmod}).

The finitized character of the irreducible module of dimension $h_\ell$ in the  graded parafermionic theory is thus
\begin{equation}
\chi_\ell^{(m)} (q)= \sum_{n_j\geq 0, \sum jn_j = m} q^{h_{{\rm mwc}(\ell)} }  \prod_{j=\frac12}^{k-\frac12}\begin{bmatrix}
p_j(\ell)+n_j\\ n_j\end{bmatrix}\;.
\end{equation}
This is a new result.

The conformal characters are recovered by taking the limit where $m\rw \y$ through $n_k\rw\y$. This yields the expression
\begin{equation}
\chi_\ell (q)= \sum_{n_j\geq 0, \frac12\leq j \leq k-\frac12} \frac{ q^{h_{{\rm mwc}(\ell)} }} { (q)_{n_\frac12} \cdots (q)_{n_{k-\frac12}} } \;,
\end{equation} 
a result that is implicit in \cite{JM.A}.

\section{Conclusion}

We have presented a new path description for the non-unitary $\M(k+1,2k+3)$ minimal models. It differs from the path representation defined by the contour of the configuration of the Forrester-Baxter RSOS model \cite{FB}. This novel description shares with  the $\M(k+1,k+2)$   paths the crucial simplifying feature  that all  vertices except the local extrema contribute $x/2$ to the  weight. But in contrast, these paths are defined  on a   lattice with half-integer spacing. Moreover, the peaks are forced to lie at integer positions -- a constraint that makes the resulting  paths different from the usual  integer-lattice paths rescaled by a factor 2.\footnote{We have then the following situation: the two classes of models $\M(k+1,\a(k+1)+1)$ for $\a=1,2$  are described by similar paths on a lattice with unit step $1/\a$, with peaks at integer positions.  This naturally calls for a quick  analysis of the naive but appealing generalization for $\a>2$, by comparing the enumeration of the corresponding paths with the $q$-expansion of candidate Virasoro characters. Unfortunately, no sensible such irreducible characters are generated in this way.
}

This new $\M(k+1,2k+3)$  path description poses an immediate problem which is to unravel the underlying statistical-model description, that is, the RSOS models whose local state-probabilities can be expressed in terms of  sums of configurations whose contours match the paths described here. In fact, it is fair to say that the  $\M(k+1,2k+3)$  paths have been introduced here without much rationale.

The similarity of the  $\M(k+1,k+2)$ and $\M(k+1,2k+3)$ path descriptions allowed us to lift  rather directly the  analysis of  \cite{OleJS,OleJSb}  and derive  finitized versions of the $\M(k+1,2k+3)$ fermionic characters. The novelty of the resulting expressions calls for another immediate natural problem  which is to demonstrate their genuine equivalence with known fermionic forms or with the bosonic expression.

The present description of the $\M(k+1,2k+3)$ minimal models has another noteworthy property that makes them very similar to that of the unitary models: by a duality transformation, their characters are related to those of a  theory of the parafermionic type. The dual models  are actually  the graded $\z_k$ parafermions \cite{CRS, JM}. Note that our result goes beyond establishing the duality at the level of the characters since  the dual $\M(k+1,2k+3)$  paths have been shown to be in a one-to-one correspondence with the graded parfermionic states. 
Note however that we have not succeeded in explaining this duality transformation in terms of some sort of level-rank duality transformation between the coset representatives of the related  conformal 
models (for the graded parafermions,  the coset representation is $\widehat{osp}(1,2)_k/\widehat{u}(1)$).

Finally, given the existence of a massless renormalization group flow
from the (ultraviolet fixed points)  $\z_k$ models to the
(infrared fixed points) $\M(k+1,k+2)$ minimal models \cite{FZa} and
that they are both related by  duality, the  present results  might
 suggest similar integrable flows between  the graded $\z_k$
 parafermionic theories and the  $\M(k+1,2k+3)$ minimal models.
 If confirmed, this would raise  the   issue of  the meaning of
(path) duality in the context of integrable flows.






\appendix{}

\section{Multiple partitions, h-partitions, jagged partitions and the Bruge correspondence}

\subsection{Multiple partitions and h-partitions}

To a multiple partition $\La^{(2k-1)}$ of total charge $\mb$, i. e., 
\begin{equation}
\mb= \sum_{i=\frac12}^{k-\frac12} j n_j
\end{equation} 
we associate a special  `partition' $(a_1,\cdots, a_{2\mb})$ with $2\mb$ non-increasing parts, some of which  being half-integer, but with the restriction that all half-integer parts must have even frequency. We call such a partition a h-partition (where the `h' reminds of possible half-integer parts). In addition, we enforce the following  difference 1 condition at distance $2k-1$:
\begin{equation} \label{difk}
 a_i\geq a_{i+2k-1}+1\;.
\end{equation}
Equivalently, if $f_{a_i}$ is the frequency of the part equal to $a_i$, we have the following frequency condition:
\begin{equation} f_{a_i} +f_{a_i+\frac12}\leq 2k-1\;. 
\end{equation}
The $\ell$-dependent part of  the boundary condition on the last part of each partition in  $\La^{(2k-1)}$, which results from combining  (\ref{loBg}) and (\ref{glal}), translates into the simple requirement that there must be at most $k-\ell$ pairs of parts equal to $\frac12$ in $(a_1,\cdots, a_{2\mb})$.

The correspondence between a multiple partition  $\La^{(2k-1)}$ and a h-partition is described as follows (which is an adaptation of the presentation in \cite{Mult}).
We start by writing the multiple partition  $\La^{(2k-1)}$ as the partition $(\la_1^{(\frac12)}\cdots  \la_{n_\frac12}^{(\frac12)})$ followed by the sequence of clusters 
 $\la_1^{(1)} \cdots \la_\ell^{(j)} \cdots \la_{n_{k-\frac12}}^{(k-\frac12)}$. Each cluster is then inserted successively  within the partition, starting with $\la_1^{(1)}$ up to $\la_{n_{k-\frac12}}^{(k-\frac12)}$ and by using the interchange rule  (\ref{com}) and treating  each part of the partition as a cluster of charge $\frac12$. Once inserted within the partition (at a position to be determined below), the cluster, say $p^{(j)}$, is unfolded into  $2j$ parts, as 
 \begin{equation} p^{(j)} \rw (a_l,\cdots , a_{l+2j-1})\;, \qquad \text{with}\qquad a_n\geq a_{n+1}\;, 
 \end{equation}
and  such that:
 \begin{enumerate}
 \item The parts sum to $p$, i.e., $\sum_{n=0}^{2j-1} a_{l+n}= p$.
 \item The parts differ at most by $\frac12$.
  \item Every half-integer part  appears with even frequency. 
   \end{enumerate}
This decomposition is unique. Indeed, setting \cite{BreL}
\begin{equation} p= js+r, \qquad \text{with} \qquad 0\leq r < j \;, \end{equation}
 the decomposition is
  \begin{equation}\label{decop}
 {p}^{(j)} \rw \big(\underbrace{\frac{s+1}{2},\cdots , \frac{s+1}{2}}_{2r},\,  \underbrace{\frac{s}{2}, \cdots, \frac{s}{2}}_{2j-2r}\big)\;.
 \end{equation}
That the parts obtained from the unfolding operation all have even frequency  ensures that it is so in particular for any resulting half-integer part. For instance, we have 
 \begin{equation}
 {13}^{(3)} \rw \l(\frac52,\frac52,2,2,2,2\r)\;.
 \end{equation}
 The only parts which are not unfolded are the clusters of charge $\frac12$ and they are necessarily integer.
 
 The position at which the cluster $p^{(j)}$ is inserted  within the partition and unfolded  is determined by two criteria: 

\begin{enumerate}
\item The resulting sequence of numbers  must be non-increasing.

\item The frequency condition $f_{a_i}+f_{a_i+\frac12}\leq 2j$ must be satisfied for all parts $a_i$ of this  resulting h-partition. Equivalently, the generated h-partition  must satisfy
 \begin{equation}\label{disj}
 a_i\geq  a_{i+2j} + 1\qquad \forall\, i\;.
 \end{equation}
 \end{enumerate}

Obviously, because $j\leq k-\frac12$, (\ref{disj})  ensures the validity of the condition (\ref{difk}) at every intermediate stage of the  construction (and recall that the inserted  clusters have charge ranging from  $j=1$ up to $k-\frac12$). Take for instance, the following multiple partition:
\begin{equation}  \la^{(\frac12)} = ( 4)\;, \qquad \la^{(1)} = ( 8,4) \;,\qquad \la^{(\frac32)} = (3)\;. \end{equation}
The corresponding h-partition is reconstructed by the following sequences of cluster insertions and unfolding:
\begin{align}
 & (4)\, 8^{(1)}\, 4^{(1)}\, 3^{(\frac32)} \rw (9^{(1)}, 3) \,4^{(1)}\, 3^{(\frac32)}\rw \l(\frac92,\frac92,3\r)\,4^{(1)}\, 3^{(\frac32)}\rw \l(\frac92,\frac92,3, 4^{(1)}\r) \, 3^{(\frac32)}\nonumber \\
&\qquad \qquad  \rw\l(\frac92,\frac92,3, 2,2\r)\, 3^{(\frac32)} \rw
\l(\frac92,\frac92,3, 2,2, 3^{(\frac32)} \r)\rw \l(\frac92,\frac92,3, 2,2,1,1,1\r).
\end{align}

This procedure is clearly  invertible. In order to associate a multiple partition to a h-partition $(a_1,\cdots , a_{2\mb})$ satisfying (\ref{difk}),  we   first  identify the sequences of $2k-1$ adjacent parts such that the first part and last part of each  sequence differ by at most $\frac12$. Each such sequence is then replaced by a cluster of charge $k-\frac12$, say $p^{(k-\frac12)}$, where $p$ is the sum of the clustered  parts. 
Once all clusters of charge $k-\frac12$ are constructed, they are moved to the right of the sequence formed by the remaining parts. This displacement is done using the interchange operation (\ref{com}),  by treating all crossed parts  as clusters of charge $\frac12$. For the resulting smaller h-partition, we redo the previous analysis but with $k-\frac12$ replaced by $k-1$. Once all clusters of charge $k-1$ are identified, they are also moved to the right of the h-partition.
This procedure is repeated for the identification of lower-charge clusters (always in decreasing value of the charge) until all clusters of charge 1 are formed and moved to the left extremity of the sequence of ordered clusters of charge $\frac32,\cdots, k-\frac12$. The remaining parts of the partition are the clusters of charge $\frac12$. The result is an ordered set of $2k-1$  partition $\La^{(2k-1)}$, where the parts of $\la^{(j)}$ are the weights of the  clusters of charge $j$.  
Here is an example:
\begin{align}
 &\l(3,3,3,\frac52,\frac52,2,\frac32,\frac32,1,1,\frac12,\frac12\r) \rw \l(14^{(\frac52)},2,\frac32,\frac32,1,1,\frac12,\frac12\r)\rw \l(3,\frac52,\frac52,2,2,\frac32,\frac32\r)\, 7^{(\frac52)}\nonumber\\
&\qquad \qquad \qquad\rw \l(3,\frac52,\frac52,3^{(2)}\r)\, 7^{(\frac52)} \rw \l(3,\frac52,\frac52 \r)\, 3^{(2)}\,  7^{(\frac52)}\rw8^{(\frac32)} \, 3^{(2)}\,  7^{(\frac52)}\;.
\end{align}

\subsection{Jagged partitions and h-partitions}

Let us now display the connection between h-partitions subject to the restriction (\ref{difk}) and those restricted jagged partitions in terms of which the graded parafermionic basis has first been formulated \cite{JM}. The relation is easily described: in a h-partition, we simply replace every pair of half-integer parts as follows:
\begin{equation}(\cdots, \frac{2r+1}2, \,\frac{2r+1}2 ,\cdots) \rw (\cdots, r,r+1, \cdots) \;.
\end{equation}
The result is a jagged partition,
namely,  a sequence of non-negative integers $(n_1,\cdots , n_{2\mb})$ such that \cite{FJM.R}:
\begin{equation}\label{jag}
 n_j\geq n_{j+1}-1\;,\qquad  \qquad  n_j\geq n_{j+2}\;, \qquad\qquad   n_{2\mb}\geq 1\;.\
\end{equation}
The restriction (\ref{difk}) is transformed into the following restrictions:
\begin{equation}\label{rjag}
n_j \geq  n_{j+2k-1} +1 \qquad{\rm or} \qquad
 n_j = n_{j+1}-1 =  n_{j+2k-2}+1= n_{j+2k-1} \;,
\end{equation}
for all values of $j\leq 2\mb-2k+1$. For instance, with $k=3$,
\begin{equation} \l(3,3,3,\frac52,\frac52,2,\frac32,\frac32,1,1,\frac12,\frac12\r) \rw (3,3,3,2,3,2,1,2,1,1,0,1) \;.
\end{equation}
We see explicitly with this example that the parts of a  jagged partition are not necessarily non-increasing: an increase of 1 between two successive parts is allowed by the  first condition in (\ref{jag})  but the second condition prevents two such successive increases.
The difference 1 condition at distance 5 $(=2k-1$) is verified in all cases except for the last one, where $n_7=n_{12}$ but then, the in-between difference 2 condition is verified: $n_8=n_{11}+2$.

The results of this section demonstrates, at the combinatorial level, the equivalence of two bases for graded parafermions previously demonstrated at the level of their generating functions \cite{JM.A}.

\subsection{Paths and h-partitions: the Bruge correspondence} 

\let\a\alpha
\let\b\beta

The  Burge correspondence \cite{Bu} provides a remarkable bijection between a partition with frequency condition and a (Bressoud-type) lattice path. We present the adaptation of this correspondence to the case of a restricted h-partition and a $\lp$ path. The correspondence  relies on the characterization of a partition in terms of non-overlapping pairs of adjacent frequencies $(f_j,f_{j+\frac12})$ with $f_{j+\frac12}>0$, starting the pairing from the largest part. For instance, for the following h-partition of 23, (with satisfies the condition (\ref{difk}) for $k\geq 3)$ we have the following pairing:
\begin{equation}\label{exi}
\l(3,3,3,\frac52,\frac52,2,\frac32,\frac32,1,1,\frac12,\frac12\r) : \quad \begin{matrix}
i: \,0&\phantom{(}\frac12&1\phantom{(}&\phantom{(}\frac32
&2\phantom{(}&\frac52&3\\
  f_i: \, 0&(2&2)&(2&1)&(2&\phantom{(}3). \end{matrix}
\end{equation}
We then define a sequence of two operations, $\a$ and $\b$ on the set of paired frequencies. This sequence will be interpreted as a binary word (with letters $\a$ and $\b$) whose graphical representation is the desired path. 

 Let us then describe these two operations.  If $(f_0,f_\frac12)$ is not a pair, we act with $\a$ defined as follows:
\begin{equation}\a: (f_j,f_{j+\frac12})\rw (f_j+1,f_{j+\frac12}-1) \qquad \forall\, j\geq \frac12 .
\end{equation}
If $(f_0,f_\frac12)=(0,f_\frac12)$ is a pair, we act with $\b$ defined as follows:
\begin{equation}\b: \left\{ \begin{matrix}
&(0,f_\frac12)\rw (0,f_\frac12-1)\phantom{\qquad\qquad\;}& \\  & (f_j,f_{j+\frac12})\rw (f_j+1,f_{j+\frac12}-1) & \forall\, j>\frac12 . \end{matrix} \right.
\end{equation}
After each operation, the pairing is modified according to the new values of the frequencies. We then act successively with $\a$ or $\b$ on the h-partition until all frequencies become zero. The ordered sequence of $\a$ and $\b$ so obtained is then reinterpreted as a $\lp$  path starting at a prescribed initial position, by considering  $\a$ to be an horizontal  or a SE step,  and $\b$  a NE step \cite{BreL}. The sequence is completed by adding 
the number of $\a$ needed to reach the horizontal axis.

For the above example (\ref{exi}), the sequence is $\a\b^3\a^4\b^4\a^{11}\b^5$ and we add $\a^5$ at the end to make the corresponding path reaches the $x$-axis, assuming that the path starts at the origin.  In terms of the sequence of peaks and their charge,  the path is 
$14^{(\frac52)}\, 6^{(2)}\, 2^{(\frac32)}$.


This procedure makes the correspondence $\La^{(2k-1)}\rw \lp$ more algorithmic than the mere statement that the clusters of $\La^{(2k-1)}$ are to be reordered such that the conditions (\ref{dist}) is everywhere satisfied. The correspondence has now an intermediate step, which is the construction of a h-partition, and it is completed by the Burge correspondence, relating the h-partition to the path.







\section{Other models described by paths on half-integer lattice  the ${\cal SM}(2,4\kappa)$ superconformal models. }

In this work, we have identified two classes of models that have a natural description in terms of paths defined on a half-integer lattice: the $\M(k+1,2k+3)$  minimal models  and the graded $\z_k$ parafermions. Although it falls a bit off our main theme, we identify yet  another class of models -- the superconformal minimal models $\SM(2,4\ka)$ -- with a similar representation.
  
    The superconformal minimal models are
labeled by the integers $p',\, p$,  with  $\tfrac12(p-p')$ and $p'$ relatively coprime. The primary fields ${\hat \phi}_{r,s}$ are further labeled by the integers $(r,s)$ with $1\leq r \leq p'-1$ and $1\leq s\leq p-1$, with the fields ${\hat \phi}_{r,s}$  and ${\hat \phi}_{p'-r,p-s}$   identified. Fields with $s+r$ even are said to belong to the  Neveu-Schwarz (NS)  sector, while those for $s+r$ odd are in the Ramond (R) sector.

For the $\SM(2,4\ka)$ models, the value of $r$ is fixed to 1 and a set of independent fields is specified by $1\leq s\leq 2\ka$.  The space of states of these models is 
described in terms of $(2\ka-1)$-restricted jagged partitions, that is, jagged partition restricted by (\ref{rjag}) but with $2k$ replaced by $2\ka-1$. This description leads to a fermionic form of the characters \cite{FJMjpa}. A second fermionic form is displayed here
which  reflects naturally the path interpretation to be described below and which relies on an expression of the generating function for restricted jagged partitions that differs from that given  in \cite{FJMjpa}.

With  $i-1$ representing the maximal number of pairs of $01$ (i.e., $2i-2$ is the maximal number of times $\tfrac12$ appears in the corresponding h-partition), the following expression for the generating function for $(2\ka-1)$-restricted jagged partitions has been found in \cite{JPath}:
\begin{equation}\label{defFabb}
J_{2\ka-1,i}(z; q)= \sum_{n_{\frac12}, n_1,n_{\frac32},\cdots,n_{\ka-1}=0}^\y  \frac{q^{\frac12( { N}_{\frac12}^2+{ N}_1^2+ { N}_{\frac32}^2+ \cdots+ { N}_{\ka-1}^2 +{ M})+{ L}_{i}} z^{{ N}}
}{ (q)_{n_{\frac12}}(q)_{n_1}\cdots (q)_{n_{\ka-1}} } \;, 
\end{equation}
where 
\begin{align}
 { N}_{j}&= n_j+n_{j+\frac12}+\cdots + n_{\ka-1} & 
{ M} &= n_{\frac12}+n_{\frac32}+\cdots +n_{\ka-\frac32}  \nonumber \\
{ L}_i &= N_i+N_{i+1}+\cdots +N_{\ka-1} &
{ N}&= n_{\frac12}+2n_1+\cdots +2(\ka-1)n_{\ka-1} \;.
\end{align}
The exponent of $z$ is the length of the jagged partition. 
Note that the  expression (\ref{defFabb}) has the following  multi-variable deformation (to be needed below): 
\begin{equation}\label{deFde}
J_{2\ka-1,i}(z_{\frac12}, z_1,\cdots, z_{\ka-1};\,q)= \sum_{n_{\frac12}, n_1,n_{\frac32}\cdots,n_{\ka-1}=0}^\y \frac{
q^{\frac12({ N}_{\frac12}^2+\cdots+ { N}_{\ka-1}^2 +M)+{ L}_{i}}\; \prod_{j=\frac12}^{\ka-1}z_j^{n_j}  }{  (q)_{n_{\frac12}}(q)_{n_1}\cdots (q)_{n_{\ka-1}} }\;. 
\end{equation}
With $z_j=z^{2j}$, the $z$ factor reduces to $z^{{ N}}$.

We now adapt this generating function  to the $\SM(2,4\ka)$  context, producing an expression for the irreducible character specified by $i$, and spell out its  lattice-path interpretation.

Let us start with the R sector. It has been demonstrated in \cite{FJMjpa} 
that  its basis of
states is specified by the
$2\ka-1$ restricted jagged partitions having  at most  $i-1$ pairs of 01 at the rightmost
end, with $1\leq i=s/2\leq \ka$ ($s$ is even in the R sector).  Equivalently, these are specified by $2\ka-1$ restricted h-partitions $(a_1,\cdots , a_{{ N}})$:
\begin{equation} \label{difkap}
 a_i\geq a_{i+2\ka-2}+1\;,
\end{equation} having  at most  $s-2$ copies of $\tfrac12$.

The corresponding (half-integer lattice) path formulation of the space of states is the following: every state can be represented by a  Bressoud-type path starting at $y_0=\ka-s/2$,  whose maximal  height is $\leq \ka-1$ and with peaks at integer positions. As before, the weight of such a path is the sum of the $x$-coordinates of the peaks. This expression of the weight is precisely  the (relative) conformal dimension of the corresponding state in the superconformal irreducible module.

The expression (\ref{defFabb}) for the  generating function for restricted jagged partitions leads to the following expression for the R characters:
\begin{equation}\label{Rca}
{\hat \chi}_{1,2i}^{(2,4\ka)}(q)= J_{2\ka-1,i}(1;q)
\end{equation}
This is precisely the second fermionic form of Melzer's  R characters (cf. the first line of eq (2.6) in \cite{Mel}).

In  the NS sector, every state is  associated to a $(2\ka-1)$-restricted jagged
partition {\it made of half-integers}, with a tail containing at most $s-1$ parts equal to $\tfrac12$ (and $s$ is now odd) \cite{FJMjpa}.\footnote{Recall that in the NS sector, the weight of a partition, that is, the sum of its parts, which is also the conformal dimension of the corresponding descendant state, can be integer or half-integer.} The associated h-partition is made of half-integers and integers but now with the frequency of integers forced to be even.

Now, how does this characterization of states can be rephrased  in terms of paths? The correspondence between paths  and h-partition obtained in Appendix A shows that the position of the peaks with half-integer charge must now be shifted to half-integer $x$-positions, while that of peaks with  integer charge remain located on integer $x$-positions.

This statement is proved as follows using the relation between a cluster $p^{(j)} $ and the parts of a h-partition that results from unfolding the cluster as in (\ref{decop}), with $p=js+r$. Let $\equiv$ means an equality modulo 1. If $s/2\equiv 0$, the frequency condition on integers forces $j-r\equiv 0$. Suppose then that $j\equiv 0$. This implies $r\equiv 0$ and as a result $p\equiv 0$.  On the other hand, if $j \equiv \tfrac12$, then $r\equiv \tfrac12$ and since $s$ is even, it follows that $p=sj+r \equiv \tfrac12$. Suppose next that $(s+1)/2\equiv 0$.  The frequency constraint requires $r\equiv 0 $. Again, if $j \equiv 0$ this  implies  that $p\equiv 0$,  while if $j \equiv \tfrac12$, since $s$ is now odd, $p \equiv \tfrac12$. In
 all cases, we see that $h\equiv p$. Now this result has been established for a generic cluster $p^{(j)}$ and not  necessarily for a peak within a path. However, since a sequence of peaks describing a path is obtained from a sequence of clusters by the interchange operation (\ref{com}) and that this operation preserves the value of the weight modulo 1, the result for clusters can be applied to peaks.

 Note finally that the initial position of the path in the NS sector  is fixed to $y_0=\kappa-(s-1)/2$.


The first three states in the vacuum module of the $\SM(2,8)$ models are presented in Fig. \ref{fig14} and the 6 states at level $\frac{15}2$ are given in Fig. \ref{fig15}.


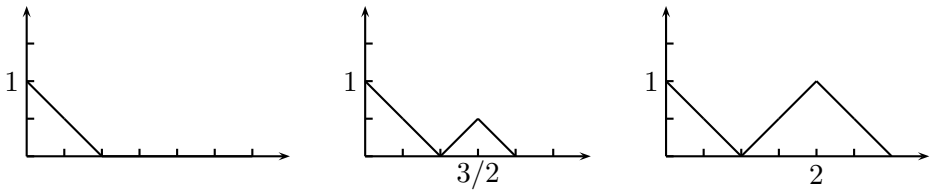
\begin{figure}[ht]
\caption{{\footnotesize The first three paths pertaining  of the vacuum module of the $\SM(2,8)$ model, representing the first three terms in  the vacuum character ${\hat \chi}_{1,1}^{(2,8)}= 1+q^\frac32+q^2+\ldots$. . }} \label{fig14}
\begin{center}
\begin{pspicture}(0,0)(11.5,3)

\psline{->}(1.0,1.0)(4.5,1.0) \psline{->}(5.5,1.0)(8.5,1.0)
\psline{->}(9.5,1.0)(13.0,1.0)
\psline{->}(1.0,1.0)(1.0,3.0) \psline{->}(5.5,1.0)(5.5,3.0)
\psline{->}(9.5,1.0)(9.5,3.0)

\psline{-}(1.5,1.0)(1.5,1.1) \psline{-}(2.0,1.0)(2.0,1.1)
\psline{-}(2.5,1.0)(2.5,1.1) \psline{-}(3.0,1.0)(3.0,1.1)
\psline{-}(3.5,1.0)(3.5,1.1) \psline{-}(4.0,1.0)(4.0,1.1)
\psline{-}(6.0,1.0)(6.0,1.1) \psline{-}(6.5,1.0)(6.5,1.1)
\psline{-}(7.0,1.0)(7.0,1.1) \psline{-}(7.5,1.0)(7.5,1.1)
\psline{-}(8.0,1.0)(8.0,1.1)
\psline{-}(10.0,1.0)(10.0,1.1) \psline{-}(10.5,1.0)(10.5,1.1)
\psline{-}(11.0,1.0)(11.0,1.1) \psline{-}(11.5,1.0)(11.5,1.1)
\psline{-}(12.0,1.0)(12.0,1.1)
\rput(7.0,0.75){{\small $3/2$}} \rput(11.5,0.75){{\small$2$}}

\psline{-}(1.0,1.0)(1.1,1.0) \psline{-}(1.0,1.5)(1.1,1.5)
\psline{-}(1.0,2.0)(1.1,2.0) \psline{-}(1.0,2.5)(1.1,2.5)
\psline{-}(5.5,1.0)(5.6,1.0) \psline{-}(5.5,1.5)(5.6,1.5)
\psline{-}(5.5,2.0)(5.6,2.0)\psline{-}(5.5,2.5)(5.6,2.5)
\psline{-}(9.5,1.0)(9.6,1.0) \psline{-}(9.5,1.5)(9.6,1.5)
\psline{-}(9.5,2.0)(9.6,2.0) \psline{-}(9.5,2.5)(9.6,2.5)
\rput(0.8,2){{\small$1$}} \rput(5.3,2){{\small$1$}}
\rput(9.3,2){{\small$1$}}

\psline{-}(1.0,2)(1.5,1.5) \psline{-}(1.5,1.5)(2.0,1)
\psline{-}(2.0,1)(2.5,1) \psline{-}(2.5,1)(3.0,1)
\psline{-}(3.0,1)(3.5,1) \psline{-}(3.5,1)(4.0,1)
\psline{-}(5.5,2.0)(6,1.5) \psline{-}(6.0,1.5)(6.5,1.0)
\psline{-}(6.5,1.0)(7,1.5) \psline{-}(7.0,1.5)(7.5,1.0)
\psline{-}(9.5,2)(10,1.5) \psline{-}(10,1.5)(10.5,1)
\psline{-}(10.5,1)(11,1.5) \psline{-}(11,1.5)(11.5,2)
\psline{-}(11.5,2)(12,1.5) \psline{-}(12,1.5)(12.5,1)

\end{pspicture}
\end{center}
\end{figure}


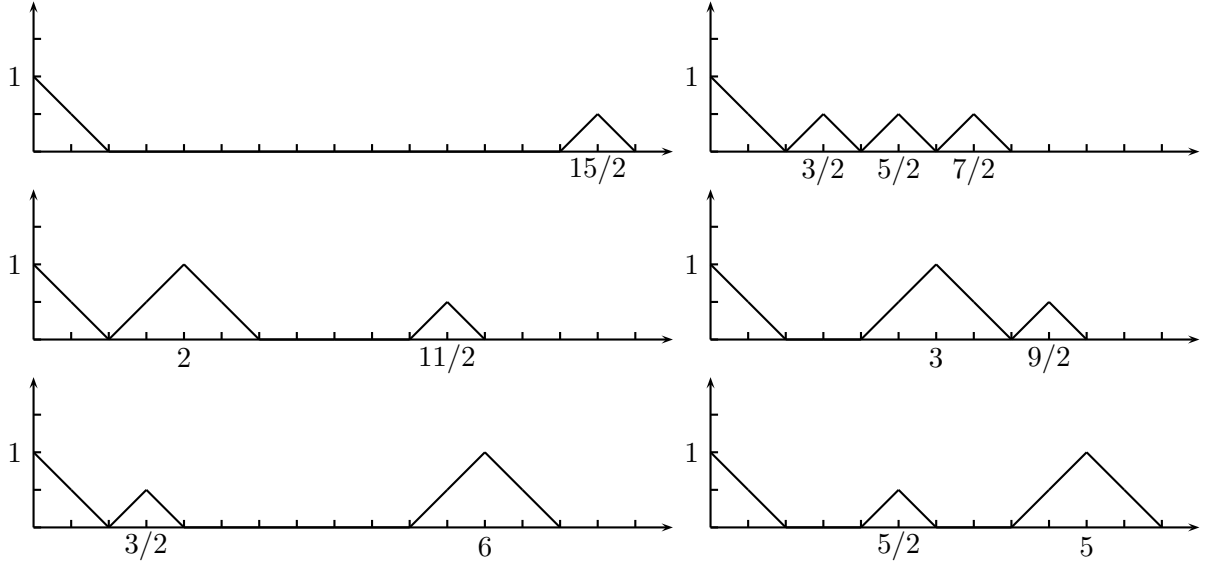
\begin{figure}[ht]
\caption{{\footnotesize The 6 paths that describe the states of the vacuum module of the $\SM(2,8)$ models at level $\frac{15}2$. }} \label{fig15}
\begin{center}
\begin{pspicture}(0,0)(15.0,8)

\psline{->}(0.5,1.0)(9.0,1.0) \psline{->}(0.5,3.5)(9.0,3.5)
\psline{->}(0.5,6.0)(9.0,6.0)

\psline{->}(9.5,1.0)(16.0,1.0) \psline{->}(9.5,3.5)(16.0,3.5)
\psline{->}(9.5,6.0)(16.0,6.0)

\psline{->}(0.5,1.0)(0.5,3.0) \psline{->}(0.5,3.5)(0.5,5.5)
\psline{->}(0.5,6.0)(0.5,8.0)

\psline{->}(9.5,1.0)(9.5,3.0) \psline{->}(9.5,3.5)(9.5,5.5)
\psline{->}(9.5,6.0)(9.5,8.0)

\psline{-}(1.0,1.0)(1.0,1.1) \psline{-}(1.5,1.0)(1.5,1.1)
\psline{-}(2.0,1.0)(2.0,1.1) \psline{-}(2.5,1.0)(2.5,1.1)
\psline{-}(3.0,1.0)(3.0,1.1) \psline{-}(3.5,1.0)(3.5,1.1)
\psline{-}(4.0,1.0)(4.0,1.1) \psline{-}(4.5,1.0)(4.5,1.1)
\psline{-}(5.0,1.0)(5.0,1.1) \psline{-}(5.5,1.0)(5.5,1.1)
\psline{-}(6.0,1.0)(6.0,1.1) \psline{-}(6.5,1.0)(6.5,1.1)
\psline{-}(7.0,1.0)(7.0,1.1) \psline{-}(7.5,1.0)(7.5,1.1)
\psline{-}(8.0,1.0)(8.0,1.1) \psline{-}(8.5,1.0)(8.5,1.1)

\psline{-}(1.0,3.5)(1.0,3.6) \psline{-}(1.5,3.5)(1.5,3.6)
\psline{-}(2.0,3.5)(2.0,3.6) \psline{-}(2.5,3.5)(2.5,3.6)
\psline{-}(3.0,3.5)(3.0,3.6) \psline{-}(3.5,3.5)(3.5,3.6)
\psline{-}(4.0,3.5)(4.0,3.6) \psline{-}(4.5,3.5)(4.5,3.6)
\psline{-}(5.0,3.5)(5.0,3.6) \psline{-}(5.5,3.5)(5.5,3.6)
\psline{-}(6.0,3.5)(6.0,3.6) \psline{-}(6.5,3.5)(6.5,3.6)
\psline{-}(7.0,3.5)(7.0,3.6) \psline{-}(7.5,3.5)(7.5,3.6)
\psline{-}(8.0,3.5)(8.0,3.6) \psline{-}(8.5,3.5)(8.5,3.6)

\psline{-}(1.0,6.0)(1.0,6.1) \psline{-}(1.5,6.0)(1.5,6.1)
\psline{-}(2.0,6.0)(2.0,6.1) \psline{-}(2.5,6.0)(2.5,6.1)
\psline{-}(3.0,6.0)(3.0,6.1) \psline{-}(3.5,6.0)(3.5,6.1)
\psline{-}(4.0,6.0)(4.0,6.1) \psline{-}(4.5,6.0)(4.5,6.1)
\psline{-}(5.0,6.0)(5.0,6.1) \psline{-}(5.5,6.0)(5.5,6.1)
\psline{-}(6.0,6.0)(6.0,6.1) \psline{-}(6.5,6.0)(6.5,6.1)
\psline{-}(7.0,6.0)(7.0,6.1) \psline{-}(7.5,6.0)(7.5,6.1)
\psline{-}(8.0,6.0)(8.0,6.1) \psline{-}(8.5,6.0)(8.5,6.1)

\psline{-}(10.0,1.0)(10.0,1.1) \psline{-}(10.5,1.0)(10.5,1.1)
\psline{-}(11.0,1.0)(11.0,1.1) \psline{-}(11.5,1.0)(11.5,1.1)
\psline{-}(12.0,1.0)(12.0,1.1) \psline{-}(12.5,1.0)(12.5,1.1)
\psline{-}(13.0,1.0)(13.0,1.1) \psline{-}(13.5,1.0)(13.5,1.1)
\psline{-}(14.0,1.0)(14.0,1.1) \psline{-}(14.5,1.0)(14.5,1.1)
\psline{-}(15.0,1.0)(15.0,1.1) \psline{-}(15.5,1.0)(15.5,1.1)

\psline{-}(10.0,3.5)(10.0,3.6) \psline{-}(10.5,3.5)(10.5,3.6)
\psline{-}(11.0,3.5)(11.0,3.6) \psline{-}(11.5,3.5)(11.5,3.6)
\psline{-}(12.0,3.5)(12.0,3.6) \psline{-}(12.5,3.5)(12.5,3.6)
\psline{-}(13.0,3.5)(13.0,3.6) \psline{-}(13.5,3.5)(13.5,3.6)
\psline{-}(14.0,3.5)(14.0,3.6) \psline{-}(14.5,3.5)(14.5,3.6)
\psline{-}(15.0,3.5)(15.0,3.6) \psline{-}(15.5,3.5)(15.5,3.6)

\psline{-}(10.0,6.0)(10.0,6.1) \psline{-}(10.5,6.0)(10.5,6.1)
\psline{-}(11.0,6.0)(11.0,6.1) \psline{-}(11.5,6.0)(11.5,6.1)
\psline{-}(12.0,6.0)(12.0,6.1) \psline{-}(12.5,6.0)(12.5,6.1)
\psline{-}(13.0,6.0)(13.0,6.1) \psline{-}(13.5,6.0)(13.5,6.1)
\psline{-}(14.0,6.0)(14.0,6.1) \psline{-}(14.5,6.0)(14.5,6.1)
\psline{-}(15.0,6.0)(15.0,6.1) \psline{-}(15.5,6.0)(15.5,6.1)

\rput(8.0,5.75){{\small$15/2$}} \rput(2.5,3.25){{\small$2$}}
\rput(6.0,3.25){{\small$11/2$}} \rput(2.0,0.75){{\small $3/2$}}
\rput(6.5,0.75){{\small$6$}} \rput(11,5.75){{\small$3/2$}}
\rput(12,5.75){{\small$5/2$}} \rput(13,5.75){{\small $7/2$}}
\rput(12.5,3.25){{\small$3$}} \rput(14,3.25){{\small$9/2$}}
\rput(12,0.75){{\small$5/2$}} \rput(14.5,0.75){{\small $5$}}

\psline{-}(0.5,1.0)(0.6,1.0) \psline{-}(0.5,1.5)(0.6,1.5)
\psline{-}(0.5,2.0)(0.6,2.0) \psline{-}(0.5,2.5)(0.6,2.5)

\psline{-}(9.5,1.0)(9.6,1.0) \psline{-}(9.5,1.5)(9.6,1.5)
\psline{-}(9.5,2.0)(9.6,2.0) \psline{-}(9.5,2.5)(9.6,2.5)

\psline{-}(0.5,3.5)(0.6,3.5) \psline{-}(0.5,4.0)(0.6,4.0)
\psline{-}(0.5,4.5)(0.6,4.5) \psline{-}(0.5,5.0)(0.6,5.0)

\psline{-}(9.5,3.5)(9.6,3.5) \psline{-}(9.5,4.0)(9.6,4.0)
\psline{-}(9.5,4.5)(9.6,4.5) \psline{-}(9.5,5.0)(9.6,5.0)

\psline{-}(0.5,6.0)(0.6,6.0) \psline{-}(0.5,6.5)(0.6,6.5)
\psline{-}(0.5,7.0)(0.6,7.0) \psline{-}(0.5,7.5)(0.6,7.5)

\psline{-}(9.5,6.0)(9.6,6.0) \psline{-}(9.5,6.5)(9.6,6.5)
\psline{-}(9.5,7.0)(9.6,7.0) \psline{-}(9.5,7.5)(9.6,7.5)

\rput(0.25,2.0){{\small$1$}} \rput(9.25,2.0){{\small$1$}}
\rput(0.25,4.5){{\small$1$}} \rput(9.25,4.5){{\small$1$}}
\rput(0.25,7.0){{\small$1$}} \rput(9.25,7.0){{\small$1$}}

\psline{-}(0.5,7.0)(1.0,6.5) \psline{-}(1.0,6.5)(1.5,6.0)
\psline{-}(1.5,6.0)(2.0,6.0) \psline{-}(2.0,6.0)(2.5,6.0)
\psline{-}(2.5,6.0)(3.0,6.0) \psline{-}(3.0,6.0)(3.5,6.0)
\psline{-}(3.5,6.0)(4.0,6.0) \psline{-}(4.0,6.0)(4.5,6.0)
\psline{-}(4.5,6.0)(5.0,6.0) \psline{-}(5.0,6.0)(5.5,6.0)
\psline{-}(5.5,6.0)(6.0,6.0) \psline{-}(6.0,6.0)(6.5,6.0)
\psline{-}(6.5,6.0)(7.0,6.0) \psline{-}(7.0,6.0)(7.5,6.0)
\psline{-}(7.5,6.0)(8.0,6.5) \psline{-}(8.0,6.5)(8.5,6.0)

\psline{-}(0.5,4.5)(1.0,4.0) \psline{-}(1.0,4.0)(1.5,3.5)
\psline{-}(1.5,3.5)(2.0,4.0) \psline{-}(2.0,4.0)(2.5,4.5)
\psline{-}(2.5,4.5)(3.0,4.0) \psline{-}(3.0,4.0)(3.5,3.5)
\psline{-}(3.5,3.5)(4.0,3.5) \psline{-}(4.0,3.5)(4.5,3.5)
\psline{-}(4.5,3.5)(5.0,3.5) \psline{-}(5.0,3.5)(5.5,3.5)
\psline{-}(5.5,3.5)(6.0,4.0) \psline{-}(6.0,4.0)(6.5,3.5)

\psline{-}(0.5,2.0)(1.0,1.5) \psline{-}(1.0,1.5)(1.5,1)
\psline{-}(1.5,1.0)(2.0,1.5) \psline{-}(2.0,1.5)(2.5,1)
\psline{-}(2.5,1.0)(3.0,1.0) \psline{-}(3.0,1.0)(3.5,1)
\psline{-}(3.5,1.0)(4.0,1.0) \psline{-}(4.0,1.0)(4.5,1)
\psline{-}(4.5,1.0)(5.0,1.0) \psline{-}(5.0,1.0)(5.5,1)
\psline{-}(5.5,1.0)(6.0,1.5) \psline{-}(6.0,1.5)(6.5,2)
\psline{-}(6.5,2.0)(7.0,1.5) \psline{-}(7.0,1.5)(7.5,1)

\psline{-}(9.5,7.0)(10,6.5) \psline{-}(10,6.5)(10.5,6.0)
\psline{-}(10.5,6.0)(11,6.5) \psline{-}(11,6.5)(11.5,6.0)
\psline{-}(11.5,6.0)(12,6.5) \psline{-}(12,6.5)(12.5,6.0)
\psline{-}(12.5,6.0)(13,6.5) \psline{-}(13,6.5)(13.5,6.0)

\psline{-}(9.5,4.5)(10,4.0) \psline{-}(10,4.0)(10.5,3.5)
\psline{-}(10.5,3.5)(11,3.5) \psline{-}(11,3.5)(11.5,3.5)
\psline{-}(11.5,3.5)(12,4.0) \psline{-}(12,4.0)(12.5,4.5)
\psline{-}(12.5,4.5)(13,4.0) \psline{-}(13,4.0)(13.5,3.5)
\psline{-}(13.5,3.5)(14,4.0) \psline{-}(14,4.0)(14.5,3.5)

\psline{-}(9.5,2)(10,1.5) \psline{-}(10,1.5)(10.5,1.0)
\psline{-}(10.5,1)(11,1.0) \psline{-}(11,1.0)(11.5,1.0)
\psline{-}(11.5,1)(12,1.5) \psline{-}(12,1.5)(12.5,1.0)
\psline{-}(12.5,1)(13,1.0) \psline{-}(13,1.0)(13.5,1.0)
\psline{-}(13.5,1)(14,1.5) \psline{-}(14,1.5)(14.5,2.0)
\psline{-}(14.5,2)(15,1.5) \psline{-}(15,1.5)(15.5,1.0)

\end{pspicture}
\end{center}
\end{figure}

Let us now  see how this shift in the position of the peaks with half-integer charge modifies $J_{2\ka-1,i}$. For this, we start with its multi-parameter deformation given in (\ref{deFde}). The simplest way of implementing a shift by 1/2 for all the peaks with half-integer height
without affecting the other ones, is to set 
\begin{align}
&z_j= 1\phantom{q^{-\frac12}} \qquad {\rm if}\quad  j \in \Z_+  \nonumber \\
&{z_j}= q^{-\frac12}\phantom{1}\qquad  {\rm if} \quad j \in \Z_++\frac12\;. 
\end{align}} 
The effect of this transformation is simply to remove the factor $ M$ in the numerator exponent of $q$ in (\ref{defFabb}) (and the $z$ dependence). But the action of repositioning a number of  peaks affects also the linear terms. We easily find it to be changed as follows:
\begin{equation}
{ L}_i \rw { L}_{i-\frac12}= m_{i-\frac12}+m_i + 2(m_{i+\frac12}+m_{i+1})+\cdots  \;. 
\end{equation} 
We end up with 
\begin{equation}\label{Rca}
{\hat \chi}_{1,2i-1}^{(2,4\ka)}(q)= J'_{2\ka-1,i}(1;q)\; ,
\end{equation}
where 
\begin{equation}
 J'_{2\ka-1,i}(1;q)= \sum_{n_{\frac12}, n_1,n_{\frac32}\cdots,n_{\ka-1}=0}^\y \frac{
q^{\frac12({ N}_{\frac12}^2+{ N}_1^2+\cdots+ { N}_{\ka-1}^2 )+{ L}_{i-\frac12} }\;  }{  (q)_{n_{\frac12}}(q)_{n_1}\cdots (q)_{n_{\ka-1}} }\; .
\end{equation}
This is  the second fermionic form of Melzer's   NS characters (cf. the first line of eq (2.6) in \cite{Mel}). 

Both forms (R and NS characters) can be written under a single formula as
\begin{equation}{\hat \chi}_{1,s}^{(2,4\ka)}(q)= \sum_{n_{\frac12}, n_1,n_{\frac32}\cdots,n_{\ka-1}=0}^\y \frac{
q^{\frac12({ N}_{\frac12}^2+{ N}_1^2+\cdots+ { N}_{\ka-1}^2 )+\frac14(1+(-1)^s){ M} +{ L}_{\frac{s}2} }\;  }{  (q)_{n_{\frac12}}(q)_{n_1}\cdots (q)_{n_{\ka-1}} }\; . 
\end{equation}
These expressions were conjectured in \cite{Mel} and were later proved in \cite{BIS,Ol}. The present approach constitutes a new proof of  these fermionic formulae starting from the basis of states obtained in \cite{FJMjpa}, in addition to dress them with a new lattice-path  interpretation.\footnote{Note that different fermionic characters for these models had  been displayed in \cite{Mel} (the first fermionic form),  in \cite{BMO,BM} and of course, in \cite{FJMjpa}.}


\vskip0.3cm
\noindent {\bf ACKNOWLEDGMENTS}

The work of PJ is supported by EPSRC and that of  PM is supported  by NSERC.
We thank O. Warnaar for very useful discussions.



\begin{thebibliography}{99}
\addcontentsline{toc}{section}{References}



 \bibitem{Andr}
 G.E. Andrews, {\it The theory of
partitions}, Cambridge Univ. Press, Cambridge, UK, (1984).

 
\bibitem{ABF}
G. E. Andrews, R. J. Baxter and P. J. Forrester,
\textit{Eight-vertex SOS model and generalized Rogers--Ramanujan-type
identities},
J. Stat. Phys. \textbf{35} (1984), 193--266.


\bibitem{BFJM}
L. B\'egin, J.-F. Fortin, P. Jacob and P. Mathieu,  {\it Fermionic characters
for graded parafermions}, Nucl. Phys. {\bf B659} (2003) 365-386.

\bibitem{BMlmp}
A. Berkovich, B. M. McCoy, {\it Continued fractions and fermionic representations for characters of M(p,p') minimal models}, Lett. Math. Phys. {\bf 37} (1996) 49-66.


\bibitem{BMO}
A. Berkovich, B.M. McCoy and  W. P. Orrick, {\it Polynomial identities, indices, and duality for the $N=1$ superconformal model $SM(2,4\nu)$}, J. Stat. Phys. {\bf 83} (1996) 795-837. 

 \bibitem{BM} 
 A. Berkovich and B.M. McCoy, {\it Generalizations of the Andrews-Bressoud identities for the $N=1$ superconformal model ${\rm SM}(2,4\nu)$},  Math. Comput. Modelling {\bf 26} (1997) 37-49.


\bibitem{BP}
A. ÊBerkovich and P. Paule, {\it Lattice paths, $q$-multinomials and two variants of the Andrews-Gordon identities},  Ramanujan J. {\bf 5} (2002) 409--425.



\bibitem{BreL}
D. Bressoud, {\it Lattice paths and Rogers-Ramanujan identities}, in
{\it Number Theory, Madras 1987}, ed. K. Alladi, Lecture Notes in
Mathematics {\bf 1395} (1987) 140-172.



 \bibitem{BIS}
 D. Bressoud, M. Ismail and D. Stanton, {\it Change of base in Bailey pairs}, Ramanujan J. {\bf 4} (2000) 435-453.

 \bibitem{Bu}
 W.H. Burge, {\it A correspondence between partitions related to generalizations of the Ramanujan-Rogers identities}, Discrete Math. {\bf 34} (1981) 9-15.



\bibitem{Byt}   
 A.G. Bytsko, {\it Fermionic representations for characters of M(3,t), M(4,5), M(5,6) and M(6,7) minimal models and related Rogers-Ramanujan type and dilogarithm identities},
{J.Phys. } {\bf A32} (1999) 8045-8058.
     
\bibitem{CRS}
J. M. Camino, A. V. Ramallo and
J. M. Sanchez de Santos, {\it Graded parafermions}, Nucl.Phys. {\bf B530} (1998) 715-741.

\bibitem{Kyoto}
 E. Date, M. Jimbo, A. Kuniba, T. Miwa and M. Okado, {\it Exactly solvable SOS models: local height probabilities and theta function identities}, Nucl. Phys. {\bf B290} (1987) 231-273.
 
 
 \bibitem{CFT}
P~Di Francesco, P. Mathieu, and D. S\'{e}n\'{e}chal,
{\em {Conformal Field Theory}},
\newblock Graduate Texts in Contemporary Physics. Springer-Verlag, New York,
  1997. 
 
 
 
\bibitem{FZa}
V.A. Fateev and Al.B. Zamolodchikov,
{\it Integrable perturbations of $Z(N)$ parafermion models and $O(3)$ sigma model}
Phys. Lett. {\bf B271} (1991) 91-100.


 \bibitem{FNO}
B~Feigin, T~Nakanishi, and H~Ooguri, {\it 
The Annihilating Ideals of Minimal Models}, 
{ Int. J. Mod. Phys.}, {\bf A7} (1992) 217--238.


 


\bibitem{FQ}
O. Foda and Y.-H. Quano, {\it Virasoro character identities from the Andrews--Bailey construction
},  Int. J. Mod. Phys. {\bf A12}  (1997) 1651-1676.


 \bibitem{FWa}
 O. Foda and T. Welsh, {\it  Melzer's identities revisited},  Contemp. Math. {\bf 248} (1999)  207-234. 
 
 

\bibitem{FLPW}
O. Foda,  K.S. M. Lee, Y. Pugai  and T. A. Welsh, {\it Path generating transforms}, in {\it q-Series from a contemporary perspective},  Contemp. Math. 254, Amer. Math. Soc., Providence, RI, 2000, 157--186.

 
 
 
\bibitem{FW}
O. Foda and T. A. Welsh, {\it 
On the combinatorics of Forrester-Baxter models},
Prog. Comb. {\bf 191} (2000) 49-103. 


\bibitem{FB}
P. J. Forrester and R. J. Baxter, 
\textit{Further exact solutions of the eight-vertex SOS model 
and generalizations of the Rogers-Ramanujan identities},
J. Stat. Phys. \textbf{38} (1985) 435--472. 


\bibitem{FJM.R}
J.-F. Fortin, P. Jacob and P. Mathieu, {\it Jagged partitions}, Ramanujan J. {\bf 10} (2005) 215-235.


\bibitem{JMO}
J.-F. Fortin,  P. Mathieu and  S. O. Warnaar, {\it Characters of graded parafermion conformal field theory}, Adv. Theo. Math. Phys. {\bf 11} (2007)
945-989.



\bibitem{JM}
P. Jacob and P. Mathieu, {\it Graded parafermions: standard and quasi-particle bases}, Nucl. Phys.
{\bf B630} (2002) 433-452.



\bibitem{FJMjpa} 
J.-F. Fortin, P. Jacob and P. Mathieu, {\it $\mathcal {SM}$(2,4$\kappa$)  fermionic characters
and restricted jagged partitions}, J. Phys. A: Math. Gen. {\bf 38} (2005) 1699-1709.

\bibitem{Huse}
D.A. Huse, {\it Exact exponents for infinitely many new multicritical points},
 Phys. Rev. {\bf B30} (1984) 3908-3915.




  \bibitem{JM.A}
  P. Jacob and P. Mathieu, {\it Parafermionic derivation of the  Andrews-type multiple sums} J. Phys. A: Math. Gen. {\bf 38} (2005) 8225-8238.
  
\bibitem{Jaca}
P. Jacob and P. Mathieu, {\it A Quasi-particle description of the $\M ( 3 , p)$
  models}, {Nucl. Phys.}  {\bf B733} (2006)  205--232.
  
  \bibitem{Jacb}
P. Jacob and P. Mathieu, {\it Embedding of bases: from the $\M( 2 , 2 \kappa + 1)$ to the $\M
( 3 , 4 \kappa + 2 - \delta )$ models},
{Phys. Lett.} {\bf B635} (2006) 350--354.


\bibitem{JPath}
P. Jacob and P. Mathieu, {\it Jagged partitions and lattice paths}, math.CO/0605551, Ann. Comb., to appear.


\bibitem{Mult}
P. Jacob and P. Mathieu, {\it  Multiple partitions, lattice paths and a Burge-Bressoud-type  correspondence}, math.CO/0609001, Discrete Math., to appear.


\bibitem{Path}
P. Jacob and P. Mathieu, {\it Paths for  $\z_k$ parafermionic models}, Lett. Math. Phys. (2007) {\bf 81} 211-226. 

\bibitem{KKMMa}
R. Kedem, T.R. Klassen, B. M. McCoy and E. Melzer, {\it Fermionic quasi-particle representations for characters of ${(G^{(1)})_1  \times (G^{(1)})_1 / (G^{(1)})_2}$}, 
Phys. Lett. {\bf B304} (1993) 263-270

\bibitem{KKMMb}
R. Kedem, T.R. Klassen, B. M. McCoy and E. Melzer, {\it Fermionic sum representations for conformal feld theory characters},  Phys. Lett. {\bf B307} (1993) 68-76.


\bibitem{Kir}
A.N. Kirillov, {\it Dilogarithm Identities}, Prog. Theo. Phys. Suppl. {\bf 118} (1995) 61-142.


 
 \bibitem{Mel}
 E. Melzer, {\it Supersymmetric analogs of the Gordon-Andrews identities, and related TBA systems}, hep-th/9412154.
 
 
\bibitem{Nahm}
 W. Nahm, A. Recknagel, M. Terhoeven, {\it 
Dilogarithm identities in conformal field teory}, Mod. Phys. Lett. {\bf A8} (1993) 1835-1848.

\bibitem{Nak}
T. Nakanishi, {\it Nonunitary minimal models and RSOS models}, 
 Nucl. Phys. {\bf B 334} (1980) 745-766.


\bibitem{Rig}
H. Riggs, {\it Solvable lattice models with minimal and nonunitary critical behavior in two-dimensions},  Nucl. Phys. {\bf B 326} (1989) 673-688.



 \bibitem{OleJS}
 S. O. Warnaar,
\textit{Fermionic solution of the Andrews--Baxter--Forrester model. I.
Unification of CTM and TBA methods},
J. Stat. Phys. \textbf{82} (1996) 657--685.

\bibitem{OleJSb}
S. O. Warnaar,
\textit{Fermionic solution of the Andrews--Baxter--Forrester model. II.
Proof of Melzer's polynomial identities},
J. Stat. Phys. \textbf{84} (1996), 49--83. 

  
   \bibitem{Ol}
 S.O. Warnaar, {\it The generalized Borwein conjecture. II. Refined $q$-trinomial coefficients}, Discrete Math. {\bf 272} (2003) 215-258.
 
 \bibitem{Wel}
T. Welsh, {\it Fermionic expressions for minimal model Virasoro characters},  
Memoirs of the American Mathematical Society, vol. {827}, AMS, RI (2005).

 
 \bibitem{ZF}
A~Zamolodchikov and V~Fateev,
{\it Nonlocal (Parafermion) Currents in Two-Dimensional Conformal Quantum
  Field Theory and Self-Dual Critical Points in $Z_N$-Symmetrical Statistical
  Systems},
 Sov. Phys. JETP, {\bf 62} (1985) 215--225.
 
 
  \end{thebibliography}
\end{document}